\documentstyle [12pt,epsfig]{article}

\makeatletter
\@addtoreset{equation}{section}
\makeatother

\newcommand{\BS}{\bigskip}
\newcommand{\SECTION}[1]{\BS{\large\section{\bf #1}}}

\begin{document}
\begin{titlepage}
\begin{center}
\vspace*{2cm}
{\large \bf Quantum Mechanics in Space--Time: the Feynman Path Amplitude Description
  of Physical Optics, de Broglie Matter Waves and Quark and Neutrino Flavour Oscillations}
\vspace*{1.5cm}
\end{center}
\begin{center}
{\bf J.H.Field }
\end{center}
\begin{center}
{ 
D\'{e}partement de Physique Nucl\'{e}aire et Corpusculaire
 Universit\'{e} de Gen\`{e}ve . 24, quai Ernest-Ansermet
 CH-1211 Gen\`{e}ve 4.
}
\end{center}
\vspace*{2cm}
\begin{abstract}
  Feynman's laws of quantum dynamics are concisely stated, discussed in 
  comparison with other formulations of quantum mechanics and applied to selected 
  problems in the physical optics of photons and massive particles as well as
 flavour oscillations. The classical wave theory of light is derived from these laws
 for the case in which temporal variation of path amplitudes may be neglected, whereas specific
  experiments, sensitive to the temporal properties of path amplitudes, are suggested. 
  The reflection coefficient of light from the surface of a transparent medium is found
  to be markedly different to that predicted by the classical Fresnel formula. Except
  for neutrino oscillations, good agreement is otherwise found with previous calculations of
  spatially dependent quantum interference effects.             
\end{abstract}
\vspace*{1cm}
\newline
{\it Keywords ;} Quantum Mechanics,
Physical Optics.
\end{titlepage}
  
\SECTION{\bf{Introduction}}
  In 1983, Richard Feynman was invited to give the first series of Alix J. Mautner Memorial
 Lectures at the University of California, Los Angeles. These lectures were subsequently
 published as the book: `QED The Strange Theory of Light and Matter'~\cite{Feyn1}. They are
 a masterpiece of popular science. The experiment Feynman chose to describe in detail was
  reflection from, and transmission of light through, glass sheets of various
 thicknesses, at near-normal incidence . Closely analogous experiments were first performed and analysed by
 Isaac Newton~\cite{Newton}\footnote{ These experiments actually studied interference
 effects of light crossing a thin air film trapped between spherical and plane 
 glass surfaces --the well-known `Newton's Rings'. The quantum mechanical analysis
 of these experiments is essentially identical to that of those discussed by Feynman in~\cite{Feyn1}.
 The jacket of the first edition of this book showed a coloured photograph of 
 thin film interference effects produced by a thin oil slick.}. After a simplified analysis,
 the complete quantum mechanical calculation for the experiments is presented in Chapter 3 
 of~\cite{Feyn1}. This is done in terms of `arrows' (vectors in the complex plane)
  that represent quantum amplitudes in one-to-one
 correspondence with paths where a photon is scattered from each atom of the sheet of glass.
 No equations are used. To the present writer's best knowledge, this is the only place, in
 Feynman's writings or elsewhere, where this calculation may be found. When Feynman had
 earlier discussed the reflection and transmission of light in `The Feynman Lectures in 
  Physics'~\cite{Feyn2} he had presented the standard text-book analysis in terms of surface
  boundary conditions on the electric and magnetic fields of electromagnetic waves, 
 leading to the Fresnel formulae 
 for the reflection and transmission coefficients. 
 \par The principal aim of the present paper is to work out, in full mathematical detail,
  the QED predictions for some of the experiments discussed in~\cite{Feyn1}. It is found, perhaps 
   surprisingly, that the reflection coefficient of light at a vacuum/glass interface, calculated
   by the path amplitude method described in Chapter 3 of~\cite{Feyn1}, is in marked disagreement
  with the Fresnel formula! 
  \par The essential ideas whose applications were developed in~\cite{Feyn1} were already
  concisely stated~\cite{Feyn3} in `The Feynman Lectures in Physics'. In describing the path
 amplitude interpretation of quantum mechanics in these lectures however~\cite{Feyn4}, Feynman
 prefered to discuss non-relativistic electrons rather than photons, in correspondence with
 his published work on the Path Integral approach~\cite{Feyn5,Feyn6} that was restricted
 to non-relativistic quantum mechanics. With the exception of the short passage cited
 above with the revelatory title `How it Works' light was always described in~\cite{Feyn2} 
 in terms of electromagnetic waves, not photons. 
 \par Feynman stated that the path integral formulation of quantum mechanics is not
 only the most fundamental, but also the easiest to understand. Why then has its impact,
 both on the teaching of physics~\footnote{An important exception to this is the work of 
   E.F.Taylor and collaborators~\cite{EFT} who have stressed the insight into classical
   mechanics and Hamilton's Principle provided by Feynman's formulation of quantum mechanics.}
   and on the presentation of quantum mechanical subjects in the research literature
  been so limited? This is true even for subjects where the path amplitude approach 
  would seem to be the most natural one, such as physical optics and quark flavour and
   neutrino oscillations in particle physics. The answer to this question is perhaps
  related to the fact that the situations in which the path amplitude formalism
  is both transparent and powerful --quantum predictions for correlated observations
  in space-time, which may be termed `quantum dynamics', is far removed from the problem
   that was confronted and solved by the founders of quantum mechanics: Bohr, Heisenberg,
   Schr\"{o}dinger and Pauli. This problem was to describe the structure of atoms and to
   calculate the rates of radiative transitions between different atomic energy levels,
   a subject which could be termed `quantum statics' since the detailed space-time
   structure of fundamental processes does not play any important role
    \footnote{Bohr made the following remark on this subject:
     `For example the experiments regarding the excitation of spectra 
    by electronic impacts and by radiation are adequately accounted for on the 
   assumption of discrete stationary states and individual transition
   processes. This is primarily due to the circumstance that in these questions
   no closer description of the space-time behaviour of the processes is required.'
   ~\cite{BohrCI}.}. Indeed the latter
   subject is not discussed at all in the vast majority of text-books on 
  quantum mechanics\footnote{ This is not the case for~\cite{LLFB}, an introductory
  text book largely based on Feynman's space-time approach. As in Section 2 below,
   Feynman's rules for constructing probability amplitudes are explicitly stated
   in this book.}.
   If future research physicists are never systematically taught Feynman's formulation
  of quantum mechanics it is not surprising that they remain forever ignorant of it. 
    When some brief discussion of, say, particle
  propagation in space-time, is included in standard text-books, concepts such as
  spatial `wave packets', relevant only to certain classes of physical problems, 
  are often introduced, in a prefectly general way, leading, in some cases, to erroneous conclusions.
  This point will be 
   considered in more detail in the following section. 
  \par Feyman's formulation of quantum mechanics is more fundamental and powerful
   than the earlier formulations of Heisenberg and Schr\"{o}dinger because, as Feynman
   showed, starting from the path integral formula Schr\"{o}dinger's
   equation and therefore all of `quantum statics' is readily derived. However 
    knowledge of  Schr\"{o}dinger's equation alone is, by far, not sufficient to obtain
   correct predictions for space-time correlated events. This is demonstated by the
   different predictions for `neutrino oscillations' to be found in the published 
   literature and text books~\cite{JHF1,JHF2,JHF3}.
   \par Before the work presented in the present paper, the present writer is aware 
   of only two papers in which the Feynman path integral formalism is formally 
   applied to optical problems. The first~\cite{BGdeG} concerns the refraction
   of light and the second~\cite{BB} the optics of beams of non-relativistic particles.
   This being said, it remains true that in the description of almost any experiment in
   quantum optics involving multiple photon paths, path amplitudes are tacitly employed,
   usually called instead `photon fields', `wave functions' or `state vectors'. The most detailed
   explicit application of the Feynman path integral method known to the present 
   writer is in atomic physics: the `Photodetachement Microscope'~\cite{BBD}.
    \par In all of the optical experiments to be described in the following, as is the 
    case in~\cite{Feyn1}, the initial state is assumed to be a single
    excited atom which subsequently decays spontaneously producing a single photon.
    This photon interacts with the experimental apparatus and later the same, 
    or a different, single photon is detected. Also as in~\cite{Feyn1}, for simplicity,
    only experiments where the effects of atomic or photon polarisation are inessential, and 
    may be neglected, are considered.
    \par The structure of the paper is as follows: In the following section Feynman's principles
    of quantum dynamics and their interpretation are reviewed. Thus the basic formulae used
   to perform the calculations presented in the subsequent sections are given. In Section 3
    the invariant propagator of a free on-shell particle is derived from the relativistic
    Feynman path integral, and the temporal propagator of a particle
    at rest from the time-dependent Schr\"{o}dinger Equation.
   In Section 4 the classical wave theory of light in which only
    spatial `waves' are considered is derived from the Feynman path integral. The calculation,
     from first principles, of the refractive index of a uniform transparent medium is 
    performed in Section 5. A new phenomenon `refraction annulment' which may occur when the
    excited atom decays promptly after production, is predicted. The laws of reflection, refraction
    and linear propagation of light are derived in Section 6. A quantitative estimation is made
   of the spread of the photon paths around the classical trajectory of ray optics corresponding
   to an extremum of the phase of the path amplitude. In Section 7 the calculation of the
    reflection coefficient of light at normal incidence at the interface of two transparent
   media, as qualitatively sketched in~\cite{Feyn1} is presented.  In Section 8 time-dependent
   variation of the fringe visiblity in a Michelson Interferometer, using as a light source
   spontaneously decaying excited atoms produced at a known time, is calculated.
   This type of prediction, testing
   essential temporal features of the path amplitude formalism, cannot be obtained using the
   classical wave theory of light. In Section 9 the{\ path amplitude method is extended to the
    description of de Broglie matter waves and heavy quark and neutrino flavour oscillations.
    In each case a `two probability amplitude' experiment is analysed in detail. These are:
    Young double slit experiments using photons or electrons, neutral kaon flavour oscillations
   in the processes $\pi^- p \rightarrow \Lambda ({\rm K}_S,{\rm K}_L)$, ${\rm K}_S,{\rm K}_L
    \rightarrow \pi^{\pm} e^{\mp} \nu$ and neutrino oscillations in the processes:
    $\pi^+ \rightarrow \mu^+ \nu$, $\nu n \rightarrow e^- p$.
    A summary and outlook are given in Section 10.
    \par As this paper is a long one, the reader is recommended to look first at the concluding
     section for an overview, returning later to any earlier sections that contain material of
     particular interest.

\SECTION{\bf{Feynman's Formulation of the Laws of Quantum Dynamics and Their
    Interpretation}}
 In Chapter 3 of Volume 3 of the `The Feynman Lectures in Physics'~\cite{Feyn2} Feynman set down
 four general principles for the quantum description of space-time processes. These are rules
  concerning the meaning and the method of construction of the {\it Probability Amplitude} 
  that provides the quantum description of any space-time experiment. Any such experiment is
  defined by a fixed (prepared) initial state $|i\rangle$ and a fixed (measured) final state
  $|f\rangle$. The probability amplitude, $A_{fi}$, is a complex number that is constructed
   by summing
  all {\it Path Amplitudes} that have {\it the same} initial state $|i\rangle$ and {\it the same}
  final state $|f\rangle$.
  The path amplitudes, $PA_{fi}$, are, in turn, constructed by multiplying together a time-ordered sequence
  of {\it Process Amplitudes}.  Feynman's four principles then define the physical meaning
  of the probability amplitude and specify how it is constructed:
  \begin{itemize}
  \item[{\bf I}] The probability, $P_{fi}$, to measure the state,  $|f\rangle$, given the
    prepared state,  $|i\rangle$, is:
   \begin{equation}
      P_{fi} = \left|A_{fi}\right|^2
  \end{equation}
  \end{itemize}
   In general the probability amplitude depends on the space-time evolution of the state $|i\rangle$, 
   which is a function of the spatial geometry and other physical characteristics of the
   experimental apparatus as well as of those of various process amplitudes where particles may be
    scattered,
   destroyed or created. It therefore does not correspond simply to any observable property
   of either the initial or final state and so should not be confused with the wavefunction
   of any quantum system.
  \par The path amplitude is constructed according to the law of {\it Sequential Factorisation}.
    If the state  $|i\rangle$  evolves in time via a series of {\it unobserved 
   intermediate states}: $|k_j\rangle,~~j=1,n$, the corresponding  path amplitude:
    $\langle f|k_n,...,k_2,k_1|i \rangle$ is given by a product of process amplitudes:  
   $\langle k_{j+1}|T_{j+1}|j\rangle$, where $T_{j+1}$ is a transition
    operator that evolves the state  $|k_j\rangle$ into  $|k_{j+1}\rangle$. These process amplitudes
    are either the space-time propagators of particles or the invariant amplitudes of particle
    decay or scattering \footnote{The scattering process may be elastic or inelastic.} processes
     that are calculated in momentum space according to the Feynman rules of QED or the
     Standard Electroweak Model. Since the latter have no space-time dependence they appear only
    as constant multiplicative factors in the path amplitudes. It is understood that all unobserved
    particle states in the process amplitudes that are not members of the chain
    of sequential states $|k_j\rangle,~~j=1,n$ that define the amplitude, are to be integrated over.
    Since the corresponding integrals are space-time independent that contribute only a 
    multiplicative constant to $P_{fi}$\footnote{In the examples discussed in the present 
   paper no account is taken of these integrations. The calculated space-time quantum interference
   effects are unaffected by this omission}.  
 \begin{itemize}
  \item[{\bf II}] Sequential Factorisation: 
   \begin{equation}
   PA_{fi} \equiv \langle f|k_n,...,k_2,k_1|i \rangle = \langle f |T_f|k_n\rangle...
    \langle k_2 |T_2|k_1\rangle \langle k_1 |T_1|i\rangle 
   \end{equation}
 \end{itemize}
    \par The third principle is {\it Quantum Mechanical Superposition}. If several different series of
    unobserved intermediate states are allowed by the experimental apparatus, the 
   probability amplitude is given
    by the sum over all such series of intermediate states:
 \begin{itemize}
  \item[{\bf III}]  Quantum Mechanical Superposition:
   \begin{equation}
    A_{fi} = \sum_{k_n}...\sum_{k_2}\sum_{k_1} \langle f|k_n,...,k_2,k_1|i \rangle
    \end{equation}
 \end{itemize} 
    \par The fourth principle applies if the initial and final states are composite,
  constructed from the tensor products of sets
 of states: $|i^{(1)}\rangle$, $|i^{(2)}\rangle$...
    $|f^{(1)}\rangle$, $|f^{(2)}\rangle$... but the path amplitudes linking the pairs 
    $|i^{(1)}\rangle$,$|f^{(1)}\rangle$; $|i^{(2)}\rangle$,$|f^{(2)}\rangle$;... have no intermediate
    states in common:
  \begin{itemize}
  \item[{\bf IV}]  Composite Factorisation:
   \begin{equation}   
    PA_{f^{(1)}f^{(2)}...i^{(1)}i^{(2)}...} = \prod_j PA_{f^{(j)}i^{(j)}} 
    \end{equation}
 \end{itemize} 
    An important special case of this principle occurs when a common prepared initial state
    gives rise to a number of separately measured final states; i.e.  $|i^{(1)}\rangle, |i^{(2)}\rangle,...
     \rightarrow |i\rangle$, in (2.4), but the paths linking $|i\rangle$ to
      $|f^{(1)}\rangle, |f^{(2)}\rangle,..$ are distinct, with no common intermediate states.
     This results, when quantum experiments containing identical particles are considered,
    in what is conventionally termed an `entangled wavefunction'. Examples are the decays
   of ortho- and para-positronium:
   \[ e^+e^-\left(1^3S_1\right) \rightarrow \gamma \gamma \gamma~,~~
    e^+e^-\left(1^1S_0\right) \rightarrow \gamma \gamma \]
     In these and similar processes quantum statistics must be respected.
     In the case of the observation of both photons from the decay of para-positronium,
     two different path amplitudes are possible:
     \begin{eqnarray}
     PA(12)_{f^{(1)}f^{(2)}i} & = & \langle f^{(1)}|T_D|\gamma_1\rangle \langle\gamma_1 |T(P_1)|\gamma_1\rangle
                           \langle\gamma_1 |T_{P}| i\rangle  \nonumber \\
                      &   & \times  \langle f^{(2)}|T_D|\gamma_2\rangle \langle\gamma_2 |T(P_2)|\gamma_2 \rangle
                           \langle \gamma_2 |T_{P}| i \rangle  \\
     PA(21)_{f^{(1)}f^{(2)}i} & = & \langle f^{(1)}|T_D|\gamma_2\rangle \langle\gamma_2 |T(P_1)|\gamma_2\rangle
                           \langle \gamma_2 |T_{P}| i\rangle  \nonumber \\
                      &   &  \times \langle f^{(2)}|T_D|\gamma_1\rangle \langle \gamma_1 |T(P_2)|\gamma_1 \rangle
                           \langle \gamma_1 |T_{P}| i \rangle  
     \end{eqnarray}
    In these formulae, $\gamma_1$ and  $\gamma_1$ are the two identical and indistinguishable decay
   photons, $|i\rangle$ the initial para-positronium state, $|f_1\rangle$ and $|f_2 \rangle$ 
   the final states of the
   photon detection processes, occuring at the ends of the paths, $P_1$ and $P_2$, through the experimental
   apparatus. $T_P$ and $T_D$ are the transition operators of the positronium decay and photon 
   detection processes respectively, while $T(P_1)$ and $T(P_2)$ are photon transition 
   operators for the paths  $P_1$ and $P_2$ respectively. Bose-Einstein statistics requires that the 
   path amplitude, $PA_{f^{(1)}f^{(2)}i}$, for the experiment is symmetric under
   the exchange $\gamma_1 \leftrightarrow \gamma_2$.
   Thus 
  \begin{equation}
    PA_{f^{(1)}f^{(2)}i} =  PA(12)_{f^{(1)}f^{(2)}i} +  PA(21)_{f^{(1)}f^{(2)}i}
   \end{equation}
   In an experiment where two identical fermions are both detected Fermi-Dirac statisics
    requires the path amplitude to be anti-symmetric under exchange of the two fermions. The similar
    path amplitudes  $PA(12)_{f^{(1)}f^{(2)}i}$ and  $PA(21)_{f^{(1)}f^{(2)}i}$ should then be subtracted rather than
   added as in (2.7), to yield the path amplitude $PA_{f^{(1)}f^{(2)}i}$ for the experiment.
 
  \par In practice, experimental set-ups do not usually correspond to the preparation
   and measurement of a unique intial state and a unique final state, but rather sets of
    such states: $|i_l\rangle,~l =1,2...$ ($I$);~  $|f_m\rangle,~m =1,2...$ ($F$) are
   prepared or measured, respectively. The appropriate prediction to be compared with
   experiment is then:
   \begin{equation} 
      P_{FI} = \sum_m  \sum_l\left| \sum_{k_n}...\sum_{k_2}\sum_{k_1}
      \langle f_m|k_n,...,k_2,k_1|i_l \rangle\right|^2
    \end{equation}
  The incoherent sums over $l$  and $m$  correspond to the density matrices of 
   initial and final states respectively.
   (2.8) is the basic formula that is used for the calculations presented below. It is
    a simple iteration of a formula first given by Heisenberg in 1930~\cite{Heis1}
   which was later adopted by Feynman as the basis for his space-time formulation
   of quantum mechanics~\cite{Feyn5}.
     \par The principles I-IV above describe the rules for constructing probability
     amplitudes, but are devoid of any dynamical content. This is provided by a fifth
     principle, the Feynman Path Integral, which gives the physical prescription
    to calculate the probability amplitude:
  \begin{itemize}
  \item[{\bf V}]  The Feynman Path Integral~\cite{FPI}:
   \begin{equation}   
     A_{fi} = \int_{\rm paths} \langle f|\exp\left\{i \frac{S[\vec{x}_p(t)]}{\hbar}\right\}
     | i\rangle \prod_p \prod_{j=1}^3 {\cal D}(x_p^j(t))
    \end{equation}
     where\footnote{Here $\vec{x}_p \equiv (x_p^1,x_p^2,x_p^3)$
    and the dot denotes a time derivative} 
    \[ S \equiv \int_{t_i}^{t_f} L(\vec{x}_1, \dot{\vec{x}}_1,\vec{x}_2,\dot{\vec{x}}_2,...
   \vec{x}_p,\dot{\vec{x}}_p,...) dt \]
    and
    \[ {\cal D}(x) = Lim(\epsilon \rightarrow 0) \int \int ...\int \int\frac{dx_0}{A}
      \frac{dx_1}{A}... \frac{dx_{j-1}}{A} dx_j \]
    \end{itemize}
     where $x_0,x_1,...$ denote successive positions along the
     path each separated by a small fixed time interval, $\epsilon$,  and  A is a normalisation constant.
     The latter will be determined for the case of a 
      free, relativistic, on-shell particle in the following Section. 
     The function $L$ is the {\it classical} Lagrangian of the system whose quantum mechanical
     behaviour is to be described, and $S$ the corresponding classical Action.
     \par In the following, (2.9) is not used {\it per se} to calculate the entire probability
      amplitude but rather the individual process amplitudes, that are combined according to
      Feynman's second principle, to obtain the complete path amplitude for the
      experiment under discussion. The most important of these process amplitudes is the Green function,
     or space-time propagator, for an on-shell photon or an excited atom.
      These are derived in the following section, the former from
     (2.9) and the latter, more conveniently, from the time-dependent
      Schr\"{o}dinger equation, shown by Feynman~\cite{Feyn5,Feyn6} to be equivalent, in
      the non-relativistic limit, to (2.9).
      In his book on quantum mechanics Dirac used, conversely, the Schr\"{o}dinger equation
      to derive (2.9)~\cite{Dirac1}.
      \par The physical interpretation of Feynman's principles I-V will now be discussed.
    In typical treatments of `The Interpretation of Quantum Mechanics' in the literature
     it is customary to attempt to reduce the problem to its bare essentials by avoiding
    explicit reference to any actual application. The physical meaning of Hilbert space
    vectors, or wavefunctions, is discussed in complete generality, the dynamical content
    being limited to the `unitary evolution' of fixed energy solutions of the
    time-dependent Schr\"{o}dinger equation. The danger of such a procedure is that of
    over-simplification. Instead of interpreting how quantum mechanics describes actual
    experiments, what is done is to instead `interpret' a simple mathematical model, that
    does not necessarily reflect the actual complexity of a real experiment.
    To avoid this danger here, the discussion focuses specifically on the type of experiment
    treated in~\cite{Feyn1}:
    \begin{itemize}
     \item[(i)] An excited atom is created.
     \item[(ii)] The atom decays spontaneously, emitting a single photon.
      \item[(iii)] The photon interacts with the experimental apparatus.
      \item[(iv)] A photon (not necessarily the decay photon) is detected.
     \end{itemize}
     First consider the simpler sequence (i), (ii) and (iv) only. The atom is created,
     it decays, and the photon propagates through free space to the detector. If the detector
     has a surface area, $S$, perpendicular to the  space vector, $\vec{r}$, drawn from the 
     excited atom to the detector, and the atom is unpolarised, there is no need to 
     use quantum mechanics (or classical mechanics either) to calculate the time-integrated
     probability that the photon will be detected. All that is needed is a knowledge
   three-dimensional spatial geometry. The answer is: $\epsilon_D S/(4 \pi r^2)$, where 
   $\epsilon_D$ is the efficiency of the photon detector. In order to calculate the 
   probabilty, $ \delta P(t_D)$, that the photon will be detected in the time
      interval $\delta t_D$ at $t_D$,
    quantum mechanics is needed, but only to calculate the mean lifetime, $\tau_S$, of the
    excited source atom. Suppose that the excited atom is produced by the passage
   of a pulsed laser beam tuned to the frequency $(E_i-E_f)/h$ where $E_i$ and $E_f$ are the
    excited and ground state atomic energy levels. The laser pulse is assumed to have a
     Gaussian form with variance $\sigma_t$. If the mean time of passage of the exciting
    pulse is $t_0$ and the production time of the excited atom is $t_P$, the above-mentioned probability is: 
 \begin{eqnarray}
  \delta P(t_D) & = & \sqrt{\frac{2}{\pi}}\left( \frac{\epsilon_D S}{4 \pi r^2}\right)
    \frac{\sigma_t}{\tau_S}\int_{-\infty}^{\infty} 
    \exp\left[-\frac{1}{\tau_S}(t_D-t_P-\frac{r}{c}) - \frac{(t_P-t_0)^2}{2\sigma_t^2}\right]
     dt_P \delta t_D  \nonumber \\
    & = & \frac{\epsilon_D S}{4 \pi \tau_S r^2}
     \exp\left[-\frac{1}{\tau_S}(t_D -t_0-\frac{r}{c}-\frac{\sigma_t^2}{2\tau_S})\right]
       \delta t_D 
  \end{eqnarray}
   The only physical postulates necessary to derive this result are an exponental decay law
   for the excited atom and a constant velocity, c, for the decay photon.
   \par It will now be instructive to re-derive this result using Feynman's principles.
    The corresponding path amplitude is, using II:
  \begin{equation}
    PA_{fi} = A_D \langle D|\gamma|\gamma\rangle\langle\gamma|T|A_i\rangle\langle\gamma|A_i|P\rangle A_P
   \end{equation}
    The five process amplitudes have the following meaning:
    \begin{itemize}
   \item $A_P$ is the production amplitude for process (i) above.
    \item $\langle\gamma|A_i|P\rangle$\footnote{The notation used for space-time propagators
    is that $\langle f |P|i\rangle$ is the amplitude for a particle (or other quantum object, such as an atom or
   molecule), $P$, initially at space-time point $x_i$, to be found at $x_f$.} is the process amplitude for
     the propagation of the excited atom from its production time $t_P$ to its decay time
    $t_{\gamma}$ (see (3.20) below):
  \begin{equation}
   \langle\gamma|A_i|P\rangle = \exp\left[-\frac{i}{\hbar}
    (E_i^0-E_f^0-i\frac{\hbar}{2 \tau_S})(t_{\gamma}-t_P)\right]
   \end{equation}
    $E_i^0$ and $E_f^0$ are the `pole center-of-mass energies' (see Section 3 below) of the atomic states
    $i$ and $f$, and $\tau_S$ is the mean lifetime of the state $i$.
    \item $\langle\gamma|T|A_i\rangle$ is the invariant amplitude for the atomic transition:
    $A_i \rightarrow A_f + \gamma$. 
     \item  $\langle D|\gamma|\gamma\rangle$  is the propagator of the photon from its 
    production time $t_{\gamma}$ to its detection time $t_D$ (see (3.11) below):
    \begin{equation}
    \langle D|\gamma|\gamma\rangle = \frac{1}{r}\exp\left[-\frac{i}{\hbar}
     [(E_{\gamma}(t_D-t_{\gamma})-p_\gamma r]\right] =  \frac{1}{r}
   \end{equation}

     The last member of (2.13) follows since the photon is a massless particle with constant velocity $c$:
     $c = E_{\gamma}/p_{\gamma} = r/(t_D-t_{\gamma})$.
 \item $A_D$ is the detection amplitude for process (iv) above.
   \end{itemize}
      Also required for the calculation is the density distribution, $\rho_P$, for 
    production of the excited 
    atom and the density, $\rho_D$ of detected final states:
  \begin{eqnarray}
   d\rho_P & = & \sqrt{\frac{2}{\pi}}\sigma_t \exp\left[ - \frac{(t_P-t_0)^2}{2\sigma_t^2}\right] dt_P  \\
       \rho_D & = &  \delta t_D
  \end{eqnarray}
   Using I, noting that $t_{\gamma} = t_D-r/c$, and integrating over $t_P$ gives:
 \begin{eqnarray}
  \delta P(t_D)^{PA} &=& \int | PA_{fi}|^2 d\rho_P \delta t_D~~~~~~~~~~~~~~~~~~~~~~~~~~~~~~~~ \nonumber \\
     &=& \sqrt{\frac{2}{\pi}}\sigma_t \frac{A_0^2}{r^2}                                
   \int_{-\infty}^{\infty}
    \exp\left[-\frac{1}{\tau_S}(t_D-t_P-\frac{r}{c})
     - \frac{(t_P-t_0)^2}{2 \sigma_t^2}\right] dt_P \delta t_D \nonumber \\
    &=&  \frac{A_0^2}{r^2}\exp\left[-\frac{1}{\tau_S}(t_D -t_0-\frac{r}{c}-\frac{\sigma_t^2}{2\tau_S})\right]
       \delta t_D    
 \end{eqnarray}
 where \[ A_0^2 \equiv |A_D|^2 |\langle\gamma|T|A_i\rangle|^2 |\langle\gamma|T|A_i\rangle|^2 |A_P|^2 \]     
   Comparing with the (2.10) shows that, for consistency:
    \begin{equation}
    |A_D|^2|\langle\gamma|T|A_i\rangle|^2|\langle\gamma|T|A_i\rangle|^2 |A_P |^2 =
    \frac{\epsilon_D S }{4 \pi \tau_S}    
  \end{equation}
     This is perfectly reasonable since $ |\langle\gamma|T|A_i\rangle|^2 \simeq  \Gamma_S = \hbar/ \tau_S$,
     and $ |A_D|^2  \simeq S$ so that the photon detection probability is
    proportional to the area of the detector.
    \par The above comparison has shown the consistency of the calculation based on Feynman's
    quantum mechanical principles with the `common sense' result (2.10). This example will
     now be compared with the typical description of the free space propagation of particles
    to be found in text books on quantum mechanics, quantum field theory, or optics.
    For example, Dirac writes down the wavefunction of a free particle in an energy-momentum
    eigenstate as~\cite{Dirac2}:
     \begin{equation}
      \psi(xyzt)= a_0 \exp[-i\frac{p \cdot x}{\hbar}] = 
      a_0 \exp[-\frac{i}{\hbar}(Et-p_x x  -p_y y - p_z z)]
    \end{equation}
    where $x =(ct,\vec{x})$ and $p = (E/c, \vec{p})$ are the space-time and energy-momentum
     4-vectors of the particle. Dirac then states that such a `plane wave' wavefunction
    has no physical significance, since it predicts the particle to have the same
   probability to be at any point in space-time\footnote{Actually, since the wavefunction
   of (2.18) is not square-integrable, the normalisation constant $a_0$ vanishes, as does also
   the probability that the particle is within any finite volume of space-time}. A similar description
   the photon in the above example would be a radial wave function of the form
    $ \exp[-i p \cdot x /\hbar]/r$ which would predict that the photon has the same 
    probability to be found at any distance from the source at any time. Such `wavefunctions'
    are evidently meaningless and bear no relation to the actual physical situation
    correctly described by (2.10) or (2.16). Dirac indeed agrees that such a wavefunction is
    meaningless, but suggests that the solution to the problem is to replace the
    plane wave in (2.18) by a spatial `wave packet'. This is then claimed to give a correct
    quantum description of actual free particles. It is argued that since the plane wave 
    in (2.18) is an eigenstate of $E$ and $p$, that it is natural from the
     Heisenberg Uncertainty Relations: $\Delta p \Delta x \ge \hbar$ and
   $\Delta E \Delta t \ge \hbar$ that that both $x$ and $t$ should be completely
     uncertain. Since it is certainly true that actual particles are never
     produced in an eigenstate of $E$ and $p$ it is further argued that they must
    therefore be produced with spatial wavepackets that are coherent superpositions
    of different $E$, $p$ eigenstates. Indeed the photons produced by the decay of the
   excited atom in the above example do have an energy distribution of non-vanishing
    width. Nevertheless, {\it there is no associated spatial wavepacket}.
     The absurdity of describing the photon in the above example by
     a wave packet of the type suggested by Dirac, becomes evident on examining the physical
     consequences of such a hypothesis. Suppose that $\tau_S$ has the typical value
     of $10^{-8}$ sec. The correponding spread, $\Delta p$, in the momentum of the photon
     is given correctly by the energy-time Uncertainty relation as $c\Delta p = \Delta E
     \simeq \Gamma_S = \hbar/\tau_S$\footnote{It is here assumed that
     the lower energy level of the atomic transition that produces the photon is
    the ground state, with infinite lifetime}. According the the momentum-space Uncertainty Relation
    invoked by Dirac, the width, $\Delta x$, of the associated spatial wavepacket should
    be: $\Delta x \simeq \hbar/\Delta p  \simeq c \tau_S \simeq 3$m. If such a wavepacket
    were actually produced in the decay of the excited atom in the above example,
    quantum mechanics would predict that it could be instantaneously detected at a distance
    of, say, 1m from the production point, in contradiction to the predictions of
    (2.10) and (2.16) above.  Using, say, a photo-diode to detect the photon, in the above     
     example, its spatial position is easily measured with a precision of $\Delta x \simeq 1$mm.
    This implies that, at the moment of detection, the product $\Delta x \Delta p$ for the
    photon is known with an accuracy $3 \times 10^3$ better than allowed by the Uncertainty
     Principle. The latter is evidently not applicable in this case. 
     \par The point is that the relation $\Delta x \Delta p \ge \hbar$ 
     applies either to {\it simultaneous measurements} of $x$ and $p$, as in the examples 
    orginally discussed by Heisenberg~\cite{Heis2}, or to the characteristic widths of the  
    spatial and momentum distributions of a bound state wavefunction, or the wave packets
   which correctly describe a beam of massive particles such as electrons or neutrons. These distributions
    are related by a Fourier transform, from which the corresponding Uncertainty Relation
    is easily derived. In the example above, knowledge of the excited state of the atom provides
    {\it prior knowledge} of the momentum uncertainty of the photon as $\hbar/(c\tau_S)$ where
    $\tau_S$ is the previously measured or theoretically calculated mean life of the
    excited atom. No measurement is required to know the photon momentum  with this 
    accuracy\footnote{Here incoherent line broadening effects due, for example, to the
    Doppler effect, are neglected, so the atom is supposed to be isolated and at low temperature.}
    \par What is not taken into account in Dirac's discussion, and that of other text book
    authors is that, in quantum mechanics,
   real (`on-shell') particles propagate over macroscopic distances in a classical manner.
   This fact is taken fully into account in the Feynman space-time formulation of
   quantum mechanics, and, as will be demonstrated in the examples worked out later in the present
   paper, is  crucial in deriving its predictions. As will be seen, many of the latter are identical
   to those of the classical wave theory of light, that are well verified experimentally.
    The conclusion is that the position
   of a particle {\it that has already been produced} must be defined in a classical manner in order
    to correctly calculate the correponding path amplitude.
     The energy-time Uncertainty Relation $\Gamma_S \tau_S = \hbar$ of course plays an essential
    role in the derivation of the common sense equation (2.10): there is quantum uncertainty
    in the time of decay. Once a photon is produced, however, it propagates over macroscopic
    distances according to the classical law: $\Delta r / \Delta t =c$ within each path amplitude.
    Adding the path amplitudes to construct the probability amplitude modifies, via quantum 
    interference, this simple behaviour, and gives rise to the purely quantum effects of refraction,
    diffraction and interference that will be discussed below.
    \par It is possible to replace Dirac's wavefunction (2.18) by one which does give 
   a meaningful space-time description of the photon produced in the decay of the excited atom. 
     Since the detection process can be considered as a probe of the spatial position of the
    photon, the `photon wavefunction' can be defined by omitting the amplitude $A_D$ in (2.11) 
    and multiplying by a suitable $\delta$-function to describe the classical space-time
    propagation property of the photon. Thus the radial wavefunction of the photon 
    is given by (2.11)-(2.13) as:
  \begin{equation}
  \psi(\vec{r},t,t_{\gamma},t_P) =  \frac{a_0}{r}
   \exp\left[-\frac{i}{\hbar}\left(E_i^0- E_f^0 -\frac{i\hbar}{2\tau_S}\right)
   \left(t-t_P-\frac{r}{c}\right)\right] \delta(r-c(t-t_{\gamma}))
   \end{equation} 
    The wavefunction of the photon  depends not only on the spatial position and the time
    as specified by $\vec{r}$ and $t$, but also on the time at which it is produced $t_{\gamma}$, as well 
   as the production time $t_P$ of the excited atom. Using the Born probabilty rule:
  \[  P = \int |\psi|^2 dV \]
    and the density distributions (2.14) and (2.15), the wavefunction of (2.19) is easily seen to give,
    with a suitable
    choice of the normalistation constant $a_0$, the same result for $\delta P(t)$ as (2.10) or (2.16).

    \par The reader may now be asking why, since the common sense formula (2.10) already
    describes the probability to observe the decay photon at any time, assuming only an
    exponential decay law and that the photon is a massless particle, what is the use of
    Feynman's formulation of quantum mechanics (that gives the same prediction via 
    a more complicated calculation) or indeed of 
    quantum mechanics in any formulation, except for calculating the mean lifetime of the 
    excited atom? The answer lies in the description of space-time experiments where
    principle III (quantum mechanical superposition) plays a crucial role. This is  the 
    case in all the examples worked out in the present paper. The essential information
    provided by the path amplitude concerns not the spatial position of the particle
    (which is calculated classically) but {\it the phase of each path amplitude
    at the instant the final state is measured}. It is these phases which control
    the observed quantum interference phenomena. However, the particle {\it does} move in a classical
    manner along each path. As Feynman strongly emphasised, the difference
    between a classical world and the actual quantum one is that the different paths
    in one-to-one correspondence with different classical histories of the quantum 
    system contribute {\it together, in parallel in time,} according to (2.3), to the
    probability amplitude that describes {\it a single detection  event}. This effect is most
    simply and graphically illustrated by the Young double-slit interference
    experiment, to be analysed in detail in Section 9 below, that Feynman chose to
     exemplify this behaviour~\cite{Feyn7}, but in reality
    not two, but an infinite number of paths each with its own distinct classical
    history, are combined, in parallel, when quantum mechanical superposition operates.
    \par It is interesting to note that the actual physical description provided by Feynman's
    formulation of quantum mechanics contains important elements of several different so-called
   `interpretations ' of quantum mechanics. These are:
    \begin{itemize}
   \item[(a)] `Consistent Histories'
   \item[(b)] Everett's `Many Worlds' interpretation
   \item[(c)] de Broglie's original `Pilot Wave' theory
   \end{itemize}
     Each of these interpretations is now briefly discussed in comparison with Feynman's
     formulation.
    \par `Consistent Histories'~\cite{Griffiths,Omnes,GMH} are usually presented in the language
   of Hilbert space projection and trace operators. In the notation of the present paper, and 
   considering the simplest non-trivial case, with a single intermediate quantum `property'~\cite{Omnes},
   consistent histories are given by summed path amplitudes similar to those in
    (2.3) above, except that the sum over intermediate states is not over the complete set
    allowed by a given experimental configuration, as is understood in (2.3), but is instead
   limited to an arbitary domain, $D_k$. Probabilities are assigned these sets of paths 
   according to a definition similar to Feynman's first principle, (2.1):
   \begin{eqnarray}
    P_{fi}(D_k) & = & |A_{fi}(D_k)|^2 \\
     P_{fi}^G (D_k) & = & \frac{|A_{fi}(D_k)|^2}{|A_{fi}|^2} 
  \end{eqnarray}
   where 
 \begin{equation}
 A_{fi}(D_k) \equiv \sum_{D_k} \langle f|T_f|k\rangle \langle k |T_k|i\rangle 
  \end{equation}

  (2.20) corresponds to the history definition of Omn\`{e}s~\cite{Omnes} and
   Gell-Mann and Zurek~\cite{GMH}, whereas (2.21) is the original definition due to 
   Griffiths~\cite{Griffiths}. The amplitude $A_{fi}$ in the denominator of (2.21)
   is given by replacing the limited domain $D_k$ of (2.22) with the complete set of allowed
   intermediate states as understood in (2.3). Histories defined by different domains
   $D'_k$, $D''_k$ are said to be `consistent' if the probabilites calculated according to
  (2.20) or (2.21) add in a classical manner:
  \begin{equation}
  P_{fi}(D'_k +D''_k) =  P_{fi}(D'_k) +  P_{fi}(D''_k) 
  \end{equation}  
   Conditions for this to be true were first given in the projection operator and
   trace formalism by Griffiths~\cite{Griffiths}. They are equivalent to requiring that
    the interference terms between the corresponding amplitudes $A_{fi}(D'_k)$ and 
    $A_{fi}(D''_k)$ vanish:
   \begin{equation}
  Re\left[A_{fi}(D'_k)A_{fi}(D''_k)^{\ast}\right] = 0
   \end{equation}
   This condition is satisfied if the difference between the phases of $A_{fi}(D'_k)$ and 
    $A_{fi}(D''_k)$ is an odd multiple of $\pi/2$, i.e. if the corresponding vectors in 
   the complex plane are orthogonal. The `consistent' histories are then those for which
   the quantum interference effects resulting from superposition vanish. The `histories'
   associated with the  Feynman path amplitudes that describe actual experiments have,
   of course, no such restriction.
   \par The relation of Feynman's formulation of quantum mechanics to Everett's
    `Many Worlds' or `Relative State' formulation, can be
   seen by comparing their descriptions of the Young double slit interference
   experiment. Suppose that, initially, photon detectors are placed immediately
   behind the two slits and an observer (with a lot of time at his disposal!) records
   the responses of the detectors when single excited atoms are created nearby as
   a photon source. The observer will either note (mostly) that no photon is detected (0)
   or, more rarely, that a photon is detected either behind Slit 1 (1) or Slit 2 (2).
   Then, as suggested by Everett, the results recorded by the observer in sucessive
   repetitions of the experiment can be stored as a string of numbers: e.g. 00010020001... .
 Another observer performing a similar series of experiments might see another sequence:
    00002000100... . For a finite number of repetitions of the experiment there are a
   countable finite number of such sequences, each of which, Everett said correponds to
   a different possible `history of the world', which will be hard-wired into the brain 
  of the observer, or his recording device, at the end of his sequence of measurements.
    The actual history of any given observer is limited to just one such sequence,
    but all of the others describe `alternative worlds' each with the same weight
    as the one actually observed..
   \par Consider now the simplest case (unlikely but not impossible) 
    when the observer in the first world notes a photon behind Slit 1 in his first
   experiment whereas the observer in the second world notes a photon behind Slit 2.
   Suppose now that the detectors are removed from behind the slits and placed in a
    position where a photon passing through either slit may be detected by them.
     There is clearly now a correspondence between the path amplitude for the photon
   to pass through Slit 1 and the `world' of the first observer, and the path
   amplitude for the photon to pass through Slit 2 and the `world' of the second
   observer\footnote{Photons passing through either slit do so whether or not there is 
   a detector immediately behind it}. Thus in a certain sense Everett's two possible
  `worlds' are {\it simultaneously present} when the corresponding path amplitudes are
   added to give the probability amplitude for the Young double slit experiment.
   Since in any actual experiment an infinite number of different path amplitudes
   contribute to the probability amplitude it is as though an infinite number
   of classically distinct `worlds' contribute, in parallel, to the probability
   amplitude. This is just what Feynman called the only real deep mystery of
   quantum mechanics.
   \par In the early days of quantum mechanics, de Broglie\footnote{An extensive discussion
   of these ideas can be found in~\cite{DeBroglie1}. In a later development
   of the theory the particle coordinates were replaced by those of a local singularity
    in a `physical' wave, $u$, that was conjectured to exist in addition to the usual
   abstract configuration-space wavefunction, $\psi$. In this article de Broglie also
  imagined, like Feynman in~\cite{Feyn1}, `clocks' moving with the particles or waves
  and recording the phase information.} proposed an interpretation called, in French `La Th\'{e}orie de la
   Double Solution' but usually now refered to as the `Pilot Wave' formulation
  of quantum mechanics~\cite{DeBroglie2}. The idea was that the space-time coordinates of
   particles should appear explicitly in the theory so as to describe as, in classical 
   mechanics, the 'real' positions of particles. There is also a purely mathematical, 
   abstract, configuration-space wavefunction, $\psi$, associated with the particle which is
  conjectured to 'guide' the motion of the particle in such a way that that the statistical
  predictions of conventional quantum mechanics are recovered. This idea was taken up by
  Bohm and Hiley and developed into a fully-fledged deterministic alternative to
  non-relativistic quantum mechanics~\cite{BH}. In this theory a `quantum potential'
  is introduced that gives rise to classical forces that reproduce observed quantum
  behaviour.
  \par However, as seen above, in Feynman's formulation, the space-time
  coordinates of particles do, in any case, occur in an essential way in the 
  equations of the theory. The physical interpretation of the photon wavefunction
  in (2.19) is of a photon moving like a classical particle according to the law
  $\Delta r/\Delta t = c$, but also carrying with it a certain phase, analogous to
 de Broglie's `Pilot Wave'. It is precisely this phase information, when
  combined with that of all other possible paths of the photon in any given 
  experimental situation, that does in fact `guide' the photon paths towards certain 
  regions of space-time and away from others due to the effect of quantum
  interference. This process is exemplified in the calculations presented
  below in the present paper.
  \par Two closing remarks in this section devoted to general principles.
   The first is that in the classical limit, $\hbar \rightarrow 0$, the sums over intermediate
  states in (2.3) reduce to a single term correponding to the classical path of the system.
  (2.1),(2.2) and (2.3) then give:
   \begin{eqnarray}
    P_{fi}^{Class} & = & |\langle f|T_f|k_n^0 \rangle ...\langle k_2^0|T_2|k_1^0 \rangle
                   \langle k_1^0|T_1|i \rangle|^2  \nonumber \\
                   & = & |\langle f|T_f|k_n^0 \rangle|^2...|\langle k_2^0|T_2|k_1^0 \rangle|^2
                         |\langle k_1^0|T_1|i \rangle|^2  \nonumber \\
                   & = & P^0_{fn}... P^0_{21} P^0_{1i}                     
  \end{eqnarray} 
   Here $|k_1^0 \rangle, |k_2^0\rangle ...$ are states along the classical path that minimises the action $S$ in
   (2.6). The last member of (2.25) is just a statement of Bayes Theorem\footnote{See, for example,~\cite{Bayes}.}
    for combining conditional
  probabilities in classical statistics.

  \par Finally, a brief mention of the `measurement problem' of quantum mechanics in
   relation to Feynman's formulation. It would seem, at first sight, that the traditional 
  `problems' of the `reduction of the wavepacket' superposition of Hilbert space vectors of
    macroscopic states~\cite{Leggett}, with the associated `Schr\"{o}dinger cat' paradox etc. are completely avoided in 
    Feynman's formulation. The essential theoretical concept, the probability amplitude
    is a complex number with a Lorentz invariant phase,
    which is completely and unambigously defined by Feynman's principles I-V and
     the experimental conditions that are to be described. On the other hand, Hilbert
   space state vectors are quite ambiguous, depending on the representation
    (Schr\"{o}dinger , Heisenberg, Interaction,...) used to specify them.
     Such a Hilbert space vector, unlike Feynman's probability
    amplitude, does not describe the results of actual experiments. 
   `Measurements' in optics do not produce photons in quantum eigenstates, but rather
   destroy them. There is typically no `wave packet' to contract, no superposition
  of Hilbert space vectors corresponding to different macroscopic final states of the `detector',
  one probability amplitude for a live cat, a completely different one for a dead one... .
   \par Feynman had very little respect for philosophers of science, and was too busy 
   doing science to spend much time on such questions himself. This being so, the
  `philosophy' of Feynman's formulation of quantum mechanics, unlike that of
   the Bohr, Heisenberg and Schr\"{o}dinger one, still remains to be written.

\SECTION{\bf{Relativistic Path Integral and the Propagator of a
   Free On-Shell Particle}}
   
   A suitable starting point for the discussion is an integral
  equation for the one-dimensional path integral~\cite{Feyn8}:
  \begin{equation}
 \psi(x,t+\epsilon) = \int_{-\infty}^{\infty}\frac{1}{A_1} \exp\left[
  \frac{i \epsilon}{\hbar} L\left(\frac{x-y}{\epsilon}\right)\right] \psi(y,t) dy
 \end{equation}
 where $\epsilon$ denotes a small time interval, $L$ is the relativistic Lagrangian of 
  a free, on-shell, particle of pole mass, $m_P$, and relativistic velocity,
  $\beta = (x-y)/(c\epsilon)$, and the normalisation constant $A_1$ is to be determined by consistency
   in the limit
  $\epsilon \rightarrow 0$, $y \rightarrow x$. This formula is now generalised to three
   spatial dimensions by introducing the variable: $\vec{\eta} = \vec{x} - \vec{y}$,
   and changing the integration variable from $\vec{y}$ to  $\vec{\eta}$:
  \begin{equation}
 \psi(\vec{x},t+\epsilon) = \int \int \int \frac{1}{A_3} \exp\left[
  \frac{i \epsilon}{\hbar} L\left(\frac{|\vec{\eta}|}{\epsilon}\right)\right]
   \psi( \vec{x}-\vec{\eta},t) d^3\eta
 \end{equation}
   Choosing the
   origin of coordinates such that $\vec{x} = (0,0,0)$, the one-axis parallel to the velocity 
  vector of the particle: $\vec{\eta} = (\eta, 0, 0)$
   and making the replacement $t+\epsilon \rightarrow t$ gives:
  \begin{equation}
 \psi(0,0,0,t) = \int d \eta_2 \int d \eta_3  \int_{0}^{\infty}\frac{1}{A_3} \exp\left[
  \frac{i \epsilon}{\hbar} L\left(\frac{\eta}{\epsilon}\right)\right] \psi(-\eta, 0,0,t-\epsilon) d \eta
 \end{equation}
 where, without loss of generality, $\eta$ has been chosen to be positive.
 The two and three axes are 
  perpendicular to the direction of particle propagation. Since $\psi$ and $L$ on the right side of (3.2)
  do not depend on $\eta_2$ and $\eta_3$ the integral over these variables may be absorbed in the
  undetermined normalisation  constant so that:
  \begin{equation}
   \frac{1}{A} \equiv \frac{\int d \eta_2 \int d \eta_3}{A_3}
  \end{equation}
   The relativistic free-particle classical Lagrangian is~\cite{Goldstein}:
  \begin{equation}
 L\left(\frac{\eta}{\epsilon}\right) = -m_P c^2\sqrt{1-\frac{\eta^2}{c^2 \epsilon^2}}
  = -m_P c^2\sqrt{1-\beta^2}  
  \end{equation}
  Making a Taylor expansion in $\eta$ and $\epsilon$, of  $\psi(-\eta, 0,0,t-\epsilon)$,
  retaining only the zeroth order term and changing the integration variable from $\eta$ to 
  $\beta$ gives:
  \begin{equation}
   \psi(0,0,0,t) =  \int_0^1 \frac{c\epsilon}{A} \exp\left[
  -\frac{i \epsilon}{\hbar}m_P c^2\sqrt{1-\beta^2}\right] \psi(0,0,0,t) d\beta
 \end{equation}
  Hence 
  \begin{equation}
 1 =   \int_0^1 \frac{c\epsilon}{A}d\beta +O(\epsilon \eta,\epsilon^2)
 \end{equation}
   or, neglecting terms of $O(\epsilon \eta,\epsilon^2)$,
  \begin{equation}
   A= c \epsilon
 \end{equation}
  Since 
 \begin{equation}
 \epsilon = \Delta t = \gamma  \Delta \tau = \frac{\Delta \tau}{\sqrt{1-\beta^2}}
 \end{equation}
 where $t$ and $\tau$ are the laboratory and proper times respectively of the propagating
  particle, then, using  (3.8) and (3.9), as well as the relation $c \epsilon = \eta/\beta$, (3.3)
  may be written as:
 \begin{eqnarray}
   \psi(0,0,0,t) & = & \int_0^{\infty} \frac{\beta}{\eta} \exp\left[
  -\frac{i m_P c^2}{\hbar} \Delta  \tau\right] \psi(-\eta,0,0,t-\epsilon) d\eta
  \nonumber \\
    & = & \int_0^{\infty} K(0,0,0,t;-\eta,0,0,t-\Delta t) \psi(-\eta,0,0,t-\Delta t)d\eta 
 \end{eqnarray}
  where $K(\vec{x_2},t_2; \vec{x_1},t_1)$ is the covariant Feynman propagator
 of the particle between the space time points $(\vec{x_1},t_1)$ and $(\vec{x_2},t_2)$.
  Exploiting the Lorentz invariance of the phase of the exponential in (3.10) the
  covariant propagator of a free, on-shell, particle may finally be written as:
   \begin{eqnarray}
    K(\vec{x_2},t_2; \vec{x_1},t_1) & = &  \frac{\beta}{|\vec{x_2}-\vec{x_1}|}
 \exp\left[-\frac{i m_P c^2}{\hbar} \Delta  \tau\right] \nonumber \\
   & = & \frac{\beta}{|\vec{x_2}-\vec{x_1}|}
 \exp\left[-\frac{i}{\hbar}\left(E(t_2-t_1)-\vec{p}\cdot(\vec{x_2}-\vec{x_1}\right)\right]
 \end{eqnarray}
  The propagator then has the form of a spherical wave, familiar from physical optics.
 Although, in the derivation, the spatial interval $\eta$ was assumed to be a `small' quantity
 it should be noted that there is no length scale in the propagator of a free on-shell
  particle. In any case, Huygen's construction which, as will be demonstrated in the following
  section, is recovered in the path amplitude formalism, guarantees spherical wave propagation
  from the combination of, in principle, infinitely short spherical `wavelets'~\cite{BW1}.
  Equation (3.11) is therefore of general validity.
 \par The mass parameter, $m_P$, in (3.11) is, by  definition, the `pole mass' of the 
  particle. In the case that the latter is unstable, with mean lifetime $\tau_M$, (3.11) is 
   modified to:
   \begin{eqnarray}
    K(\vec{x_2},t_2; \vec{x_1},t_1) & = &  \frac{\beta}{|\vec{x_2}-\vec{x_1}|}
 \exp\left[-\frac{i c^2}{\hbar}( m_P-\frac{i \Gamma}{2 c^2})\Delta  \tau\right] 
 \end{eqnarray}
  where the decay width $\Gamma \equiv \hbar/\tau_M$ has been introduced.
 The physical mass, $m$,  of the particle is then smeared around the value
  $m_P$ according to a distribution of width $\simeq \Gamma/c^2 $. This distribution
  (the momentum-energy space propagator of the unstable particle) is given by the Fourier
  transform with respect to $\Delta  \tau$ of the exponential factor in (3.12):
  \begin{equation}
  K(E;E_0,\Gamma) = \int_0^{\infty}
   \exp\left[-\frac{i}{\hbar}( E_0-\frac{i \Gamma}{2})\Delta  \tau \right]
    \exp\left(\frac{i E \Delta  \tau}{\hbar}\right) d(\Delta  \tau)
   \end{equation}
 where $E_0 \equiv m_P c^2$ is the center-of-mass energy corresponding to the
  pole mass and  $E \equiv m c^2$ is the physical center-of-mass energy of
   the particle. The integral in (3.13) yields:
  \begin{equation}
   K(E;E_0,\Gamma) = \frac{\hbar}{i(E-E_0) -\Gamma/2}
 \end{equation}
  This propagator corresponds to the well-known Lorentzian `natural'
  optical line shape in the case that the `particle' is identified
   with the initial atomic state in a spontaneous radiative decay process.
   For the decay of an unstable elementary particle it is the equally familiar
   Breit-Wigner amplitude. 
   \par It will be important, for the discussion of the damping of
   interference effects in Section 8 below, to note that the imaginary part
   of the argument of the exponential in the space-time propagator
   (3.12) is a function of the pole mass, not the physical mass, of
    the particle.
    \par The dimension, [$L^{-1}$], of the space-time propagator is a necessary consequence of its definition
     in (3.10). Although the phase of this amplitude is Lorentz-invariant, the entire amplitude is not.
     The propagator may also be written as:
      \begin{equation}
   K( \Delta \tau, m_P, \gamma)  =   \frac{1}{ c \gamma \Delta \tau}
 \exp\left[-\frac{i m_P c^2}{\hbar} \Delta  \tau\right]
       \end{equation}
   where the appearence of the relativistic parameter $\gamma$ makes manifest the non-Lorentz-invariant
  character of the propagator.
  In the rest frame of the particle,  (3.15) becomes:
     \begin{equation}
   K( \Delta \tau, m_P, 1)  =   \frac{1}{ c \Delta \tau }
 \exp\left[-\frac{i m_P c^2}{\hbar} \Delta  \tau\right]
       \end{equation}
  which differs from Feynman's~\cite{Feynprop} asymptotic Lorentz-invariant space-time propagator:
     \begin{equation}
   K^{asym}( \Delta \tau, m_P)  =   \frac{m_P}{ 8 \pi i c \Delta \tau }
 \exp\left[-\frac{i m_P c^2}{\hbar} \Delta  \tau\right]
       \end{equation}
  only by the factor $m_P$ and a normalisation constant.
  Neither (3.16) nor (3.17) is suitable to describe the time
 evolution of the wavefunction of a particle at rest, due to the manifestly 
  unphysical behaviour of the factor $(c\Delta  \tau)^{-1}$ originating in the spatial
 integral in (3.10) which is no longer defined for a particle at rest.
  In seems that the one dimensional propagator describing the time evolution 
  of the state of a particle at rest can not be obtained as a simple
  limit of the four-dimensional space-time propagator. Instead, the propagator
 of a particle at rest may be simply and directly derived from
  the time-dependent Schr\"{o}dinger Equation. For a particle at rest
  the non-relativistic limit necessary for the validity of the Schr\"{o}dinger Equation
  is actually the same as the exact kinematical result, so that the propagator found
 in this way is both exact, and Lorentz-invariant.
   However the `Schr\"{o}dinger Equation' used here, though universally so-called 
  in discussions of particle flavour oscillations, is in fact a relativistic 
  version of this equation since the non-relativistic kinetic energy term 
 of the conventional equation is replaced by the sum of this term and the rest
  energy of the particle. In the use of such an equation to describe atomic
  transitions the rest mass term contributes an overall multiplicative phase factor
  in the solutions of the equation, without any physical consequences, since this phase 
  factor cancels in all atomic transition matrix elements. In physical optics, as aleady 
  mentioned above, the crucial dynamical element in the probability amplitude
  is the temporal propagator of the excited atom that constitutes the photon source,
  or, more precisely, the time dependence of the decay amplitude of such an atom.
  Although this formula may be found, for example, in Dirac's book
  on quantum mechanics~\cite{Diracprop}, for clarity and completeness it 
  is rederived here. 
   \par Denoting the mass of the atomic ground state by $m_0$ and the pole excitation
  energy of the state $|\psi_i\rangle$ by $E^{0}_i$
  the  Schr\"{o}dinger Equation for this quasi-stationary state is:
  \begin{equation}
  i\hbar \frac{\partial |\psi_i\rangle}{\partial \tau} = (m_0 c^2+ E^{0}_i)|\psi_i\rangle
  \end{equation}
  with the solution:
 \begin{equation}
|\psi_i(\tau)\rangle = |\psi_i(0)\rangle\exp\left[-\frac{i( m_0 c^2 + E^{0}_i)}{\hbar} \tau \right] 
  \end{equation}
 Hence the time evolution of the transition amplitude between the states  $|\psi_i\rangle$ and
 $|\psi_f\rangle$ is give by the relation:
 \begin{equation}
 \langle \psi_f(\tau)|T|\psi_i(\tau)\rangle = \langle \psi_f(0)|T|\psi_i(0)\rangle 
   \exp\left[-\frac{i(E^{0}_i- E^{0}_f-i\Gamma_i/2)}{\hbar} \tau \right] 
   \end{equation}
  where an imaginary part $-i\Gamma_i/2$ has been added to $E^{0}_i$ to take into account, as discussed
  above, the lifetime damping of the amplitude of the excited state. In path amplitude 
  language, (3.20) may be interpreted as the product as the amplitude for the excited state to evolve 
  from proper time zero to proper time $\tau$ (the exponential factor) times the time independent
  decay amplitude of the state.  The formula (3.20) is directly applicable to nuclear $\beta$ transitions
   on making the relacement $E^{0}_i- E^{0}_f \rightarrow E^{0}_{\beta}$ where the latter
   quantity is the total energy release in the nuclear transition. As will be discussed in 
    Section 9 below, the temporal  propagator of the neutrino source state is predicted
    by the Feynman path amplitude formalism to make an important contribution to the
    neutrino oscillation phase. The time dependence of the decay amplitude of a particle
  that is destroyed in the transition is given by the replacements $E^{0}_i \rightarrow m_P c^2$,
   $E^{0}_f \rightarrow 0$, in (3.20) where $m_ P$ is the pole mass of the particle.
   That is, the `excitation energy' of the particle is equivalent to its mass, while the mass-energy 
   of the corresponding 'ground state' vanishes.

\SECTION{\bf{Derivation of the Classical Wave Theory of Light from the Path Amplitude
             Formalism Of Quantum Mechanics}}

 \begin{figure}[htbp]
\begin{center}
\hspace*{-0.5cm}\mbox{
\epsfysize7.0cm\epsffile{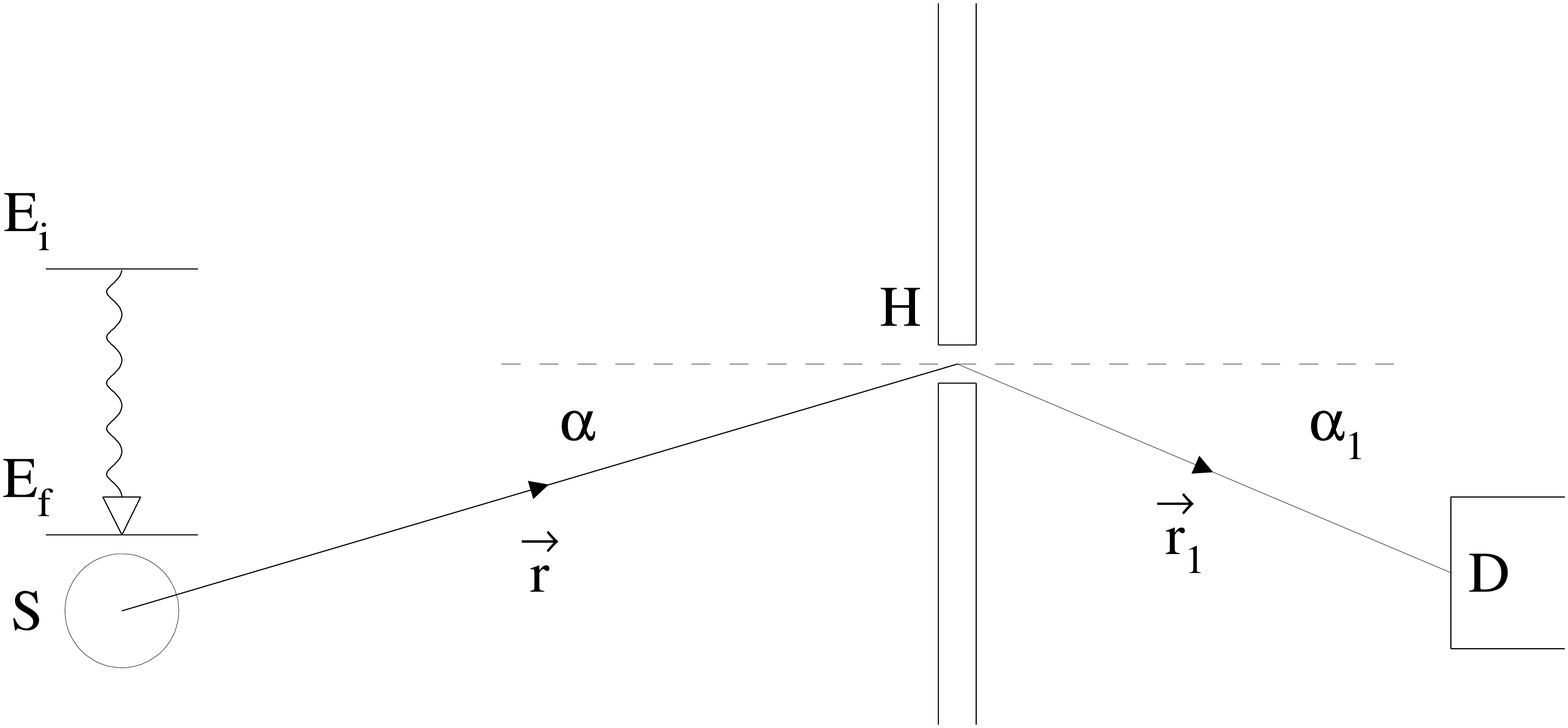}}
\caption{{\sl Diffraction of a photon, of energy $E_i-E_f$, produced in the spontaneous 
  decay: $A_i \rightarrow A_f+ \gamma$ of an excited atom at S, by a small hole, H, in an opaque
  screen. The photon is detected at D. The vectors $\vec{r}$, $\vec{r_1}$ lie in the plane of the figure}} 
\label{fig-fig1}
\end{center}
\end{figure}

 \par  The path amplitude $\langle f_m|k_n,...k_2,k_1|i_l \rangle$ defined in (2.2)
  above  is specified by the initial and final states
  $i_l$ and $f_m$ respectively, and by the set of unobserved intermediate states: $k_n,...k_2,k_1$.
   The latter may refer to different space-time positions of the same particle, in which case,
   the `process amplitude'
    $ \langle k_i|T_i|k_j\rangle$ is the space-time propagator between $k_j: \vec{x}_j , t_j$ and
   $k_i: \vec{x}_i , t_j$ or, alternatively, to the transition amplitude of some quantum
   process in which a particle is scattered or new particles are created. Of particular importance
   for the discussion of physical optics is the case of the creation of a photon in the spontaneous
  decay of an excited atom.  In this case 
  the first process amplitude (the rightmost one in (2.2)) describes the production of the
  excited atomic state, ${\cal A}_i$, from the ground state, ${\cal A}_0$, at time $t_0$:
  \begin{equation}
   \langle k_1|T_1|i_l\rangle =  \langle {\cal A}_i|T_P| {\cal A}_0 \rangle
 \end{equation} 
 where $T_P$ denotes an atomic transition operator. The second process amplitude  $\langle k_2|T_2|k_1\rangle$  
 describes the space-time propagation of the excited atom from its production time $t_0$, to the production time,
  $t_{\gamma}$ of the photon.
  The third process amplitude:
   \begin{equation}
   \langle k_3|T_3|k_2\rangle = \langle \gamma  {\cal A}_f|T| {\cal A}_i\rangle   
 \end{equation} 
  describes the decay of the excited atom into a photon, $ \gamma$, of energy 
  $E_{\gamma} = E_i-E_f$ where $E_i$ and $E_f$ denote the energies of the initial and final
  atomic states respectively. The appropriate formula to describe the product of process
  amplitudes  $\langle k_3|T_3|k_2\rangle\langle k_2|T_2|k_1\rangle$, is given by (3.20):
    \begin{equation}
\langle k_3|T_3|k_2\rangle\langle k_2|T_2|k_1\rangle =  \langle \gamma  {\cal A}_f|T| {\cal A}_i\rangle
   \exp\left[ -\frac{i}{\hbar}(E_i^0-E_f^0-i\Gamma_i/2)(t_{\gamma}-t_0)\right]
  \end{equation} 
    The first factor on both sides of (4.3) is the (time-independent) decay amplitude of the excited
   atom  in (4.2). The second is the temporal propagator of the excited atom, assumed to be
   at rest.
  \par The remaining product of process amplitudes $\langle k_{n}|T_n|k_{n-1}\rangle...\langle k_4|T_4|k_3\rangle$
  denote either free space propagation of photons according to (3.11) or, in the case of
  diffraction, refraction or reflection, photon scattering amplitudes. The final (left-most)
  process amplitude describes the photon detection process:
   \begin{equation}
  \langle f_m|T_f|k_{n}\rangle =  \langle {\cal D}_m|T_D| {\cal A}_n \gamma_n \rangle 
  \end{equation}
   Here ${\cal D}_m$ denotes the final state of the photon detection process
   $\gamma_n {\cal A}_n \rightarrow {\cal D}_m$ that is the same in all the path amplitudes
   with final state label $m$. This may be an activated atom in a photographic plate,
   a photoelectron, a conduction electron in a photo-diode etc. The detection process
   does not discriminate the state $\gamma_n$ of the detected photon, and 
   occurs at the same time, $t_D$, in all the path amplitudes with final state 
   label $m$.
   \par To illustrate the application of the path amplitude formula (2.2), a simple case
    is considered where an exited atom is produced at S and the decay photon is detected at D after
    diffraction by a small hole, H, in an opaque screen (Fig. 1). The corresponding 
    path amplitude is:
   \begin{eqnarray}
   \Delta A & = & \langle {\cal D}|T_D| \gamma \rangle
               \langle D|\gamma|H\rangle {\cal A}_{diff}\Delta S
              \langle H|\gamma|\gamma\rangle \nonumber \\
           &   &    \langle \gamma  {\cal A}_f|T| {\cal A}_i\rangle
                \langle\gamma |{\cal A}_i| 0\rangle 
                \langle {\cal A}_i|T_P|{\cal A}_0 \rangle
     \end{eqnarray}
 The process amplitudes $\langle\gamma |{\cal A}_i|0 \rangle$ , 
 $\langle H|\gamma|\gamma \rangle$ and 
 $\langle D|\gamma|H \rangle$ are, respectively, the space-time
  propagators of: the excited atom from its production time $t_0$
  to its decay time $t_{\gamma}$, the photon, before diffraction at the hole at time $t_H$,
  and the photon
  after diffraction. The product of the three space-time independent process
 amplitudes , $\langle {\cal A}_i|T_P|{\cal A}_0 \rangle$, $\langle \gamma  {\cal A}_f|T| {\cal A}_i\rangle$
  and $\langle {\cal D}|T_D| \gamma \rangle$ described above, is denoted below as $\tilde{{\cal A}}$.
   The diffractive scattering
  amplitude, $ {\cal A}_{diff}$, will be derived below. The parameter $\Delta S$ is the area of the hole 
  H, which is used as a weighting factor proportional to the number of space time paths 
  crossing the hole.
  \par Since, for a massless photon, $E\Delta t- p \Delta x = m \Delta \tau = 0$, there is no
  contribution to the phase of the path amplitude (4.5) from the photon propagators\footnote{
   This was noted by Feynman on P103 of~\cite{Feyn1}.`Once a photon has been emitted there is no
   further turning of the arrow as the photon goes from one point to another in space-time.'}. Using (4.2) and
  (4.3) the propagator of the excited atom is given as\footnote{This is the physical origin of the
  `stopwatch hand' discussed by Feynman in Fig 67 of~\cite{Feyn1}:`But when we construct a {\it monochromatic}
  source, we are making a device that has been carefully arranged so that the amplitude for a photon to
  be emitted at a certain time is easily caculated: it changes its angle at a {\it constant} speed like a
  stopwatch hand.' It is clear from (4.6) that the hand actually rotates in a clockwise direction, 
  not an anticlockwise one as in  Fig 67 of~\cite{Feyn1}. This change of sign has no effect on the
   predictions of the path amplitude calculations.}
  \begin{equation}
  \langle P|{\cal A}_i| 0 \rangle =  \exp\left[ -\frac{i}{\hbar}(E_i^0-E_f^0-i\Gamma_i/2)(t_{\gamma}-t_0)\right]
 \end{equation}
  Noting now that
   \begin{equation}
     t_D = t_{\gamma} + \frac{r+r_1}{c}
 \end{equation}
 the path amplitude of (4.5) may be written as :
  \begin{equation}
  \Delta A = \frac{\tilde{{\cal A}}}{r r_1} {\cal A}_{diff} \Delta S
   \exp\left[ -\frac{i}{\hbar}(E_i^0-E_f^0-i\frac{\Gamma_i}{2})(t_D-t_0- \frac{r+r_1}{c})\right]  
 \end{equation}
   Here the factors $1/r$, $1/r_1$ originate (see (3.11)) from the photon propagators. 
  \par Introducing the parameters:
    \begin{equation}
    \kappa \equiv \frac{E_i^0-E_f^0}{\hbar c} =  \frac{(M_P^i-M_P^f)c}{\hbar}
   \end{equation}
   and
  \begin{equation}
  \rho   \equiv \frac{\Gamma_i}{2\hbar c}
   \end{equation}
   where $M_P^i$ and $M_P^f$ are the `pole masses' of the initial and final
   atomic states, 
 the path amplitude may be written as the product of four factors. The first has no
  spatio-temporal dependence, the second depends only on $r$, 
 the third only
 on $r_1$ and the fourth only on the times $t_D$ and $t_0$:
 \begin{equation}
  \Delta A = \tilde{{\cal A}}\cdot \frac{e^{(i \kappa+ \rho)r}}{r} 
    \cdot \frac{e^{(i \kappa+ \rho)r_1}{\cal A}_{diff}\Delta S }{r_1}
    \cdot e^{-c(i \kappa+ \rho)(t_D-t_0)}
 \end{equation}
In the description of diffraction or interference effects on the right of the screen in Fig. 1, at a given
 detection time, $t_D$, and for sufficiently large values of $t_D-t_0$, only the
 $r_1$ variation of $\Delta A$ is of importance. If the finite decay width 
  of the excited atom may also be neglected, i.e., if $ \rho  \ll  \kappa$, the $r_1$ dependence of the
  path amplitude is that of a spherical wave field:
  \begin{equation}
   U(r_1) =  \frac{e^{i \kappa r_1}}{r_1}
 \end{equation}
 which satisfies the Helmholtz equation:
  \begin{equation}
  \nabla^2 U+\kappa^2 U = 0 
\end{equation}
 In this case, the description provided by the path amplitude formalism is mathematically
 identical to the classical wave theory of light. Diffraction is
  described by Kirchoff's equation~\cite{BW2,MW1} and all the well-known interference 
  phenomena of light, such as Huygen's construction~\cite{BW1}, the Young double slit experiment and
  the Michelson Interferometer, may also be described purely spatially,
  like analogous wave phenomena of classical physics, such as transverse waves on a
  string or surface
  waves on a liquid subjected to the force of gravity. In coherent laser optics where the
   effective lifetime of the photon source is very long\footnote{Defining a `coherence
   length': $L_{coh} = c \hbar/\Gamma_i = 1/(2\rho)$, typical values of $L_{coh}$ for 
  a spontaenously decaying atom and a coherent laser source are 3m and 30km respectively~\cite{MW2}.},
   $\rho \ll \kappa$,
   and so the classical wave theory of light is expected to be a very good approximation. It is therefore
  necessary to look elsewhere for the specifically temporal effects that distingish
  quantum mechanics from the classical wave theory.
  \par The diffractive scattering amplitude, ${\cal A}_{diff}$, in (4.5) is derived by applying Green's 
   Theorem, with suitable boundary conditions,
    to (4.13). This yields the well-known Fresnel-Kirchoff diffraction formula~\cite{BW2}, 
   and the expression:
       \begin{equation}
  {\cal A}_{diff}(\alpha, \alpha_1) = -\frac{i \kappa}{4 \pi}[\cos \alpha+ \cos\alpha_1]
 \end{equation} 
  where the angles $\alpha$ and $\alpha_1$ between the normal to the screen and the vectors
  $\vec{r}$ and  $\vec{r_1}$ are shown in Fig 1. It is interesting to note that, in order to 
   derive the  Fresnel-Kirchoff diffraction formula in the classical wave theory of light, it is
   necessary to assume, in addition to Green's theorem, that the wave field $U$ vanishes at
   large distances from the point under consideration. In the path amplitude description, this
    damping occurs automatically due to the factor: $\exp[-\rho(c(t_D-t_0)-r-r_1)]$ in (4.11).
     Since (see Fig.1) $c(t_D-t_0) \ge c(t_D-t_{\gamma})=r+r_1$, the argument of the exponential factor is 
    always zero or negative, producing strong damping of the amplitude for diffractive scattering
      at distances much larger than $r$ or $r_1$ since, in this case, $c(t_D-t_0) \ge c(t_D-t_{\gamma})
       \gg  r+r_1$. 

\begin{figure}[htbp]
\begin{center}
\hspace*{-0.5cm}\mbox{
\epsfysize7.0cm\epsffile{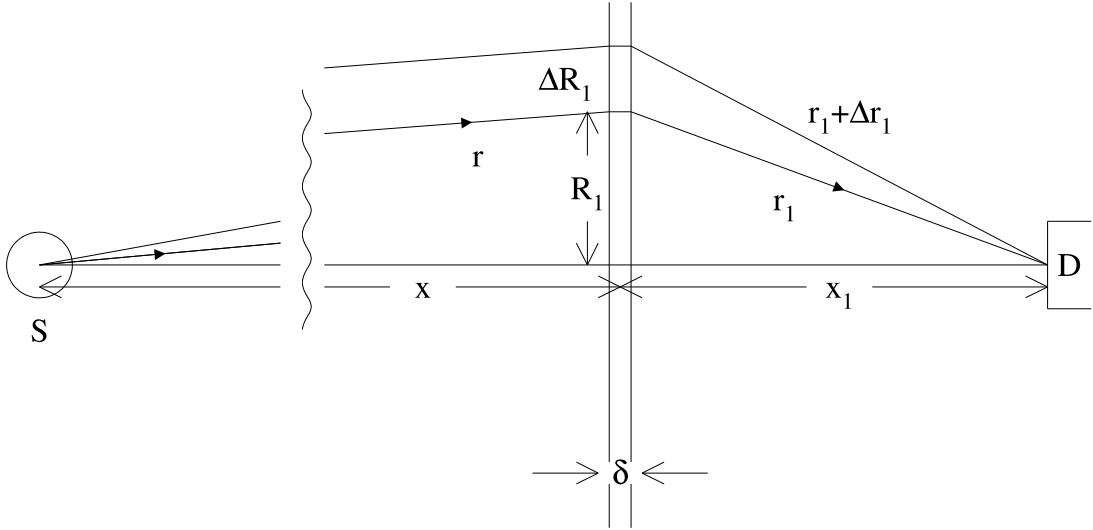}}
\caption{{\sl A photon produced by spontaneous decay of an atom at S is detected
  at D. The integral of the path amplitudes of photons diffracted or scattered in a ring
 of width $\Delta R_1$ and radius $R_1$ in a plane perpendicular to SD is constructed.
 In the case of diffraction, (4.16), the ring lies within an imaginary surface at distance $x$
  from the source. In the case of scattering (refraction), (5.1), the ring lies in a thin
  sheet of transparent material of thickness $\delta$.}} 
\label{fig-fig2}
\end{center}
\end{figure}
  
  \par The limit of the formula (4.14) for forward diffraction, can also be be derived,
    directly, using the path amplitude
    method. This is done by
  considering rectilinear photon progagation (propagation
   of plane waves in the classical theory of light). Consider a photon produced by a distant
   excited atom at O and observed in a photon detector at D as shown in Fig. 2. The x-axis lies along the line 
   joining the atom and the detector. Following Feynman~\cite{Feyn9} a plane surface perpendicular
   to the x-axis, distant $x$ from the atom and $x_1$ from the detector, where $x \gg x_1$, is 
   imagined to be divided up into small areas, from each of which the photon trajectories
    from the atom may be diffracted towards the detector. In particular, diffraction is considered
    from a ring of radius $R_1$ and width $\Delta R_1$, centered on the x-axis. Using (4.8) above
    and noting that $c(t_D-t_0)-r-r_1 = c(t_{\gamma}-t_0)$
   \begin{equation}
  \Delta A = \frac{\tilde{{\cal A}}}{r r_1}
   \exp \left[ - i \kappa \left(c(t_D-t_0)-r-r_1 \right)-\rho c(t_{\gamma}-t_0)\right]
  {\cal A}_{diff} 2 \pi R_1 \Delta R_1
 \end{equation} 
   Since (see Fig. 2), $r_1^2 = x_1^2+R_1^2$, it follows that $R_1 \Delta R_1 = r_1  \Delta r_1$.
   Summing over
   all paths diffracted at the plane, the probability amplitude at the photon detector is:
    \begin{equation}
  A =  \frac{\tilde{{\cal A}}}{r}
   \exp\left[ -i \kappa\left( c(t_D-t_0)- r\right)\right] 2 \pi {\cal A}_{diff}
   \int_{x_1}^{r_1^{max}} \exp\left[ i\kappa r_1)-\rho c (t_{\gamma}-t_0) \right]dr_1 
 \end{equation} 
   where an arbitary upper limit, $r_1^{max}$ has been assumed for the $r_1$ integral. 
   Noting the inequality:
     \begin{equation}
       c (t_{\gamma}-t_0) \ge r+r_1- x -x_1
   \end{equation} 
    it can be seen that large values of $r_1$  will give an exponential 
     suppression of the integrand in (4.16). In this case the $r_1$ integral is readily evaluated
 by use of the Huygens-Fresnel\ Principle~\cite{BW1}. This states that the value of the 
  integral is one half of the contribution due to the first half-period zone. With the change
   of variable; $\varphi = \kappa r_1$, (4.16) may then be written as:
 \begin{eqnarray}
A  & = & \frac{\tilde{{\cal A}}}{r}
   \exp\left[ -i \kappa\left( c(t_D-t_0)- r \right)-\rho c (t_{\gamma}-t_0) \right]\frac{\pi {\cal A}_{diff}}
    {\kappa}
   \int_{\kappa x_1}^{\kappa x_1 +\pi} \exp\left[ i\varphi \right]d\varphi
  \nonumber \\
  & = & \frac{\tilde{{\cal A}}}{r}
    \exp\left[ -i \kappa \left( c(t_D-t_0)- r-x_1\right)-\rho c (t_{\gamma}-t_0) \right]
    \left[\frac{ 2 \pi i {\cal A}_{diff}}{\kappa}\right] \nonumber \\
 & \simeq & \frac{\tilde{{\cal A}}}{x_{SD}}
    \exp\left[ -i \kappa \left( c(t_D-t_0)- x_{SD} \right)-\rho c (t_{\gamma}-t_0) \right]
    \left[\frac{ 2 \pi i {\cal A}_{diff}}{\kappa}\right]
 \end{eqnarray}
    where, because of the small variation of $t_{\gamma}$ over the first half-period zone, the factor
     $\exp\left[-\rho c (t_{\gamma}-t_0)\right]$ has been taken outside the integral
  and in the last line, the approximation $r +x_1 \simeq  x +x_1 = x_{SD}$ where $x_{SD}$ is the
   source-detector distance, has been made in the exponential factor
   and $r  \simeq  x_{SD}$ in the denominator.
    The path amplitude for photon propagation from the excited atom to the detector is
    given by an equation analogous
   to (4.5), with the fifth and sixth  process amplitudes (in temporal order) omitted.
   The corresponding probability amplitude is:
    \begin{equation}
  \ A = \frac{\tilde{{\cal A}}}{x_{SD}}
   \exp\left[ -i \kappa ( c(t_D-t_0)- x_{SD})-\rho c (t_{\gamma}-t_0) \right] 
 \end{equation}
  Consistency with (4.18) then requires:
    \begin{equation}
   {\cal A}_{diff} = -\frac{i \kappa}{2 \pi} = -\frac{i}{\lambda_{\gamma}^0}
  \end{equation}
  where $\lambda_{\gamma}^0$ is the de Broglie wavelength of the photon
   \footnote{Note that $\lambda_{\gamma}^0$ is defined, according to (4.9) and (4.20), in terms of the
    difference of the  {\it pole energies} $E_i^0$ and $E_f^0$ of the atomic states $A_i$ and $A_f$,
    not their physical energies  $E_i$ and $E_f$.}, in agreement with (4.14)
   for $\alpha = \alpha_1 = 0$.
  .
 
.
  \SECTION{\bf{ The Refractive Index of a Uniform Transparent Medium}}
 
  \par The calculation of the diffractive scattering amplitude presented above is easily adapted
   to calculate the refractive index of a uniform transparent medium in terms of the atomic
   density, ${\cal N}$, and the elastic scattering amplitude, ${\cal A}_ {scat}$, of a photon from
    an atom of the 
   material. Thus, in Fig. 2, the plane in which the photons are considered to be diffracted is 
   replaced by a thin, uniform, sheet of transparent medium of thickness $\delta$. Assuming an
   isotropic angular distribution for the scattered photons, the contribution to the path amplitude
    of photons scattering once from the atoms in the sheet
    \footnote{Note a similar calculation of the radiation field from a `sheet of oscillating
     charges' in Section 30-7 of~\cite{Feyn2}, in particular the contribution of the upper
    limit of the $r$ integral. See also the analogous calculation of the refractive index of neutrons
 interacting in matter in Section 5.3.4 of~\cite{LLFB}.}is, by analogy with (4.16),:
   \begin{equation}
  A(\delta) =  \frac{\tilde{{\cal A}}}{r}\exp\left[ - i \kappa \left(c(t_D-t_0)- r - r_1
   \right)-\rho c(t_{\gamma}-t_0)\right] {\cal N} {\cal A}_{scat} \delta
    \int_{x_1}^{r_1^{max}} \int_0^{2 \pi}  d r_1  d\phi_1
\end{equation}
   where the integration variable has been changed from $R_1$ to $r_1$ and where the azimuthal
   angle, $\phi_1$, around the x-axis, has been introduced. If the sheet is infinite in 
   transverse extent the integral may be evaluated using the Huygens-Fresnel principle, as done
   for the case of diffraction above. As in all actual experiments the transparent material
   has finite transverse dimensions it will be found interesting, for the following discussion,
   to assume that the limit $r_1^{max}$ is a function of a set of geometrical parameters, $d_i$,
   that define the transverse extent of the sheet, as well as of $\phi_1$. The examples of
   rectangular and circular geometry are considered in Appendix A. In this case the equation
   analgous to (4.18) above is:
   \begin{eqnarray}
 A(\delta) & = & \frac{\tilde{{\cal A}}}{r}\exp\left[ - i\kappa \left(c(t_D-t_0)- r)
   \right)-\rho c (t_{\gamma}-t_0)\right]\frac{i{\cal N} {\cal A}_{scat} \delta}{\kappa} \nonumber \\ 
      &   & \times \int_0^{2 \pi}\{ \left[ \exp[i \kappa x_1]- \exp[i\kappa
    r_1^{max}(d_i,\phi_1) \right]\} d\phi_1  \nonumber \\
      & = & \frac{\tilde{{\cal A}}}{x_{SD}}\exp\left[ -i \kappa \left(c(t_D-t_0)- x_{SD})
   \right)-\rho c (t_{\gamma}-t_0)\right] \frac{ 2 \pi i {\cal N} {\cal A}_{scat}}{\kappa} \delta
  \end{eqnarray}
  since the  $\phi_1$ average of $\exp[i\kappa r_1^{max}]$ vanishes due to rapid phase variation. 
  Adding the contribution from the path amplitudes of the unscattered photon from (4.19) gives:
   \begin{equation}
  A' = A + A(\delta) = A \left[ 1+\frac{ 2 \pi i {\cal N} { \cal A}_{scat} \delta}
  {\kappa}\right] \simeq A \exp[i \delta \phi] 
 \end{equation}
   where 
   \begin{equation} 
\delta \phi = \frac{  2 \pi {\cal N} {\cal A}_{scat }\delta}{\kappa}
  =  \frac{  2 \pi {\cal N} {\cal A}_{scat } f x}{\kappa}
 \end{equation}
  The last member of (5.3) is valid providing that $\delta$ is chosen sufficiently small
  that $\delta \phi \ll 1$. In (5.4) $f$ denotes the ratio $\delta/(x_1+x) \simeq \delta/x$, i.e. 
  the fraction of the total distance between the exited source atom and the detector 
  filled with the transparent material.
   \par It is interesting to note that the result (5.2) may also be obtained by restricting the $r_1$
   integral to the contribution of the first half-period zone. Thus the effect of random geometrical
  boundaries in a medium of finite spatial extent is the same as the use of the Huygens-Fresnel Principle
  in an infinite medium. This equivalence is used to simplify the calculation of the reflection
   coefficient at the surface of a transparent medium in Section 7 below. 
  \par Since the detection time $t_D$ and the production time $t_0$ of the excited atom are fixed, as is the
  distance between the photon source and the detector, the change of phase of the path amplitude induced
  by paths that scatter once on the atoms of the transparent material, implies that the apparent
  velocity, $v(f)$, of the light, between the source and the detector, when the fraction $f$ of the
  space between the source and the detector is filled with the transparent medium, is less than the
  speed of light in vacuum.
 \par In vacuum:
     \begin{equation}
   \phi(A) =  - \kappa c\left(t_D-t_0-\frac{x}{c} \right)
 \end{equation}
  On adding the transparent material:
   \begin{eqnarray}
 \phi(A') & = & - \kappa c\left(t_D-t_0- \frac{x}{c} \right) + \delta \phi
   =  - \kappa c\left(t_D-t_0 - \frac{1}{c}(x+\frac{ \delta \phi}{\kappa}) \right) 
  \nonumber \\
   & = & - \kappa c\left(t_D-t_0-\frac{x}{v(f)} \right)
 \end{eqnarray}
   where, combining (5.4) and (5.6):
   \begin{equation}
  \frac{c }{v(f)} = 1+  \frac{2 \pi {\cal N} {\cal A}_{scatt} f }{ \kappa^2}
 \end{equation}
   Defining the refractive index of the transparent
    medium, $n$, as:
     \begin{equation}
    n \equiv \frac{c}{v} = \frac{c}{v(1)}   
 \end{equation}
   and using the last menber of (4.20), gives the well-known formula~\cite{Sakurai}:
      \begin{equation}
    n  = 1 + \frac{1}{2 \pi} (\lambda_{\gamma}^0)^2 {\cal N} {\cal A}_{scatt}   
 \end{equation} 
   It is interesting to note that in the calculation, based on the the geometrical
  configuration shown in Fig.2, leading to (5.6), the photon corresponding to each 
  individual path amplitude propagates at the vacuum speed $c$, but when the path amplitudes 
   corresponding to single scattering processes  are combined with those of unscattered 
  photon the phase of the total amplitude is the same as that of a photon
  propagating in free space, but with a reduced velocity. This is clearly seen by inspection
  of (5.5) and (5.6) above.
 The apparent velocity, $v(f)$, has a linear dependence
 on the filling fraction $f$:
     \begin{equation}
 v(f) = (1-f)c+fv
 \end{equation}
This equation may be re-arranged to give:
     \begin{equation}
  \frac{\Delta v}{\Delta v_{max}} = \frac{c- v(f)}{c- v} = f
\end{equation}
 which states that the change of effective velocity normalised 
to its maximum value is  equal to $f$, i.e. it is proportional to the number of
 scattering processes contributing to the probability amplitude.
 \par It may be remarked that the above calculation of the refractive index has been
 carried out entirely in the language of particles moving in space--time. The de Broglie
 wavelength  $\lambda_{\gamma}^0$ has been introduced in (5.9) only to show that the
  usual wave-mechanics result is recovered in the path amplitude approach.
  \par The calculation of the refractive index just presented was performed on the
  assumption that the the sheet of transparent medium is sufficiently thin that
  the approximation of the last member of (5.3) is valid, and that only a single 
   scattering event needs to be considered. Combining (5.4) and (5.9) the phase shift 
   produced by the sheet is:
    \begin{equation}
  \delta \phi = \frac{2 \pi (n-1) \delta}{\lambda_{\gamma}^0}
    \end{equation}
 Thus, for the approximation $1+i\delta \phi \simeq \exp i \delta \phi$ to be valid,
  $ \delta $ is required to be a tiny fraction of $\lambda_{\gamma}^0$, say less than $6 \times 10^{-7}$ cm,
  for the case of a Sodium D-line where $\lambda_{\gamma}^0$ =  $5.9 \times 10^{-5}$ cm. 
  \par The calculation of the refractive index is now repeated, first considering the case of a 
   slab of transparent medium of thickness much greater than  $\lambda_{\gamma}^0$, so that 
   the approximation of the last member of (5.3) is no longer valid, but assuming that
   the value of $t_D-t_0$ is appreciably larger than the typical photon flight time between the source
   and the detector, and secondly, also considering a thick slab, but assuming that $t_D-t_0$ may take
   any value, including ones near to the minimum value fixed by the source-detector
   distance and the velocity of light in vacuum. In both cases $\rho = 0$ is assumed, so that damping
   effects due to the finite lifetime of the source atom are neglected.

\begin{figure}[htbp]
\begin{center}
\hspace*{-0.5cm}\mbox{
\epsfysize9.0cm\epsffile{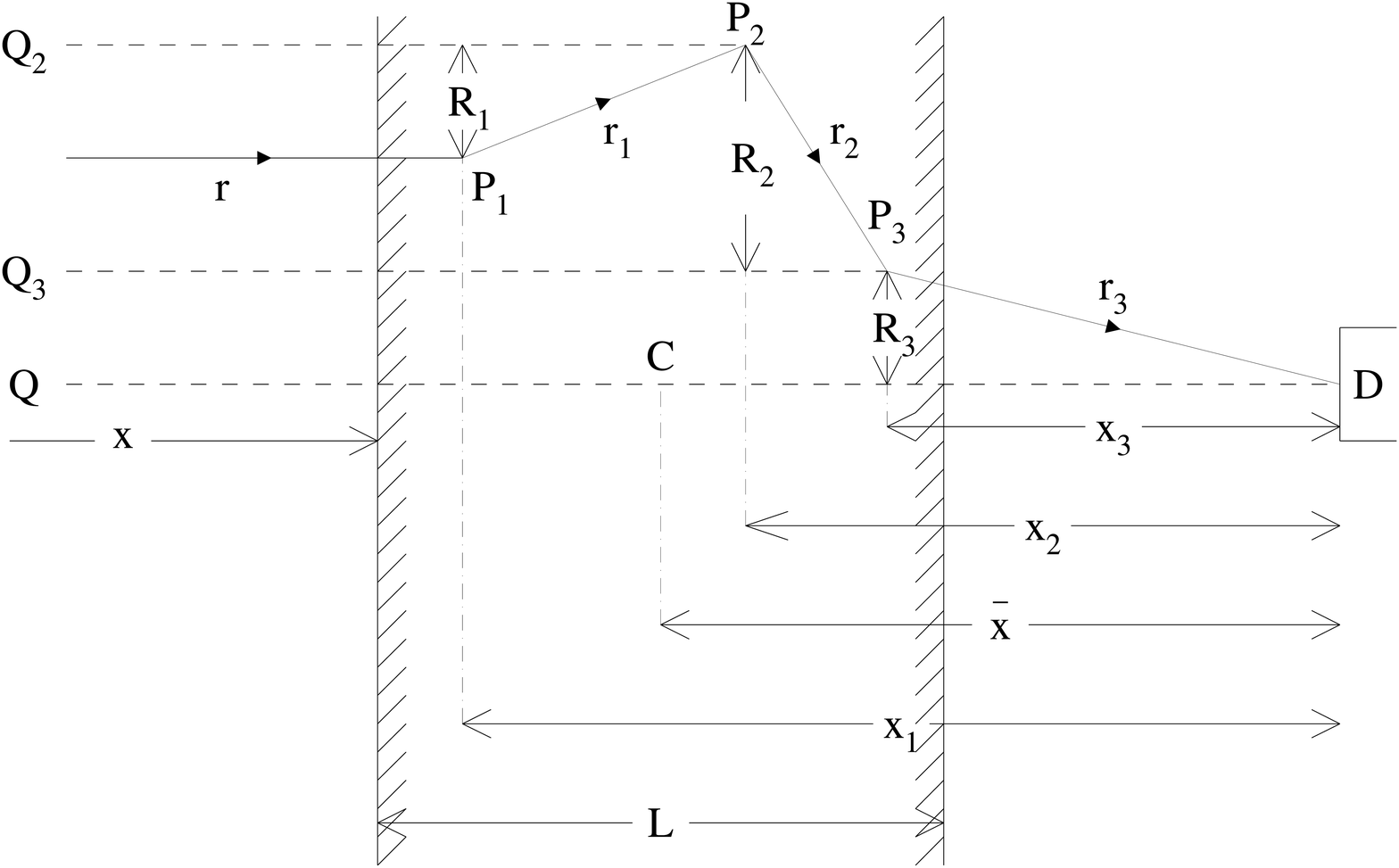}}
\caption{{\sl A photon produced by spontaneous decay of an atom at S, far to the left
 of the figure scatters at the points P$_1$, P$_2$  and P$_3$  inside a block
 of transparent material before being detected at D. The path SP$_1$P$_2$P$_3$D
  lies in the plane of the figure. Various geometrical parameters used in (5.13)
 are defined.}} 
\label{fig-fig3}
\end{center}
\end{figure}

    \par In order to calculate the
   refraction of light in a block of transparent material of thickness, $L$, much greater than $\lambda_{\gamma}^0$,
   it will be found necessary to sum over all possible configuations of multiple scattering of the
    photon from the atoms of the medium. In Fig.3 the source atom is situated far to the left,
     so that the path of the photon from the atom to the front surface of the block may be taken to
    be a constant, equal to the distance, x, from the atom to the front face of the block.
    As in Fig.2, the photon detector, D, lies on the x-axis, which is perpendicular to the
  faces of the block. The distance from the center of the block to the detector is $\bar{x}$. In 
   Fig.3, the photon scatters at the points $P_1$, $P_2$ and $P_3$ which lie in a plane
  containing the x-axis, QD, and are at distances from D, along this axis, of $x_1$, $x_2$ and $x_3$.
  The path lengths between the points $P_1$ and $P_2$,  $P_2$ and $P_3$ and  $P_3$ and $D$ are
   $r_1$, $r_2$ and $r_3$.
  Other possible paths of the photon in the block are generated by rotating the point $P_1$ 
  about the axis $Q_2 P_2$ to give the series of points: $P_1',P_1'',,P_1'''...$ all distant
  $r_1$ from $P_2$ and lying on a circle of radius $R_1$ in the yz plane. By varying the value
  of $R_1$ the integral is performed over the positions of all possible atomic scattering events 
  between $x_1$ and $x_1+\Delta x_1$. Similarly, by considering the points $P_2',P_2'',,P_2'''...$
   and $P_3',P_3'',,P_3'''...$  lying on circles of radius $R_2$ and $R_3$ respectively, 
  centered on the axes  $Q_3 P_3$ and $Q D$, and integrating over all values of 
  $R_2$ and $R_3$, all possible atomic scattering events in the intervals from  $x_2$ to $x_2+\Delta x_2$ and
   $x_3$ to $x_3+\Delta x_3$ are summed. Including all paths with radii in the range
   $R_i$ to $R_i +\Delta R_i$, (i =1,2,3) the single scattering formula analgous to (4.15),
   generalises, for the three-scattering case, to\footnote{Note that in (5.13) only {\it forward}
   scattering processes are included. Back-scattering between two atoms separated by distances 
   $\ge \lambda_{\gamma}^0$ results results in rapid variation of the phase of the path amplitude
    strongly suppressing such contributions. The vanishing of the amplitude for backward
     scattering in a uniform transparent medium can also be understood as an effect of destructive 
     interference between different path amplitudes, as described in Section 7 below.}:

  \begin{eqnarray}
     \Delta A_3 & = & \frac{\tilde{{\cal A}}}{(x +\bar{x} +\frac{L}{2}-x_1)}
   \exp\left[ - i \kappa\left(c(t_D-t_0)-(x + \bar{x} +\frac{L}{2}-x_1)\right)\right]
      \nonumber \\
      &\times & \frac{\exp[i \kappa r_1]}{r_1} {\cal N}{\cal A}_{scatt} R_1 \Delta R_1 \Delta \phi_1 \Delta x_1
      \nonumber \\
       &\times  & \frac{\exp[i \kappa r_2]}{r_2} {\cal N}{\cal A}_{scatt} R_2 \Delta R_2 \Delta \phi_2 \Delta x_2
     \nonumber \\
      & \times & \frac{\exp[i \kappa r_3]}{r_3} {\cal N}{\cal A}_{scatt}  R_3 \Delta R_3  \Delta \phi_3 \Delta x_3
 \end{eqnarray} 
    
     Changing the integration variables to $r_1$, $r_2$ and $r_3$, as in
     deriving (4.16) from (4.15)  and performing the integrations,
     neglecting, as in (5.2), the contributions from the upper limits of the integrals, on 
      performing the azimuthal integrations,
     due to their rapid phase variation, gives:
   \begin{eqnarray}
     \Delta A_3 & = & \frac{\tilde{{\cal A}}}{(x +\bar{x} +\frac{L}{2}-x_1)}
   \exp\left[ - i \kappa\left(c(t_D-t_0)-(x +\bar{x} +\frac{L}{2}\right)\right]
      \nonumber \\
      &\times & \exp \left [i \kappa \left( -x_1 + (x_1-x_2) + (x_2-x_3) +x_3 \right)\right]
      \left(\frac{i 2 \pi {\cal N}{\cal A}_{scatt}}{\kappa}\right)^3 \Delta x_1 \Delta x_2 \Delta x_3
      \nonumber \\
       & \simeq  &\frac{\tilde{{\cal A}}}{x_{SD}}
   \exp\left[ - i \kappa\left(c(t_D-t_0)-x_{SD}\right)\right]
      \left(\frac{i 2 \pi {\cal N}{\cal A}_{scatt}}{\kappa}\right)^3 \Delta x_1 \Delta x_2 \Delta x_3    
 \end{eqnarray} 
    Where the approximation $x +\bar{x} +\frac{L}{2}-x_1 \simeq x +\bar{x} +\frac{L}{2} = x_{SD}$
  has been made in the denominator, where $x_{SD}$ is the source-detector distance. Integrating now over $x_1$, $x_2$ 
  and $x_3$ gives:
     \begin{eqnarray}
      A(L)_3 & = & \frac{\tilde{{\cal A}}}{x_{SD}}
   \exp\left[ - i \kappa\left(c(t_D-t_0)-x_{SD}\right)\right]
      \left(\frac{i 2 \pi {\cal N}{\cal A}_{scatt}}{\kappa}\right)^3
      \int_{x_2}^{\bar{x} +\frac{L}{2}} d x_1  \int_{x_3}^{\bar{x} +\frac{L}{2}} d x_2
   \int_{\bar{x} -\frac{L}{2}}^{\bar{x} +\frac{L}{2}} dx_3
   \nonumber \\
     & = & \frac{\tilde{{\cal A}}}{x_{SD}}
     \exp\left[ - i \kappa\left(c(t_D-t_0)-x_{SD}\right)\right]
     \frac{1}{3!} \left(\frac{i 2 \pi {\cal N}{\cal A}_{scatt} L}{\kappa}\right)^3  
 \end{eqnarray}
  The limits of the $x_1$, $x_2$ and $x_3$ integrals are determined by the inequalities
  (see Fig.2) $x_1 \ge x_2 \ge x_3$. In (5.15) the general expression for the 
  nested $n$-fold integral: 
  \begin{equation}
  {\cal I}_n =   \int_{x_2}^{\bar{x} +\frac{L}{2}} d x_1  \int_{x_3}^{\bar{x} +\frac{L}{2}} d x_2
   ...\int_{\bar{x} -\frac{L}{2}}^{\bar{x} +\frac{L}{2}} d x_n = \frac{L^n}{n!}
   \end{equation}
   is used. This formula is derived in Appendix B.
   For the case of $n$ scatterings of the photon in the block, (5.15) generalises to:
   \begin{equation}
     A(L)_3  =  \frac{\tilde{{\cal A}}}{x_{SD}}
   \exp\left[ - i \kappa\left(c(t_D-t_0)-x_{SD}\right)\right]
    \frac{1}{n!} \left(\frac{i 2 \pi {\cal N}{\cal A}_{scatt} L}{\kappa}\right)^n 
   \end{equation}
    Adding the contribution of the unscattered photon from (4.19) gives, for the fully integrated
    path amplitude (i.e. the probability amplitude) at the detector:
   \begin{eqnarray}
      A(L)_{tot} & = & A +  A(L)_1 +  A(L)_2 + ...  \nonumber \\
          &  = & \frac{\tilde{{\cal A}}}{x_{SD}}
   \exp\left[ - i \kappa\left(c(t_D-t_0)-x_{SD}\right)\right] \nonumber \\
        & \times & \left[ 1+\frac{i 2 \pi {\cal N}{\cal A}_{scatt} L}{\kappa}
        +  \frac{1}{2!} \left(\frac{i 2 \pi {\cal N}{\cal A}_{scatt} L}{\kappa}\right)^2 
         +...\right]  \nonumber \\
    & = & \frac{\tilde{{\cal A}}}{x_{SD}}
   \exp \left[ - i \kappa\left(c(t_D-t_0)-x_{SD}-\frac{2 \pi {\cal N}{\cal A}_{scatt} L}{\kappa^2}
    \right)\right] \nonumber \\
    & = & \frac{\tilde{{\cal A}}}{x_{SD}}
   \exp  \left[ - i \kappa\left(c(t_D-t_0)-(x_{SD}-L+nL)\right)\right]  
   \end{eqnarray}
    Where the refractive index, $n$, defined in (5.8), has been introduced. Defining, as in
    (5.4) above, $f$ as the fraction of the total distance between the source atom and the
    photon detector occupied by the transparent medium:
  \begin{equation}
  f = \frac{L}{x_{SD}}
  \end{equation}
   (5.18) may be written as:
     \begin{equation}
   A(L)_{tot} = \frac{\tilde{{\cal A}}}{x_{SD}}
   \exp\left[ - i \kappa\left(c(t_D-t_0)-\frac{c x_{SD}}{v(f)}\right)\right]
    \end{equation}
    where:
        \begin{equation}
     v(f) = \frac{c}{1+(n-1)f}
       \end{equation}
     which may be compared with (5.6) and (5.7) above.
      The phase of the amplitude in (5.20) agrees with that in (5.6) above,
      derived by considering single scattering of the photon in a thin sheet.
     \par  The calculation of the refractive index is now repeated taking properly into 
      account the time interval $\Delta t \equiv t_D-t_0$ between the time of production 
      of the excited atom $t_0$ and the time of photon detection $t_D$. Referring 
      to Fig.3 it can be seen that, for the configuration of excited atom and detector 
      shown there, the minimum possible value of $\Delta t$ is equal to the 
      time of flight in vacuum, $x_{SD}/c$, of the photon between the excited atom and the 
      detector and correponds to the straight line path SQCD between the source, S,
     and the detector. It also corresponds to the case when the excited atom decays promptly after
      production. For later detection times, the photon may follow
       paths displaced from
      the axis SQCD but, considering, for example, the case of triple scattering,
      the upper limits on the $r_1$, $r_2$ and $r_3$ integrals in (5.14) will now be
      restricted by the value of $\Delta s+x_{SD}$, which is the maximum
      path length allowed by the actual values of $t_D$ and $t_0$ and the source-detector
      distance  $x_{SD}$:
      \begin{equation}
      \Delta s \equiv c(t_D-t_0)-x_{SD} = c\Delta t-x_{SD} 
       \end{equation}
        Note that a path of length $\Delta s+x_{SD}$ still corresponds to prompt decay. Decays
     occuring at later times, for the same value of $t_D$, have path lengths less than
      $\Delta s+x_{SD}$.
       Inspection of Fig.3 shows that $r_1$, $r_2$ and $r_3$ must satisfy the inequality:
       \begin{equation}
       \Delta s \ge r_1 +  r_2 +  r_3- x_1  
       \end{equation}
        Noting also that;
      \begin{eqnarray}
       r_1 & \ge & x_1-x_2 \\
       r_2 & \ge & x_2-x_3 \\
       r_3 & \ge & x_3
       \end{eqnarray}
      the following inequalities may be derived from (5.23):
      \begin{eqnarray}
   x_1-x_2 & \le & r_1 \le  \Delta s-r_2-r_3+x_1 \\
   x_2-x_3 & \le & r_2 \le  \Delta s-r_3+x_2 \\
   x_3 & \le &  r_3 \le  \Delta s+x_3
    \end{eqnarray}
    In the present case, the definite integral derived from (5.13) above is:
    \begin{eqnarray}
   A(L)_3 & = & \frac{\tilde{{\cal A}}}{x_{SD}}
   \exp\left[ - i \kappa\left(c(t_D-t_0)-x_{SD}+x_1 \right)\right](2 \pi {\cal N}{\cal A}_{scatt})^3
   \int_{x_2}^{\bar{x} +\frac{L}{2}} d x_1  \int_{x_3}^{\bar{x} +\frac{L}{2}} d x_2
   \int_{\bar{x} -\frac{L}{2}}^{\bar{x} +\frac{L}{2}} dx_3 
   \nonumber \\
      & \times &  \int_{x_1-x_2}^{\Delta s-r_2-r_3+x_1} \exp[i\kappa r_1]  d r_1
   \int_{x_2-x_3}^{\Delta s-r_3+x_2} \exp[i\kappa r_2]  d r_2
    \nonumber \\
     & \times & \int_{x_3}^{\Delta s+x_3} \exp[i\kappa r_3]  d r_3 
  \end{eqnarray}
  Performing the integrals over $r_1$, $r_2$, $r_3$, $x_1$, $x_2$ and $x_3$, as shown in
   Appendix C, yields the result:
   \begin{eqnarray}
     A(L)_3 & = & \frac{\tilde{{\cal A}}}{x_{SD}}
      \exp\left[ - i \kappa\left(c(t_D-t_0)-x_{SD}\right)\right] \nonumber \\
       & \times  & \frac{1}{3!}\left(\frac{i 2 \pi {\cal N}{\cal A}_{scatt} L}{\kappa}\right)^3
        \left[\left(-\frac{(i\Delta \Phi)^2}{2!} + i \Delta \Phi -1\right)\exp[i \Delta \Phi]+1\right]
    \end{eqnarray}
     where:
     \begin{equation}
     \Delta \Phi \equiv \kappa  \Delta s
     \end{equation}
     The nested $r_1$, $r_2$ and $r_3$ integrals in (5.30) generalise in a straightforward
     manner to an arbitary number of photon scatterings. The expression for the probability amplitude
     in (5.20) above then replaced by:
  \begin{eqnarray}
      A(L)_{tot} & = &  \frac{\tilde{{\cal A}}}{x_{SD}}
   \exp\left[ - i \kappa\left(c(t_D-t_0)-x_{SD}\right)\right] \nonumber \\
        & \times & \left\{ \exp[i\beta L] + \exp[i\Delta \Phi]\left[ -i\beta L +
       \frac{(i\beta L)^2}{2!}\left(i\Delta \Phi-1\right).\right.\right. \nonumber \\
        & + & \frac{(i\beta L)^3}{3!}\left(-\frac{(i\Delta \Phi)^2}{2!}
        + i\Delta \Phi-1\right)  \nonumber \\
        & + & \left .\left .  \frac{(i\beta L)^4}{4!} \left(\frac{(i\Delta \Phi)^3}{3!}
         -\frac{(i\Delta \Phi)^2}{2!} 
        + i\Delta \Phi-1\right)+ ...  \right] \right \}
   \end{eqnarray}
      where 
      \begin{equation}
      \beta \equiv \frac{ 2 \pi {\cal N}{\cal A}_{scatt}}{\kappa}
  \end{equation}
    For vanishingly small values of $\Delta \Phi$,
    the term containing the large square brackets tends to the limiting value:
    $1-\exp[i\beta L]$ so that $ A(L)_{tot}$ reduces to the vacuum path amplitude
    between the source and the detector and the refractive effect of the block of
    transparent medium disappears.
    \par In order to discuss the probability amplitude  for arbitary values of
   $\Delta \Phi$, it is convenient to re-write the infinite series in the curly
   brackets of (5.33) separated out into real and imaginary parts. The algebraic manipulations
   necessary for this are discussed in Appendix D. The following expression
   is obtained:
    
 \begin{eqnarray}
      A(L)_{tot} & = &  \frac{\tilde{{\cal A}}}{x_{SD}}
   \exp\left[ - i \kappa\left(c(t_D-t_0)-x_{SD}\right)\right] \nonumber \\
        & \times & \left\{1 + \beta L[S+i(1-C)] +  \frac{(\beta L)^2}{2!}\left[[
         C-1+\Delta \Phi S  + i[S-\Delta \Phi +\Delta \Phi (1-C)]\right] \right. \nonumber \\
        & + & \frac{(\beta L)^3}{3!}\left[\Delta \Phi - S +\Delta \Phi (C-1) +\frac{(\Delta \Phi)^2}{2!}S
          \right. \nonumber \\
       & + & \left. i[C-1+\frac{(\Delta \Phi)^2}{2!}+ \Delta \Phi(S-\Delta \Phi)
            +\frac{(\Delta \Phi)^2}{2!}(1-C)]\right]  \nonumber \\
        & + &  \frac{(\beta L)^4}{4!} \left[
            1-\frac{(\Delta \Phi)^2}{2!}-C+ \Delta \Phi(\Delta \Phi-S)+
            \frac{(\Delta \Phi)^2}{2!}(C-1)+\frac{(\Delta \Phi)^3}{3!}S \right.  \nonumber \\
         &+& \left. i[ \Delta \Phi - \frac{(\Delta \Phi)^3}{3!} -S +
         \Delta \Phi (C-1 +\frac{(\Delta \Phi)^2}{2!} )- \frac{(\Delta \Phi)^2}{2!}(\Delta \Phi-S)
          + \frac{(\Delta \Phi)^3}{3!}(1-C)]\right] \nonumber \\
        & + & \left. ...\right\}
   \end{eqnarray}
    where
  \[ S\equiv \sin \Delta \Phi~~~,~~~ C\equiv \cos \Delta \Phi \]
    This formula for the probability amplitude may be written in a more compact fashion by introducing
    a notation for truncated series expansions of trigonometric functions: 
 \begin{eqnarray}
     S_0 & = & 0,~~ S_1 \equiv  \Delta \Phi,~~ S_2 \equiv \Delta \Phi-  \frac{(\Delta \Phi)^3}{3!},~~
      S_3 \equiv \Delta \Phi-  \frac{(\Delta \Phi)^3}{3!}+ \frac{(\Delta \Phi)^5}{5!} ,...\\
   C_0 & = & 1,~~ C_1 \equiv  1- \frac{(\Delta \Phi)^2}{2!},~~ C_2 \equiv 1- \frac{(\Delta \Phi)^2}{2!}
       + \frac{(\Delta \Phi)^4}{4!},~~ , ...
   \end{eqnarray}
    The expression in the large curly brackets of (5.35) is denoted as $F(\Delta \Phi,\beta L)_{ref}$.
    It is the complex amplitude that multiplies the vacuum probability amplitude $A$
    to take into account the interaction of
    the photon with the atoms of the refractive medium. Using the definitions in (5.36) and (5.37), 
    the real and imaginary parts of  $F_{ref}$ may be written as:
   \begin{eqnarray}
     Re F_{ref} & = & 1+\beta L S +  \frac{(\beta L)^2}{2!}(C+SS_1-1) + \frac{(\beta L)^3}{3!}(C S_1-S C_1 )
      \nonumber \\
      & + &  \frac{(\beta L)^4}{4!}(1-CC_1-SS_2) + \frac{(\beta L)^5}{5!}(SC_2-CS_2 ) + ... \nonumber \\
     & + & (-1)^{n} \frac{(\beta L)^{2n}}{2n!}(1-CC_{n-1}-SS_{n}) +(-1)^{n} \frac{(\beta L)^{2n+1}}{(2n+1)!}
       (S C_{n}-CS_{n}) \nonumber \\
     &  & + ...
 \end{eqnarray}
  \newpage
   \begin{eqnarray}
        Im F_{ref} & = & \beta L (1-C)+  \frac{(\beta L)^2}{2!}(S-C S_1) + \frac{(\beta L)^3}{3!}(CC_1+SS_1-1)
      \nonumber \\
      & + &  \frac{(\beta L)^4}{4!}(C S_2-S C_1)
        + \frac{(\beta L)^5}{5!}(1-CC_2-SS_2) \nonumber \\
       & + &   (-1)^{n}\frac{(\beta L)^{2n}}{2n!}(CS_{n}-S C_{n-1}) +(-1)^{n} \frac{(\beta L)^{2n+1}}{(2n+1)!} 
       (1-CC_{n}-SS_{n})   \nonumber \\
      &   &  + ... 
     \end{eqnarray}
     (5.38) may also be written as:
    \begin{eqnarray}
     Re F_{ref} & = & 1+\beta L S +\frac{\beta L}{2}(\beta L)(C+SS_1-1)
       + \frac{\beta L}{3}\frac{(\beta L)^2}{2!}(C S_1-S C_1)
      \nonumber \\
      & + &  \frac{\beta L}{4} \frac{(\beta L)^3}{3!}(1-CC_1-SS_2)
       + \frac{\beta L}{5} \frac{(\beta L)^4}{4!}(S C_2- CS_2) \nonumber \\
       &   & +... \nonumber \\
       & + & \frac{\beta L}{2n}\left[(-1)^{n} \frac{(\beta L)^{2n-1}}{(2n-1)!} 
       (1-CC_{n-1}-SS_{n})\right] \nonumber \\
       & + & \frac{\beta L}{2n+1}\left[(-1)^{n} \frac{(\beta L)^{2n}}{(2n)!}
       (S C_{n}-CS_{n}))\right]\nonumber \\
       &   &  + ... 
     \end{eqnarray}  
     As will be discussed below, the values of quantities $\beta L$ and $\Delta \Phi$ typically 
     satisfy the condition $\beta L \gg \Delta \Phi$. Inspection of successive terms in the series
   (5.39) and (5.40) shows that there is a one-to-one correspondence between terms with
    similar trigonometric coefficients (e.g. $1-CC_1-SS_1$ in (5.39) and $1-CC_1-SS_2$ in
     (5.40)). However, in (5.40), the $(\beta L)^n/n!$ terms of (5.39) are multiplied by the numerically
     large factor $-\beta L/n$. The contribution of not-too-large values of $n$ to $ Re F_{ref}$
      is then expected to be much larger than the corresponding contribution to $Im F_{ref}$.
     For large values of $n$ both series are expected to converge rapidly, since in this case,
     all the trigonometric factors tend to zero:
    \[ C C_n+S S_n-1 \simeq C^2+S^2-1 = 0 \]
   \[ S C_n- C S_n \simeq SC-CS = 0 \]  
     It is then to be expected that, in general:
     \[ | Re F_{ref}| \gg | Im F_{ref}| \]
         which implies that the phase, $\Phi_F$, of the amplitude $ F_{ref}$
       is close to zero, independently of the values of  $\beta L$ and  $\Delta \Phi$. 
     In this case the phase of the vacuum path amplitude remains unchanged by the 
     interaction of the photon with the atoms of the refractive medium and
     a phenomenon that may be called `refraction annulment' is expected to occur.    
      Refraction, i.e. propagation of light in the medium with an apparently reduced
     velocity $c/n$, only occurs when, as in (5.14), the contributions from
     the upper limits of the $r$ integrals vanish on integration
    over the azimuthal angles due to their rapid phase variation.
     The latter is induced by the geometric boundaries of the refractive
     medium that determine the phase of the integrand.
    These are effectively randomly placed at the distance scales of the order 
     of the wavelength of light. This point
     is further discussed in Appendix A. However, for small values of $\Delta s$, the photon paths
     always remain away from the transverse boundaries of the refractive material so that the latter
     have no effect on the value of the integral over paths. This is the case in (5.33).
 
\begin{figure}[htbp]
\begin{center}
\hspace*{-0.5cm}\mbox{
\epsfysize7.0cm\epsffile{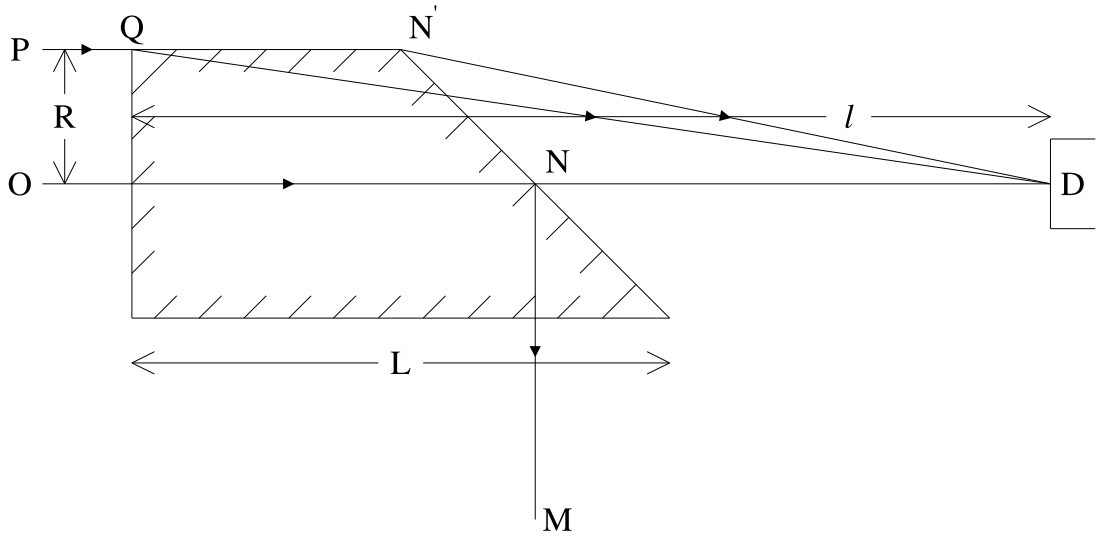}}
\caption{{\sl A simple experiment to search for refraction annulment.
 The side view is shown of a cylindrical glass block of
   refractive index $n = 1.5$, radius R and length L 
 with an oblique face
 containing the points N and N'. Far to the left is 
 an intense photon source consisting of excited sodium atoms.
 Photons produced by spontaneous decay of the atoms
 corresponding to the minimum detection time $t_D$ for a given production
 time $t_0$ of the excited atom,
 can reach the detector D only by paths near ON and for decays occuring
 promptly after production of the excited atom. In this case the phase
 change of the probability amplitude due to interactions with the atoms of 
 the block is strongly suppressed, the effective refractive index is very low, and so
 the photon
 can be transmitted through the oblique face and detected at D. For later detection times, photons
 may be produced
 at later decay times, and all photon paths between ON and PQN' are available.
 Normal refraction then occurs in the glass, and each photon is 
 totally internally reflected at the oblique face (e.g. the path ONM).}} 
\label{fig-fig4}
\end{center}
\end{figure}

     \par To estimate the importance of the refraction annulment effect consider
      the simple experimental configuration sketched in Fig.4. A prism is constructed
     by cutting obliquely the face of a glass cylinder of length, $L$,
     and radius, $R$, at 45$^\circ$ to the axis of the 
     cylinder. In Fig.4, the normal to the oblique plane face
     of the prism is in the plane of the diagram. The source atom is assumed to be 
     at a large distance to the left along the x-axis, which coincides with that 
     of the cylinder, so that all photons from the source have essentially the 
     same time-of-flight to the front face of the cylinder. The photon detector,
     $D$, is situated on the x-axis at distance ${\it l}$ from the front face of
     the prism. The largest possible value of $\Delta s$ for all the $r$ integrals
     to be limited by the production time of the excited atom, rather than by
     the geometrical boundaries of the prism, corresponds to a photon
     trajectory such as PQD. A simple geometrical calculation yields:
     \begin{equation}
      \Delta \Phi^{max} = \frac{2 \pi}{\lambda_{\gamma}^0}\Delta s^{max}
        =  \frac{2 \pi}{\lambda_{\gamma}^0}\frac{R^2}{2 {\it l}}+ O(\frac{R^4}{\lambda_{\gamma}^0{\it l}^3})
     \end{equation}
     If the exited atom is sodium in the upper state of the D-line transition,
      $\lambda_{\gamma}^0 = \lambda_{D} = 5.9 \times 10^{-5}$ cm. With $R = 5$cm
     and ${\it l} = 200$cm, $\Delta s^{max} = 625\mu m$ and $ \Delta \Phi^{max} = 
     6.66 \times 10^3$ rad or 1060 periods. Assuming a refractive index for
     the sodium D-line of $n = 1.5$ and using (5.9) and (5.34) with
     $L = 40$cm gives
    \begin{equation}
     \beta L = \frac{2\pi(n-1) L}{\lambda_D} = \frac{\pi L}{\lambda_D} = 2.12 \times 10^6
    \end{equation}
     Thus, in this case, $\beta L \gg  \Delta \Phi^{max} \gg 1$, so that,
       according to the argument given above,
     the phase of $F_{ref}$, and hence the corresponding refractive index, is expected to be
     very small. For a given
     geometrical configuration, the ratio of $\beta L$ to $\Delta \Phi^{max}$
     is a constant, independent of the dimensions of the apparatus.
     In order for the upper limits of the $r$ integrals to be independent of
     the geometrical boundaries of the prism, the excited atom
     must decay earlier than a time $\Delta s^{max}/c = 2.2 \times 10^{-12}$sec
     after production. Since the lifetime of the excited state producing
     the sodium D-lines is $5.4 \times 10^{-8}$sec, a fraction $\simeq   4 \times 10^{-4}$ 
     of all D-line photons emitted into the solid angle of the 
     detector is then expected to cross the prism following an unrefracted
     trajectory close to OND, as if they were propagating in free space.
      These are the atoms that decay during the
     time interval: $0 < t < \Delta s^{max}/c$ after production.
     A simple experiment using photographic or
     photon counting techniques would be sufficient to confirm of invalidate 
     this prediction. For larger values of $\Delta s$ (i.e. larger values of $t_D-t_0$)
     such that the upper limits of the $r$ integrals are determined by the geometrical
     boundaries of the prism, as in (5.14), the usual refraction phenomenon will
    occur in the medium and all the photons from the source will be internally
    reflected from the oblique face of the prism\footnote{The critical angle of
    incidence for total internal reflection is 42$^{\circ}$ for $n = 1.5$.} 
    so that the photons will follow trajectories close to ONM in Fig.4.

\SECTION{\bf{The Laws of Reflection, Refraction and Rectilinear Propagation
  of Light, and Fermat's Principle}}

\begin{figure}[htbp]
\begin{center}
\hspace*{-0.5cm}\mbox{
\epsfysize7.0cm\epsffile{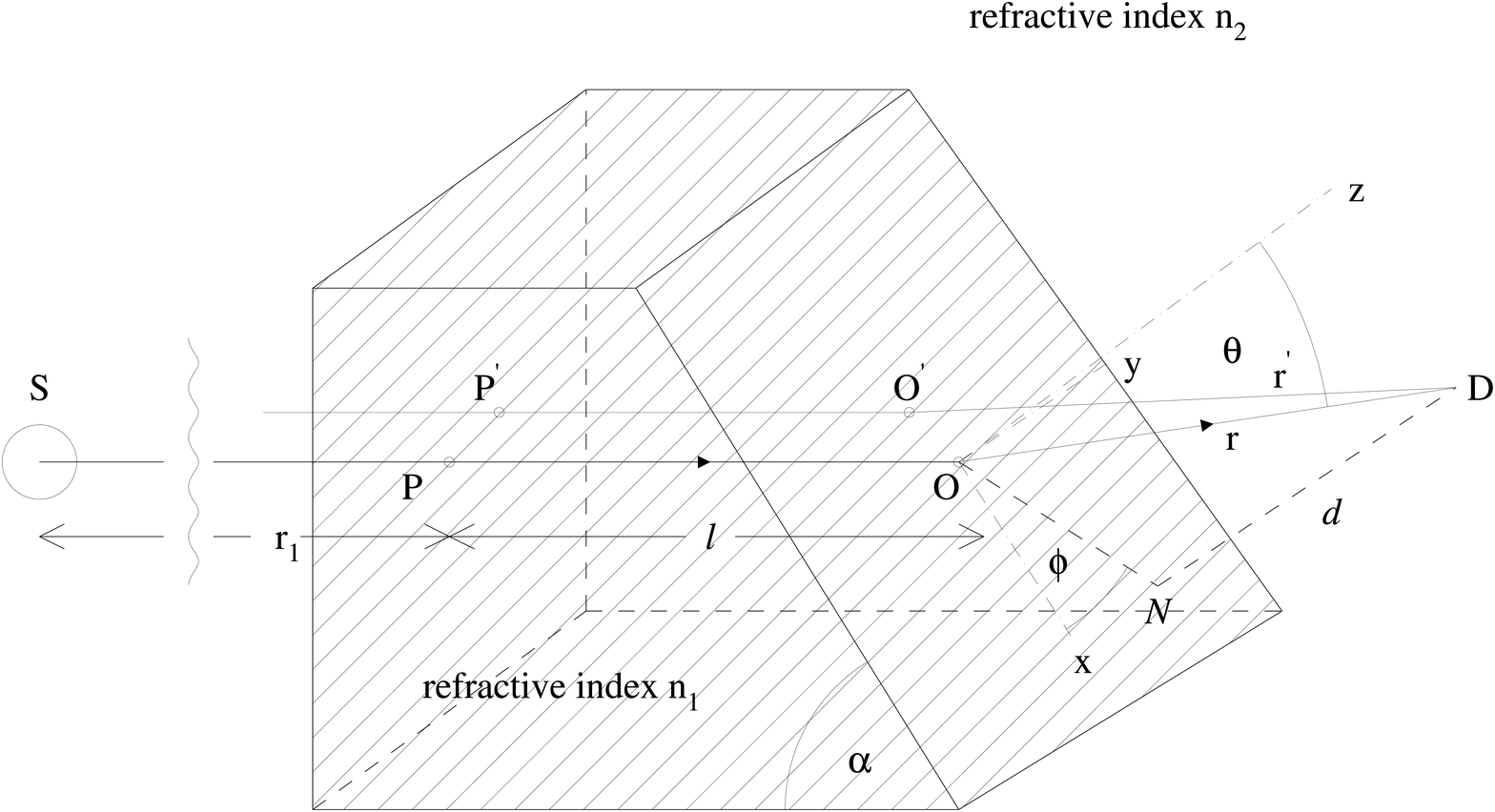}}
\caption{{\sl A photon produced by the source S is observed in a detector at D, which
 is movable in a plane distant $d$  from the oblique face of a block of
 transparent material of refractive index $n_1$ surrounded by a medium of 
 refractive index $n_2$. The position of D that 
 corresponds to an extremum of the phase of the probability amplitude for
 photon detection at D is found to correspond to Snell's law of 
 refraction.}} 
\label{fig-fig5}
\end{center}
\end{figure}

\begin{figure}[htbp]
\begin{center}
\hspace*{-0.5cm}\mbox{
\epsfysize7.0cm\epsffile{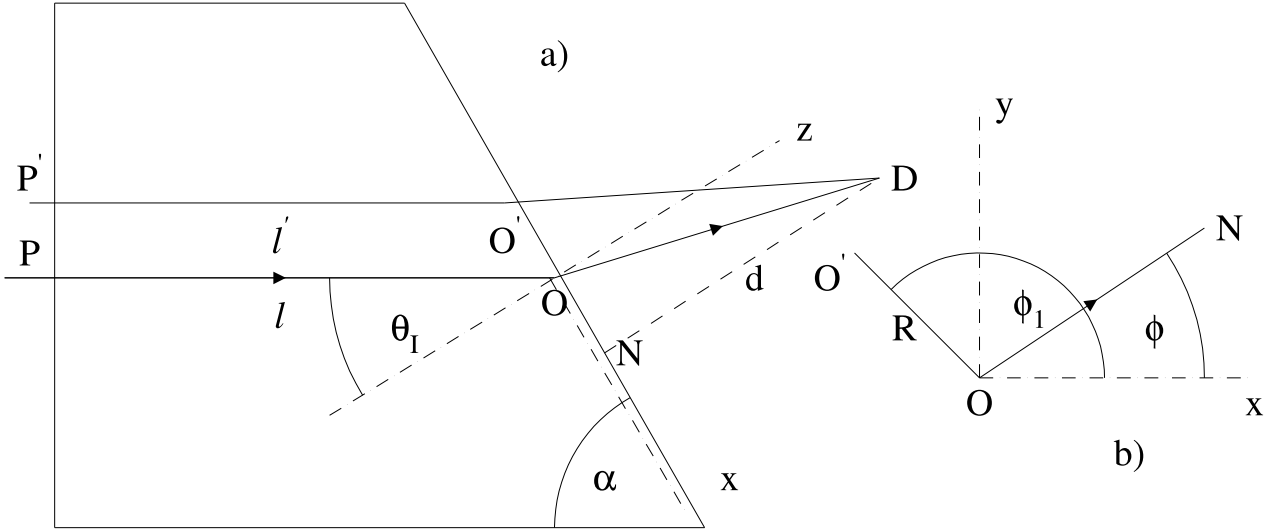}}
\caption{{\sl a) $xz$ projection, b) $xy$ projection, of the experimental
 layout of Fig.5.}} 
\label{fig-fig6}
\end{center}
\end{figure}

  In Fig.5 is shown the trajectory, SPOD, of a photon from the decay of an excited atom, S,
  at a point sufficiently far to the left of the figure that all straight line paths
  from the atom may be considered parallel to the line PO. Over the path segment, PO,
   of length, {\it l}, the photon propagates in a transparent medium of
   refractive index $n_1$. The surrounding space, back to the position of the excited atom, is 
    assumed to be filled with a transparent medium of refractive index $n_2$. The interface
 between the two media, passing through O, is a plane surface whose normal is at an angle
 $\theta_I = \pi/2-\alpha$ to the segment PO (see Fig. 6a), where, in the plane
  defined by PO and the normal, the angle between the projection of the interface and PO is
   $\alpha$. The photon detector, D,  is at a fixed distance,
  $d$, from the plane interface at a position defined by spherical polar coordinates $r$, $\theta$, $\phi$.
 The cartesian coordinate axes are defined so that the
  $z$ axis lies along the normal to the interface and the path segment PO is in the
  $xz$ plane.
 \par By analogy with (5.18) the probability amplitude corresponding to detection of the
  photon at D at time $t_D$ is\footnote{ In this section it is assumed that $c(t_D-t_0) -x_{SD}$
  is sufficiently large that there is no refraction annulment effect.}:
   \begin{equation}
{\cal A }    =  \frac{\tilde{\cal A}}{r_1 {\it l} r }
   \exp\left[ -\kappa c \left(t_D-t_0- \frac{n_2 r_1}{c}\right)+ i\Phi(r,{\it l}) \right]
 \end{equation}
    where
   \[ \Phi(r,{\it l}) \equiv \kappa (n_2 r +n_1 {\it l}) \] 
   The position of the detector D that maximises the photon detection probability is now determined by
  requiring that $ \Phi(r,{\it l})$ is stationary. As first pointed out by
   Dirac~\cite{Dirac2}, and much emphasised later by Feynman~\cite{Feyn5}, this condition ensures
   that neighbouring trajectories have almost the same phase, giving an amplitude, by vector addition
   in the complex plane, that maximises the modulus of the probability
   amplitude $A_{fi}$ in (2.1) above. 
   The phase is varied by parallel displacement of the trajectory to P'O',
   where the point O', in the $xy$ plane is specified by the coordinates: $x = R \cos \phi_1$,
   $y = R \sin \phi_1$ so that (see Figs.5 and 6) $r \rightarrow r'$ and ${\it l} \rightarrow {\it l}'$ where
  \begin{eqnarray}
   r'^2 & = & (d \tan \theta \cos \phi -R \cos \phi_1)^2
      + (d \tan \theta \sin \phi -R \sin \phi_1)^2+d^2 \\
   {\it l}' & = &  {\it l} +R \cos \phi_1 \cos \alpha
  \end{eqnarray}
    The parameters $\theta$, $\phi$, $R$ and $\phi_1$ are now varied so that the $r$, {\it l}
 dependent part, $\Phi$, of the phase in (6.1) is stationary. This requires that:
   \begin{equation}
  n_2 \frac{\partial r'}{\partial R}+ n_1\frac{\partial {\it l}' }{\partial R}
  =   n_2\frac{\partial r'}{\partial \phi_1}+ n_1\frac{\partial {\it l}' }{\partial \phi_1} = 0
  \end{equation}  
  Differentiating (6.2) and (6.3):
  \begin{eqnarray}
   \frac{\partial  r'}{\partial  R}  & = &  -\frac{d \tan \theta}{r'}(\cos \phi \cos \phi_1+ \sin \phi \sin \phi_1)
    +\frac{R}{r'} \\
  \frac{\partial  r'}{\partial \phi_1}  & = &  \frac{R d \tan \theta}{r'}(\cos \phi \sin \phi_1
    - \sin \phi \cos \phi_1) \\
  \frac{\partial {\it l}' }{\partial  R}  & = &   \cos \phi_1 \cos \alpha  \\
 \frac{\partial {\it l}' }{\partial \phi_1}  & = &   -R \sin \phi_1 \cos \alpha 
  \end{eqnarray} 
   Substituting (6.5) and (6.7) into the first member of (6.4) gives:
   \begin{equation}
  -\frac{n_2 d \tan \theta (\cos \phi \cos \phi_1+ \sin \phi \sin \phi_1)}{r'} + n_2 \frac{R}{r'}
   + n_1 \cos \phi_1 \cos \alpha = 0
 \end{equation}
  The solution to this equation with $R= \phi = 0$ is:
   \begin{equation}
  (-\frac{n_2 d \tan \theta }{r} + n_1 \cos \alpha) \cos \phi_1  = 0
 \end{equation}
 Since $d/r = \cos \theta$, (6.10) gives:
  \begin{equation}
   \sin \theta \equiv \sin \theta_O = \frac {n_1}{n_2} \cos \alpha =  \frac {n_1}{n_2} \sin \theta_I
\end{equation}
 which is Snell's law of refraction.
 Substituting (6.6) and (6.8) into the second member of (6.4) and setting $\phi = 0$ gives:
 \begin{equation}
  R(\frac{n_2 d \tan \theta }{r}- n_1 \cos \alpha) \sin \phi_1 = 0
\end{equation}
 which is verified both by the condition $R = 0$ and (6.11). (6.10) and (6.12) 
  show that the phase is stationary for an arbitary value of $\phi_1$ provided that 
  $R = \phi =0$.
  \par When $\alpha = \pi/2$, so that the photon is incident  normally on the plane interface,
   so that $\theta_I =0$,
   (6.11) gives $\theta_O = 0$. The stationary phase condition then 
 requires rectilinear propagation of the photon between S and D. 
  \par The above calculation is easily adapted to the case of reflection at the interface,
   within the medium of refractive index $n_1$, by the replacements in (6.12):
  \[ \theta \rightarrow \pi -\theta_R,~~~n_2 \rightarrow n_1 \]
   where $\theta_R$ denotes the angle of reflection relative to the inward normal at the interface.
   In this case (6.10) gives:
  \begin{equation}
 n_1(-sin \theta_R + \sin  \theta_I) \cos \phi_1 = 0
\end{equation}
 i.e., $ \theta_R =  \theta_I$, the law of reflection.
 \par The solution, (6.11), for the stationary phase with $R = \phi = 0$ and with 
  $\theta$ determined by Snell's law, implies that the photon trajectory
  corresponding to the stationary phase of the path amplitude lies in 
   the $xz$ plane defined by the incident trajectory PO and the normal to the surface
    separating the refractive media.
  \par The sizes of the deviations to be expected from the classical trajectory corresponding
  to the stationary  phase condition, as well as the spatial extent, within the media, of the regions where
  photon scattering processes contribute significantly to refraction or reflection will now be
   estimated using the path amplitude formalism.
  \par In the region of the stationary point, the variation of the phase is determined by the
   second order partial derivatives of the phase with respect to the trajectory parameters.
   For the angle of refraction $\theta_O$, Taylor's theorem gives:
   \begin{equation}
 \Delta \Phi =  \Phi - \Phi_{stat}= \frac{1}{2}\left[\frac{\partial^2 \Phi}{\partial R^2}
  \left(\frac{d R}{d \theta}\right)^2+
 \frac{\partial^2 \Phi}{\partial\phi_1^2}
  \left(\frac{d \phi_1}{d \theta}\right)^2 \right]_{stat} (\Delta \theta_O)^2
 \end{equation}
 From (6.1):
 \begin{eqnarray}
 \frac{\partial^2 \Phi}{\partial R^2} & = & \kappa \left[n_2  \frac{\partial^2 r'}{\partial R^2}
  + n_1  \frac{\partial^2 {\it l}'}{\partial R^2}\right] \\
 \frac{\partial^2 \Phi}{\partial\phi_1^2} & = & \kappa \left[n_2  \frac{\partial^2 r'}{\partial \phi_1^2}
  + n_1  \frac{\partial^2 {\it l}'}{\partial \phi_1^2}\right]
\end{eqnarray}
 Taking the R derivative of (6.5):
  \begin{equation}
 \frac{\partial^2 r'}{\partial R^2} = \frac{1}{r'^2}\frac{\partial r'}{\partial R} d \tan \theta
 (\cos \phi \cos \phi_1+ \sin \phi \sin \phi_1)+\frac{1}{r'}-\frac{R}{r'^2}\frac{\partial r'}{\partial R}
 \end{equation}
 Choosing now $\phi_1 = 0$ so that the photon trajectory is in the $xz$ plane, and substituting the parameters
  at the stationary point: $R = \phi =0$ in Eqns (6.5 ) and (6.17) gives:  
\begin{eqnarray}
 \left.\frac{\partial r'}{\partial R}\right|_{stat} & = &   -\frac{d \tan \theta_O}{r} \\
 \left.\frac{\partial^2 r'}{\partial R^2}\right|_{stat} & = &  \frac{d}{r}
 \left.\frac{\partial r'}{\partial R} \right|_{stat}  \tan \theta_O + \frac{1}{r} 
\end{eqnarray}
Combining (6.18) and (6.19):
\begin{eqnarray}
 \left.\frac{\partial^2 r'}{\partial R^2}\right|_{stat} & = &  -\frac{d^2}{r^3}\tan^2 \theta_O
   + \frac{1}{r} \nonumber \\
 & = &  \frac{1}{r} \left[1- \left(\frac{d}{r}\right)^2 \tan^2 \theta_O \right] \nonumber \\
 & = &  \frac{\cos^2 \theta_O}{r}
\end{eqnarray}
 The R derivative of (6.7) gives:
 \begin{equation}
  \frac{\partial^2 {\it l}'}{\partial R^2} = 0 
 \end{equation}
 (6.15), (6.20) and (6.21) may be combined to obtain:
  \begin{equation}
 \left.  \frac{\partial^2 \phi}{\partial R^2} \right|_{stat} =
  \kappa n_2 \frac{\cos^2 \theta_O}{r}
 \end{equation}
 Since $\cos \theta = d/r$,
   \begin{equation}
  \left. \left. \left. \left.\frac{d \cos \theta}{d R}\right|_{stat} = -\sin  \theta_O 
   \frac{d \theta}{d R}\right|_{stat} = - \frac{d}{r^2}
  \frac{d  r}{d  R}\right|_{stat} =  - \frac{d}{r^2}\frac{\partial r'}{\partial R}\right|_{stat} 
    = \frac{\sin  \theta_O  \cos  \theta_O }{r}
 \end{equation}
  That is,
  \begin{equation}
   \left. \frac{d R}{d \theta}\right|_{stat} = -\frac{r}{ \cos  \theta_O}
 \end{equation}
  Combining (6.14), (6.22) and (6.24) and noting that, since $\phi_1$ is chosen to be zero,
  the second term in square bracket in (6.14) vanishes,
    \begin{equation}
  \left. \Delta \Phi = \frac{1}{2} \frac{\partial^2 \Phi}{\partial R^2}\right|_{stat}
   \left.\left(\frac{d  R}{d \theta}\right|_{stat}\right)^2 (\Delta \theta_O)^2
   = \frac{\kappa n_2 r}{2} (\Delta \theta_O)^2
 \end{equation}
 With $\Delta \Phi = \pi$, the spread of the angle $\theta_O$ about its classical value 
   is then estimated to be:  
   \begin{equation} 
  \Delta \theta_O = \sqrt{\frac{2 \pi}{\kappa n_2 r}} = \sqrt{\frac{\lambda_{\gamma}^0}{n_2 r}}
 \end{equation}
\par In a similar way, the displacements parallel to the $x$ and $y$ axes, 
  $\Delta x_0$  and $\Delta y_0$, corresponding
 to $\Delta \Phi = \pi$ are found to be:
  \begin{eqnarray} 
  \Delta y_O & = &  \sqrt{\frac{\lambda_{\gamma}^0 r}{n_2}} \sec \theta_O \\
   \Delta x_O & = &  \sqrt{\frac{\lambda_{\gamma}^0 r}{n_2}}  
 \end{eqnarray}
  So, for the Sodium D-lines with $\lambda_{\gamma}^0 =5.9\times10^{-5}$ cm,
  refractive index of 1.5 and $r = 1$m, and normal incidence,
   $\Delta x_O =   \Delta y_0 = 6.3 \times 10^{-4}$ cm. If the transverse dimensions
 of the interface are larger than this the photon paths are well represented by
  the classical light-ray corresponding to the stationary phase condition. 
 For significantly smaller dimensions this is no longer the case and the photon paths
  are smeared over a much wider region. This is the diffraction-dominated domain of the classical
  wave theory of light.
   \par Introducing the effective velocity, $v$, of the photon in the refractive medium: $v = c/n$ according
 to (5.6) the phase $\Phi(r,{\it l})$ defined after (6.1) may be written as:
  \begin{equation}
   \Phi(r,{\it l}) = \kappa(n_2 v_2 t_2^{eff}+n_1 v_1 t_1^{eff}) =  \kappa c (t_2^{eff}+t_1^{eff})
    =  \kappa c T^{eff}
 \end{equation}
   where 
  \[ t_1^{eff} = \frac{{\it l}}{v_1},~~~t_2^{eff} = \frac{r}{v_2} \]
  and $ T^{eff} = t_2^{eff}+t_1^{eff}$ is the total effective propagation time of a photon from P to D in Fig.5.
   Since the
  propagation time from S to P' is a constant, independant of the position of P', a stationary value
   of $\Phi$ correponds  also to a stationary value of the total effective propagation time from the source to
   D. Thus the classical photon trajectory passing through D correponds to a stationary value of the
   photon's total effective propagation time, which is Fermat's Principle.

  \SECTION{\bf{The Reflection Coefficient of Light at Normal
  Incidence from a Plane Interface between Transparent Media}}

\begin{figure}[htbp]
\begin{center}
\hspace*{-0.5cm}\mbox{
\epsfysize10.0cm\epsffile{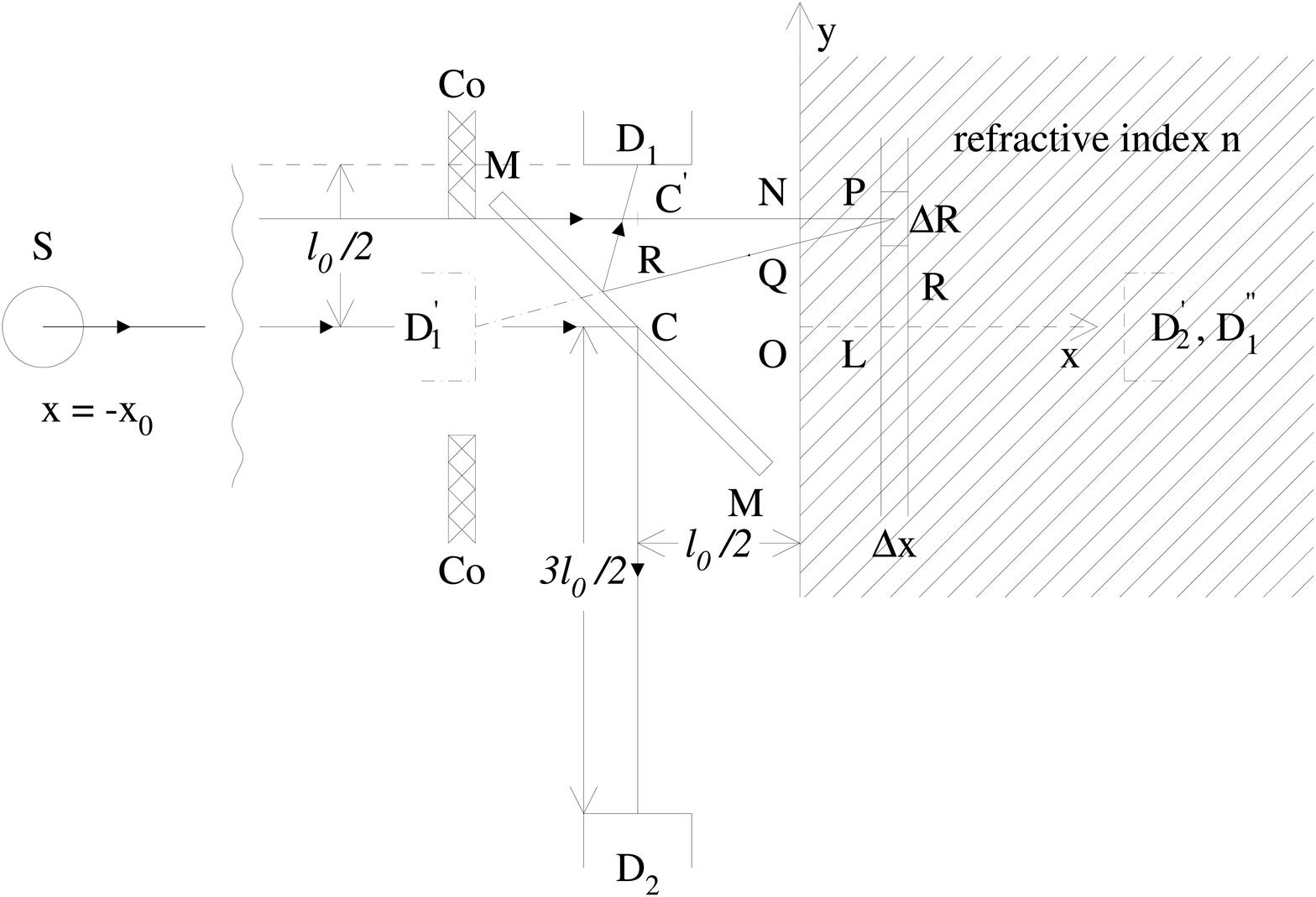}}
\caption{{\sl An apparatus to measure the reflection coefficient at
 normal incidence from a plane interface between vacuum and a block of uniform
 transparent material of refractive index $n$ (see text).}} 
\label{fig-fig7}
\end{center}
\end{figure}

 A practical experimental set-up to measure the reflection coefficient of light is shown in Fig 7. A plane
 interface between vacuum and a medium of refractive index $n$ lies in the $yz$ plane of a
  Cartesian coordinate system. $S$ is a light source (as previously a single excited atom 
  produced at time $t_0$) at a large distance $x = -x_0$ ($x_0 > 0$) from the interface.
  In an actual experiment this could conveniently be replaced by a small source in the focal plane of a converging
  lens or by a parallel laser beam.
  At $(x,y,z)= (-{\it l}_0/2, 0, 0)$ is a half-silvered mirror (HSM) (or an equivalent optical
  component) MM, with the normal to its surface in the $xy$ plane at an angle of 45$^{\circ}$ to the $x$-axis.
  A circular collimator CoCo  restricts the photon paths to a cylindrical region around the $x$-axis. Two small photon
   detectors $D_1$ and $D_2$ are situated at ($-{\it l}_0/2, {\it l}_0/2, 0$) and
  ($-{\it l}_0/2, -3{\it l}_0/2, 0$) respectively. The virtual images of $D_1$ and $D_2$ generated by 
  reflection in the HSM, as viewed from O and S respectively,
  are denoted as  $D'_1$ and $D'_2$, and that generated by reflection of $D_1$ 
   in both the HSM and the surface of the transparent medium, as viewed from S, by $D''_1$ .
   Since  $D'_2$ and $D''_1$ coincide
    (see Fig 7)  it can be seen that, with this arrangement,
   equal solid angles are subtended at the detectors by the source $S$. Photons detected in
   $D_1$ cross the HSM, are 
   backscattered from the transparent medium and reflected at the back surface of the HSM (e.g. path SNPQR$D_1$).
   Photons detected in  $D_2$ are reflected from the HSM directly into
   $D_2$ (e.g. path SC$D_2$). 
   \par It will be assumed that the decay width of the excited atom may be neglected (i.e. $\rho = 0$ in
    (4.11)). The time dependence of the path amplitude for a given detection time $t_D$ is then
    $\simeq \exp[-i \kappa c(t_D-t_0)]$. Assuming that the back surface of the HSM is half-silvered, as typically
   done in Michelson interferometers, which have a similar geometry, the probability amplitude for the 
  photon to be observed in $D_2$ is\footnote{Note that a photon detected in $D_2$ crosses the HSM twice}:
   \begin{equation}
   {\cal A}(D_2) = {\cal A}_0 {\cal A}_D T_{HSM}^2 R_{HSM}\exp[i(2\phi_{HSM}^T+\phi_{HSM}^R)]
    \exp\left[i\frac{3 \kappa {\it l}_0}{2}\right]
   \end{equation} 
   where ${\cal A}_0 {\cal A}_D $ denotes the probability amplitude for photon
   detection if  $D_2$ were situated in the plane $x_0 = - {\it l}_0/2$, in
   the absence of the HSM, (including also the time dependent factor mentioned above),
    and ${\cal A}_D$ is the process amplitude
   for photon detection. $T_{HSM}\exp[i\phi_{HSM}^T]$ and  $R_{HSM}\exp[i\phi_{HSM}^R]$
    are the transmisson and reflection amplitudes of the HSM. 
   \par The path amplitude, $\Delta {\cal A}(D_1)$  for the photon to be scattered at some point, P,
   in the interior of
   the transparent medium, and subsequently be detected in $D_1$ is given by the product of
    seven process amplitudes:
   \begin{eqnarray}
  \Delta {\cal A}(D_1) & = &   {\cal A}_D \langle D|\gamma|R \rangle R_{HSM}\exp[i\phi_{HSM}^R]
      \langle R|\gamma|P\rangle \nonumber \\
      &  & \times ({\cal N} {\cal A}_{scat} R \Delta R \Delta x \Delta \phi) \langle P|\gamma|C'\rangle
      T_{HSM}\exp[i\phi_{HSM}^T]{\cal A}_0
      \end{eqnarray}
       Here C' is the intersection with the plane $x = -{\it l}_0/2$ of the photon path SN, and ($R,\phi, x$) 
    are cylindrical coordinates specifying the position of P. Using (5.18) to write explicitly the triple
   product of photon propagators in (7.2), and integrating over all possible positions, P, of
   the scattering process  gives
    \begin{equation}
  {\cal A}(D_2) = \int \int \int \tilde{ {\cal A}}_1 \exp \left[i \kappa[\frac{{\it l}_0}{2}+{\it l}- r +n(x+r)]\right]
         \frac{{\cal N} {\cal A}_{scat} R dR dx d\phi}{{\it l}}
   \end{equation}
    where
 \[ \tilde{ {\cal A}}_1 \equiv {\cal A}_0 {\cal A}_D R_{HSM} \exp[i\phi_{HSM}^R] T_{HSM}\exp[i\phi_{HSM}^T] \] 
    In (7.3) the following spatial intervals have been introduced (see Fig 7):
    \[  {\it l} \equiv PD'_1,~~~ r \equiv PQ,~~~ x \equiv NP = OL \]
   Noting the relation:
    \begin{equation}
      \frac{{\it l}_0}{x} = \frac{{\it l} -r}{r}
    \end{equation}
     which is a consequence of the similarity of the triangles $D'_1QO$ and  $D'_1PL$, the distance $r$ may be
     eliminated from (7.3) to give:
    \begin{equation}
  {\cal A}(D_2) = \int \int \int \tilde{ {\cal A}}_1 
     \exp \left[i \kappa[\frac{{\it l}({\it l}_0+nx)}{{\it l}_0+x}+\frac{{\it l}_0}{2}+nx]\right]
         \frac{{\cal N} {\cal A}_{scat} R dR dx d\phi}{{\it l}}
   \end{equation}

   As, according to the Huygens-Fresnel Principle, only values of $x$ of order $\lambda_{\gamma}^0$ 
   contribute to the $x$ integral in (7.3), it follows that, in the integrand of (7.3),
    $x,r \ll {\it l}_0$ so that ${\it l}^2 =  R^2+({\it l}_0+x)^2 \simeq  R^2+{\it l}_0^2$ is a good
    approximation. Thus $RdR = {\it l} d{\it l}$. Using this relation to 
   eliminate $R$ in favour of {\it l} in (7.5) and performing the {\it l} integration,
    in a similar way to the $r_1$ integration in (5.1), gives:
   \begin{eqnarray}
  {\cal A}(D_2) & = & \int \int \frac{ \tilde{ {\cal A}}_1 ({\it l}_0+x)}{i \kappa ({\it l}_0-nx)}{\cal N} {\cal A}_{scat}
       dx d\phi \nonumber \\
      &   & \times \left\{ \exp \left[i \kappa[\frac{{\it l}_{max}(\phi)({\it l}_0+nx)}{{\it l}_0+x}+\frac{{\it l}_0}{2}-nx]\right]
                  -  \exp \left[i \kappa [\frac{3{\it l}_0}{2}+2nx] \right]\right\}   
   \end{eqnarray}
    The lower limit of the integral is ${\it l}_{min} = {\it l}_0+x$. On performing the $\phi$ integration
    the contribution from the upper limit of the {\it l} integral vanishes due to the rapid phase
     variation resulting from irregularities of size $\lambda_{\gamma}^0$ or larger in the shape of the
    collimator Co that determines ${\it l}_{max}(\phi)$.
    \par Performing the $x$ integral according to the Huygens-Fresnel Principle (the integral is equal to
     one half of the contribution of the first half-period in $x$), and neglecting $x$ relative to ${\it l}_0$,
    except in the phases of the path amplitudes, gives the final result for ${\cal A}(D_2)$:
   \begin{eqnarray}
  {\cal A}(D_2) & = & - \tilde{ {\cal A}}_1 \left(\frac{\pi {\cal N} {\cal A}_{scat}}{n \kappa^2}\right)
    \exp[i\frac{3 \kappa {\it l}_0}{2}]  \nonumber \\
         & = & -\frac{\tilde{ {\cal A}}_1}{2n}\left(\frac{1}{2\pi} (\lambda_{\gamma}^0)^2
      {\cal N} {\cal A}_{scat} \right) \exp[i\frac{3 \kappa {\it l}_0}{2}] \nonumber \\
      & = & -\tilde{{\cal A}}_1 \left(\frac{n-1}{2n}\right) \exp[i\frac{3 \kappa {\it l}_0}{2}]       
     \end{eqnarray}
   where, in the last line, (5.9) has been used. (7.7) gives the reflection coefficient at normal
    incidence, calculated using the Feynman path ($FP$) method, $\rho_R^{FP}$:
   \begin{equation}
 \rho_R^{FP} = \left|\frac{n-1}{2n}\right|^2
    \end{equation}
    It is measured by observing the ratio of the counting rates of the similar detectors (assumed
    to be equally efficient), $D_1$ and $D_2$:
    \begin{equation}
     \frac{Rate(D_1)}{Rate(D_2)} = \frac{|{\cal A}(D_1)|^2}{|{\cal A}(D_2)|^2} =\frac{\rho_R^{FP}}{T_{HSM}^2}     
\end{equation}
The modulus of the transition amplitude of the HSR, $T_{HSM}$  is readily measured by replacing the block of
 transparent material with  a plane specular reflector of known reflectivity.
  \par In should be noticed that $\rho_R^{FP}$ differs markedly from the Fresnel prediction
  for the reflection coefficient at normal incidence:
    \begin{equation}
  \rho_R^{Fresnel} = \left|\frac{n-1}{n+1}\right|^2
    \end{equation}
 For example, for a glass/vacuum interface, with $n_{
 glass} = 1.5$, $\rho_R^{FP} = 0.028$, 
   $\rho_R^{Fresnel} = 0.040$, a difference of 43$\%$. A simple experiment similar to that sketched
   in Fig 7 could easily discriminate between these predictions. Another interesting difference
   with respect to the Fresnel formula is related to the minus sign on the RHS of (7.7). This sign,
     for the Fresnel prediction, is, in fact, ambiguous. Taking the limit as the angle of incidence 
     tends to zero for photons linearly polarised perpendicular to the plane of incidence gives for the reflected
     amplitude $(n-1)/(n+1)$, whereas the similar limit for photons polarised in the plane of
     incidence is $-(n-1)/(n+1)$~\cite{BW3}. Of course, at normal incidence, the terms `perpendicular'
    and `parallel' polarisation become meaningless as the `plane of incidence' is no longer
    defined. However, the actual phase shift for reflection at normal incidence is certainly
    measurable, and the path amplitude calculation, unlike the Fresnel formula, gives the
    definite prediction, $\pi$, for the phase shift. 
   \par This phase shift could be measured by replacing the the detector $D_2$ by a suitably placed
    plane specular mirror and adding an optical attenuator so that the light reflected from 
    this mirror and detected in $D_1$, and the light scattered by the transparent medium, reflected
    by the HSR and detected in $D_1$ have a similar intensity so as the maximise interference
    effects  for a photon passing along either of the two possible paths. Such an experiment,
    using a suitably calibrated optical attenuator could also measure the reflection coefficient
    with  a single photon detector. The so-modified apparatus is in fact a Michelson interferometer
    with one of the specular mirrors relaced by a vacuum/glass interface. The Feynman path amplitude
    description of a conventional  Michelson interferometer is given in the next section.
    \par The result (7.7) for the probabilty amplitude is easily adapted to the case when the vacuum
    in Fig 7 is replaced by a uniform transparent medium of refractive index $n_1$ while the refractive
    index of the original block of transparent medium is denoted by $n_2$. Suitably modifying
     the photon propagators in (7.2), using (5.18) and calculating the contribution to the 
     probability amplitude from photon paths scattered from the atoms of the medium of refractive
     index $n_1$, gives, instead of (7.7), the expression:
  \begin{eqnarray}
  {\cal A}(D_2) & = & -\tilde{{\cal A}}_1\left(\frac{n_2-1}{2n_1n_2} - \frac{n_1-1}{2n_1^2}\right)
   \exp[i\frac{3 \kappa {\it l}_0}{2}] \nonumber \\
   & = & -\tilde{{\cal A}}_1\left(\frac{n_2-n_1}{2n_1^2 n_2}\right) \exp[i\frac{3 \kappa {\it l}_0}{2}]
  \end{eqnarray}
  which may be compared with the Fresnel formula, which gives, instead, for the quantity in the 
  large curved brackets of (7.11): $\pm(n_2-n_1)/(n_2+n_1)$.
  The minus sign of the second term in the large curved brackets in the first line of (7.11) results
   from a reversal of the order of the limits in the $x$ integration over the first half-period zone,
   since the atoms of the medium of refractive index $n_1$ are situated at negative $x$.
     Setting $n_2 = n$ , $n_1 = 1$ in (7.11) recovers the previous result (7.7). If $n_1 > n_2$ there is 
    no phase shift on reflection  in accordance with the well-known result of the classical
     wave theory of light. If $n_1 = n_2$ ${\cal A}(D_2)$ vanishes so there is no back-scattered
     light in a uniform refractive medium.
      According to (7.11) this may be interpreted as the result of perfect destructive
    interference between the path amplitudes of photons  back-scattered scattered from atoms
     with positive and negative $x$.
     \par Interference phenomena closely related to those occuring in `Newton's Rings'~\cite{Newton}
      that are extensively discussed in ~\cite{Feyn1} are simply analysed
     by suitably modifying the $x$ integration limits in (7.6) above. Replacing the block of transparent
    medium by a thin sheet of thickness $\lambda_{\gamma}^0(1+2p)/4n$, where $p$ is a positive integer,
    gives a reflection coefficient  four times larger than in (7.8). If the sheet has a thickness
    $\lambda_{\gamma}^0(1+p)/2n$,  $\rho_R^{FP}$ vanishes due to perfect destructive interference, in 
    this case,
    of the path amplitudes corresponding to the scattering of the photon from each atom of the
    medium comprising the sheet.

  \SECTION{\bf{Spatio-Temporal Interference Effects in the Michelson Interferometer}}

\begin{figure}[htbp]
\begin{center}
\hspace*{-0.5cm}\mbox{
\epsfysize15.0cm\epsffile{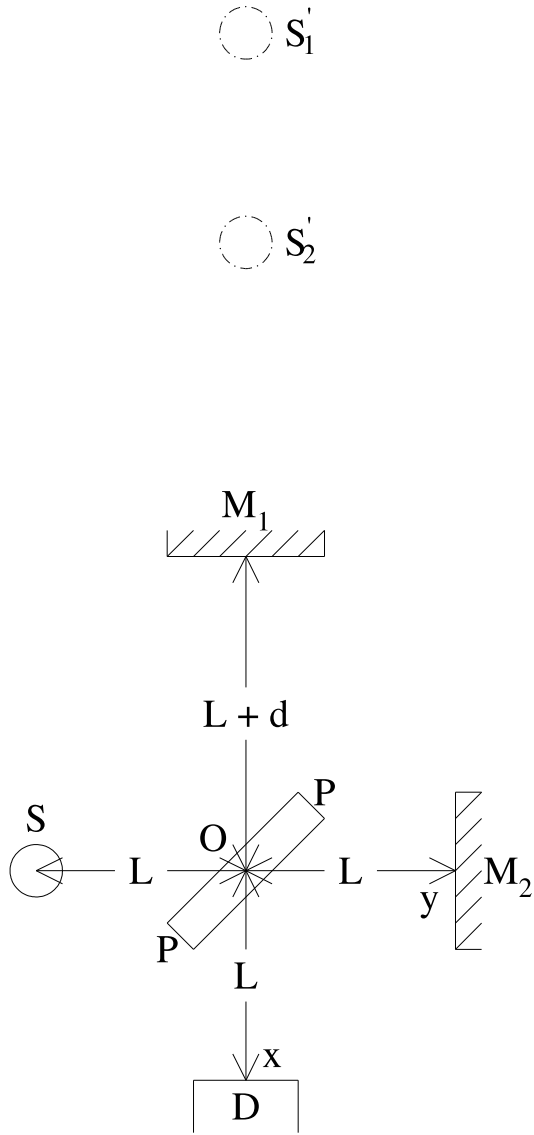}}
\caption{{\sl The Michelson interferometer. A photon from the decay of an excited
  atom at S, produced at time $t_0$, follows the paths SOM$_1$OD or  SOM$_2$OD in alternative histories of the decay
  process and subsequent space-time propagation of the photon, before being detected at D at time $t_D$. PP
   is a glass plate half-silvered on its back surface. $S'_1$ is the virtual image of S after reflection 
   first in PP then in the plane mirror $M_1$.  $S'_2$ is the virtual image of S after reflection 
   first in the plane mirror $M_2$ then in  PP. The path diffrence $2d$ is chosen to be $\geq c \tau_S$ where 
   $\tau_S$ is the mean lifetime of the excited source atom. Note that the atom must decay 
    at different times in its alternative histories in order that the photon arrives at the 
   detector at the fixed time $t_D$. It is just this time difference, determining the phase advance 
   of the propagator of the excited atom, that determines the size of the quantum interference
   term generated by the amplitudes corresponding to the two alternative paths. }} 
\label{fig-fig8}
\end{center}
\end{figure}

  \par A schematic layout of a Michelson interferometer~\cite{JW1} is shown in Fig 8.
    The photon source, $S$,
   a single excited atom, is produced at a known time, $t_0$. For concretness, as discussed in Section 2 above, this can be
  taken to be an atom of, say, sodium vapour, resonantly excited by a tuned, pulsed, laser beam.
   The pulse width should be at most a few nanoseconds and the time of passage should be known with a precision
  of 1ns or better.
  \par The photon produced by decay of the excited atom may follow the paths $SOM_1OD$ (reflection in by the
    mirror $M_1$) or $SOM_2OD$ (reflection in by the mirror $M_2$) before being detected, at time $t_D$, in the 
   photon detector at $D$. The glass plate PP situated at the origin of a Cartesian coordinate system,
   contains the $z$-axis of the latter and is inclined at 45$^{\circ}$ to the $x$- and $y$-axes. The back face
   of the plate is partially silvered to increase the moduli of the path amplitudes associated with
   the paths  $SOM_1OD$ and  $SOM_2OD$. In a practical interferometer, focussing lenses would be installed in
    the paths $SO$ and $OD$ to render the photon trajectories in the interferometer parallel to the $x$- and
    $y$-axes and increase the efficiency of photon detection. They would contribute a constant multiplicative
    factor to the interfering path amplitudes and so have no effect on the interference phenomena 
    discussed here. 
    \par The simplest way to understand and analyse the operation of the interferometer is to consider the virtual
   images $S'_1$ and $S'_2$ of the source, as reflected in the partially silvered plate and the mirrors
    $M_1$ and $M_2$, respectively, and viewed from the position of the photon detector~\cite{JW1} (see Fig 8).
    The following length intervals are defined:
    \[ SO = OM_2 = OD \equiv L,~~~OM_1 \equiv L+d  \]
    It then follows that:
    \[ S'_1D \equiv L_1 = 4L+2d,~~~S'_2D \equiv L_2 = 4L \]
     The probability amplitudes, ${\cal A}(1)$, ${\cal A}(2)$ for photons reflected at $M_1$,$M_2$ and detected 
     by D at time $t_D$ are given by (5.18) as:
     \begin{eqnarray}
   {\cal A}(1) & = & \frac{\tilde{{\cal A}_1}}{L_1} \exp[-i(\kappa c- \frac{i}{2 \tau_S})(t_D-t_0-\frac{L_1}{c})] \\
  {\cal A}(2) & = & \frac{\tilde{{\cal A}_2}}{L_2} \exp[-i(\kappa c- \frac{i}{2 \tau_S})(t_D-t_0-\frac{L_2}{c})]
    \end{eqnarray}
     where the the space-time independent amplitudes $\tilde{{\cal A}_1}$ and $\tilde{{\cal A}_2}$ include the
     production and detection process amplitudes as well as the amplitudes describing all transmission
     and reflection processes in the arms of the interferometer. In (8.1) and (8.2) the mean life
     $\tau_S = \hbar/\Gamma_S$ of the source atom has been introduced as in (3.12). For the Sodium
      D-lines $\tau_S \simeq 10$ns. This value will be taken for the quantitative predictions presented
     below.  Using (2.1) and (2.3), the probability, $P(t_D<t_D^{max})$ to detect the photon in D 
     during the time interval
      $t_0 < t_D < t_D^{max}$ is given by the following expressions:
       \begin{itemize}
    \item[(i)] For: $ t_D^{max} \le t_0 +L_2/c$
       \end{itemize}
     \begin{equation}
    P(t_D < t_D^{max}) = 0
     \end{equation}
     \begin{itemize}
    \item[(ii)] For: $t_0 +L_2/c < t_D^{max} \le t_0 +L_1/c$
       \end{itemize}
   \begin{equation}
   P(t_D < t_D^{max}) = \int_{t_0 +L_2/c}^{t_D^{max}} | {\cal A}(2)|^2 d t_D = \frac{\tau_S |\tilde{{\cal A}_2}|^2}
    {L_2^2}\left[1 - \exp[-\frac{1}{\tau_S}(t_D^{max}-t_0-\frac{L_2}{c})]\right]
    \end{equation}
      \begin{itemize}
    \item[(iii)] For: $t_0 +L_1/c < t_D^{max}$
     \end{itemize} 
 \begin{eqnarray}
   P(t_D < t_D^{max}) & = &   \int_{t_D^{min}}^{t_D^{max}} | {\cal A}(1)+{\cal A}(2)|^2 
   d t_D 
   \nonumber \\
  & = &  \int_{t_0 +L_2/c}^{t_D^{max}} | {\cal A}(2)|^2 d t_D +
     \int_{t_0 +L_1/c}^{t_D^{max}}(|{\cal A}(1)|^2 +2{\rm Re}[{\cal A}(1) {\cal A}(2)^{\ast}]) d t_D
  \nonumber \\
   & = & \tau_S \left\{ \frac{|\tilde{{\cal A}_2}|^2}{L_2^2}
 \left[1 - \exp[-\frac{1}{\tau_S}(t_D^{max}-t_0-\frac{L_2}{c})]\right] \right.
 \nonumber \\
  & + & \left[1 - \exp[-\frac{1}{\tau_S}(t_D^{max}-t_0-\frac{L_1}{c})]\right] \nonumber \\
   &\times & [ \frac{|\tilde{{\cal A}_1}|^2}{L_1^2} + 2 \frac{|\tilde{{\cal A}_1}||\tilde{{\cal A}_2}|}{L_1 L_2}
     \exp \left( -\frac{(L_1-L_2)}{2 c \tau_S}\right)
 \nonumber \\
 &\times & \left. \cos\{\kappa (L_1-L_2)+\phi_{12}\} ] \right\}
  \end{eqnarray}
    where
  \[ \phi_{12} = {\rm phase}(\tilde{{\cal A}_1}) - {\rm phase}(\tilde{{\cal A}_2}) \]
    For condition (i) the photon can arrive at the detector by neither path, so its detection
    probability vanishes. For condition (ii) the photon can arrive at the detector only
     via the path$SOM_2OD$ so that there is no interference phenomenon. For condition (iii) the photon
    may arrive at the detector via either path so interference is possible. 
    With the aid of optical compensators in the arms $OM_1$, $OM_2$, $\tilde{{\cal A}_1}$ and 
     $\tilde{{\cal A}_1}$ may be chosen so that:
    \begin{equation}
     \frac{\tilde{{\cal A}_1}}{L_1} =  \frac{\tilde{{\cal A}_2}}{L_2} = K
    \end{equation}
    In this case (also setting $t_0= 0$) (8.5) simplifies to :
    \begin{eqnarray}
     P(t_D < t_D^{max}) & = & \tau_S K^2\{ 2-f(t_D^{max})(1+ e^{\frac{2d}{c \tau_S}}) \nonumber \\
        & + & 2(e^{-\frac{d}{c \tau_S}}-f(t_D^{max})e^{\frac{d}{c \tau_S}})\cos(2 \kappa d + \phi_{12})\}
    \end{eqnarray}
     where 
  \[   f(t) \equiv \exp\left[ -\frac{1}{\tau_S}(t-\frac{4 L}{c}) \right] \]
    Neglecting small corrections due to the exponential terms in (8.7)  the maximum and minimum values of 
    $P(t_D < t_D^{max})$ are:
    \begin{eqnarray}
       P_{max(min)} & = & \tau_S K^2\{ 2-f(t_D^{max})(1+ e^{\frac{2d}{c \tau_S}}) \nonumber \\
        & +(-) &2( e^{-\frac{d}{c \tau_S}}-f(t_D^{max})e^{\frac{d}{c \tau_S}})\}
    \end{eqnarray}
     and the `fringe visibility' ${\it V}(t_D < t_D^{max})$ is:
     \begin{equation} 
  {\it V}(t_D < t_D^{max}) = \frac{ P_{max}- P_{min}}{ P_{max}+ P_{min}}
       = \frac{2( e^{-\frac{d}{c \tau_S}}-f(t_D^{max})e^{\frac{d}{c \tau_S}}) }
   {2-f(t_D^{max})(1+ e^{\frac{2d}{c \tau_S}})}
    \end{equation}
    Curves of ${\it V}(t_D < t_D^{max})$ as a function of $ t_D^{max}$ are presented in Fig 9 for:
    $\tau_S = 10$ns, $L = 50$cm and $d=$ 12.5, 25 and 50cm. Such interference effects as a function of
 $t_D^{max}-t_0$, are readily observable. It is sufficient to adjust the interferometer for maximum
 destructive or constructive interference near some fixed value $d = d_0$ and measure ${\it V}$, by
   small variations of $d$ around $d_0$, for different values of $t_D^{max}$.
   If $N_{max}$ ($N_{min}$) are the numbers of photons recorded with
   $t_D<t_D^{max}$  when the interferometer is adjusted for maximum constructive (destructive) 
   interference then:
    \begin{equation}
     {\it V}(t_D < t_D^{max})_{exp} = \frac{N_{max}-N_{min}}{N_{max}-N_{min}}
    \end{equation}
     which may be compared with the prediction of (8.9). Fitting this prediction to the 
    data determines the the only parameter, $\tau_S$, that is not fixed by the geometry of the
    experiment.\

\begin{figure}[htbp]
\begin{center}
\hspace*{-0.5cm}\mbox{
\epsfysize10.0cm\epsffile{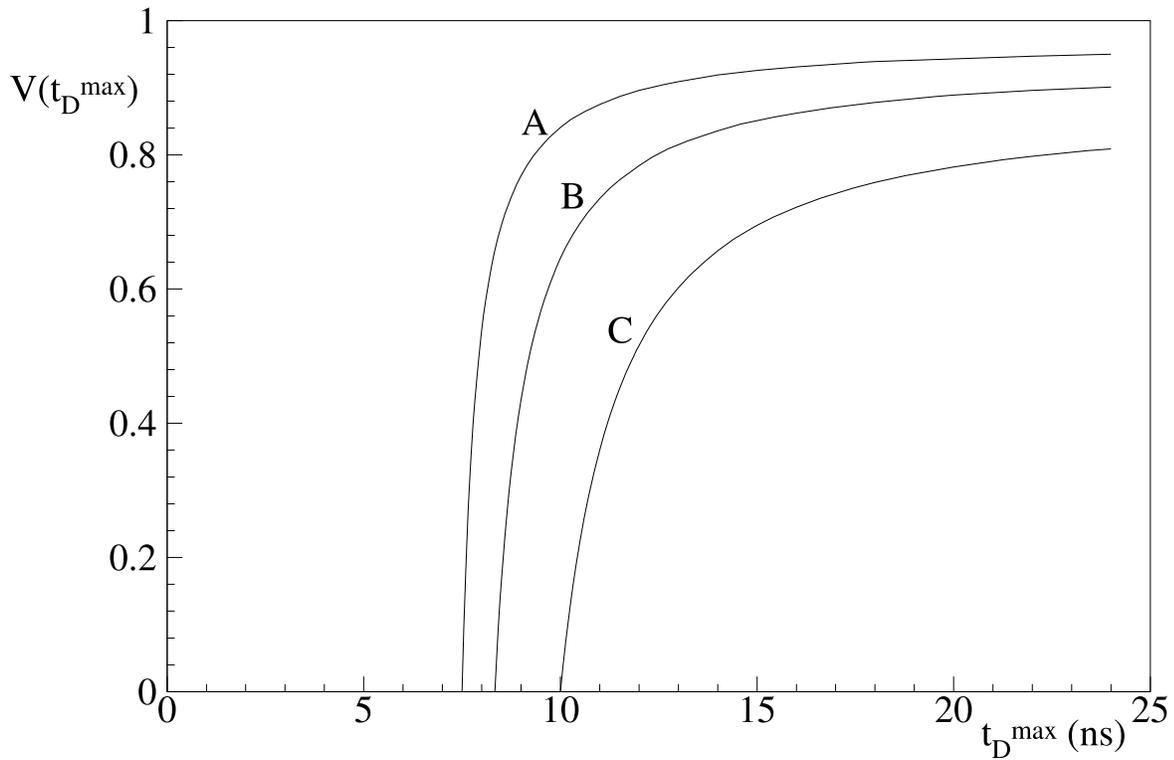}}
\caption{{\sl Fringe visibilty  ${\it V}$ in the Michelson interferometer of Fig.8 as a function
  of $t_D^{max}$ the maximum observation time of the photon at D. The excited source atom 
  is produced at time $t=0$. $L= 50$cm and $\tau_S = 10$ns. The curves A,B and C correspond to $d= 12.5, 25$ and 50cm.
  }} 
\label{fig-fig9}
\end{center}
\end{figure}

    \par Choosing a value of $t_D^{max} \gg c \tau_S$ results in $f(t_D^{max})\simeq 0$. The time-integrated
    fringe visibility as a function of $d$, ${\it V}_{\infty}(d)$, is then qiven 
    by (8.9) as 
    \begin{equation}
    {\it V}_{\infty}(d) = {\it V}(t_D < \infty ) = \exp\left(-\frac{d}{c \tau_S}\right)
    \end{equation} 
     Thus the time integrated visiblity is predicted to decrease exponentially with the path
    difference $2d$. Exactly the same behaviour is predicted by the classical wave theory of
    light~\cite{MW3}. In this calculation, similar to that of Michelson~\cite{Michelson1},
    The visibility was calculated using the equation obtained by making the 
    substitutions $\kappa \rightarrow 2 \pi c \nu_{\gamma}$ and  $f(t_D^{max}) \gg c\tau_S$
   in (8.7) above, and by weighting the cosine interference term with
   a Lorentzian distribution in the freqency, $\nu_{\gamma}$, of the photon. 
     In the Feynman path amplitude calculation $\kappa$ is defined by the pole masses
    of the initial and final states (see (4.9) above), which, unlike the photon
    frequency, do not vary on an event-by-event basis. Thus, from the point-of-view of
    the Feynman path amplitude calculation, the classical wave calculation is incorrect,
    even though it predicts the same result.
    \par In the above treatment, two potentially important physical effects have been neglected.
    \par The first effect is `pressure broadening' of the atomic linewidth. This results in an observed
    value of $\tau_S$ less then that, $\tau_S^{nat}$, corresponding to the `natural' linewidth of
    an isolated, freely decaying, atom:
      \begin{equation}
       \frac{1}{\tau_S}=  \frac{1}{\tau_S^{nat}}+ \frac{1}{\tau_P}
     \end{equation}
      where $\tau_P$ is a characteristic lifetime parameter that tends to infinity as the 
      pressure of the source of excited atoms tends to zero. The physical origin of this effect
    is easily understood in the path amplitude language. In order for the source atom to retain
     the coherent phase relationship between different path amplitudes, following from
      (4.3), that is assumed in (8.1) and (8.2) above, the atom must remain in the excited state sufficiently
      long that unhindered spontaneous decay can occur. If this is not the case, for example,
    if the atom is dexcited, or excited into a different state by inter-atomic collisions, after the time
    corresponding to the `earlier'  path amplitude but before the time corresponding to the
    `later' one, no interference will be possible. In the case that all inter-atomic collisions
     destroy the excited state then the parameter $\tau_P$ is simply related to the mean time
   between such collisions. There is then competition between decay and inter-atomic collisions
   for destruction of the excited state. On the assumption that these are independent processes:
      \begin{equation}
       \Gamma_S  =   \Gamma_S^{nat} +   \Gamma_P 
     \end{equation}
    Which is equivalent to (8.12) above. In fact, the actual situation is much more complicated, since the
    inter-atomic collisions do not always result in destruction of the excited atom and may modify thee
    energy level or the decay probability of the excited atom.
     Many attempts were made during the '30s and '40s of the last century to calculate $\Gamma_P$
     from phenomenological atomic models~\cite{Linewidth,Lineshape}.

    \par The second effect is the possible correction due to motion of the source atom, 
    which has previously been assumed, throughout the present paper,
     to be at rest. An analogous calculation has recently been performed~\cite{JHF2} by the present
    author for the related `neutrino oscillation' problem. It is adapted to the present case in 
     Appendix E below. The result is markedly different from `Doppler effect' formulae found in 
     the previous literature and text books. There is no damping effect at any order in $v/c$, 
    only a phase shift.
     The classical wave theory of light predicts a Gaussian dependence of the fringe visiblity
     on the path difference due to the first order Doppler Effect (DE):
      \begin{equation}
       {\it V}_{\infty}(d)^{DE} = \exp \left[-\pi\left(\frac{2 \pi d}{\lambda_{\gamma}^0}\right)^2 \frac{kT}{M} \right]
      \end{equation}
       This formula, due to Lord Rayleigh, is obtained by weighting the cosine interference term in the visibility
      function by a Gaussian distribution of $\nu_{\gamma}$ derived, using the Doppler effect, from a Maxwellian
     distribution of source-atom velocities. As in the classical wave derivation of (8.11) this requires the
     substitution  $\kappa \rightarrow 2 \pi c \nu_{\gamma}$ in (8.7) above, which is incorrect for
    the Feynman path amplitude calculation.

   \begin{table}
   \begin{center}
   \begin{tabular}{|c|c c c c c c c|} \hline  
      Transition   & $\lambda_{\gamma}^0$ (\AA)  & $\Delta^{exp}$ (cm)  &  $\Delta^{DE}$ (cm) & $\Delta^{nat}$ (cm) 
    & $\tau_S$ (ns) & $\tau_S^{nat}$ (ns) &  $\tau_P$ (ns)  \\
   \hline 
     H$_r$ 3p-2s & 6563 & 19.0 & 14.3 & 225 & 0.46 & 5.4  & 0.50 \\
    H$_b$ 4p-2s & 4861 & 8.5 & 10.6 & 516 & 0.204 & 12.4 & 0.207 \\
    Na D 3p-3s & 5893 & 80.0 & 65.0 & 225 & 1.92 & 12.4 &  2.98 \\
  \hline
  \end{tabular}
   \caption[] { Measurements from ~\cite{Michelson1} of path lengths $\Delta^{exp}$ yielding a fringe 
    visiblity of 50$\%$ in comparison with theoretical expectations $\Delta^{nat}$ derived using
    (8.11) and $\Delta^{DE}$ from (8.14). The corresponding values of the lifetime
     prameters $\tau_S$ and $\tau_S^{nat}$ are
     also shown as well as the pressure-broadening parameter $\tau_P$ derived from (8.12).} 
  \end{center}
  \end{table} 

      \par The prediction of (8.14) was compared by Michelson in ~\cite{Michelson1} to a number of experimental
   measurements of fringe visiblity. The results obtained for some transitions in Hydrogen and Sodium are 
   summarised in Table 1. The experimental observable, $\Delta$, is the path difference $2d$ in the
   Michelson interferometer for which ${\it V}_{\infty}(\frac{\Delta}{2}) = 0.5$. Also shown in
   Table 1 are the measured values of $\tau_S = \Delta/(2cln2)$ derived from (8.11), $\Delta^{DE}$ from (8.14), 
    theoretical values of $\tau_S^{nat}$
   \footnote{The values of $\tau_S^{nat}$ for the hydrogen lines are taken from P136 of ~\cite{CS}.
    The value for the sodium D-lines is calculated using the formula: $1/\tau_S^{nat} = 4 \pi r_e \nu_{mn}2 f/(3c)$
    ~\cite{Ditchburn} where $r_e$ is the classical electron radius and $f = 0.9755$.} and 
   values of the pressure-broadening lifetime $\tau_P$ calculated from  $\tau_S$ and
     $\tau_S^{nat}$ using (8.12). At the time that Michelson performed the measurements shown
    in Table 1, Quantum Mechanics had yet to be invented so no theoretical values of $\tau_S^{nat}$
    were available. The values of $\tau_P$ shown in Table 1 show that pressure-broadening effects
    are very important for the Hydrogen lines, but less so for the Sodium D-lines. Michelson had observed
   that  line widths increase with increasing pressure, but compared his measurements only with the
    Doppler formula (8.14). For this comparison, in the case of the Hydrogen lines, Michelson 
    chose a temperature of 485$^{\circ}$K in (8.14) corrsponding to a R.M.S. velocity, $\sqrt{kT/M}$,
    of $2.0 \times 10^3$ m/sec. Equation (8.14) was used to predict values of $\Delta/\lambda_{\gamma}^0$
    for different atomic transitions, which were compared with the corresponding experimental quantities.
    Rough agreement was claimed although some positive deviations of up to 30$\%$ and negative 
    deviations up to 40$\%$ were observed in some cases. No uncertainties were quoted on the
   experimental measurements. Given the importance of
   pressure broadening effects for the measurements presented in~\cite{Michelson1}, and the
    difficulty to distinguish between the former effect and that due to source motion,
    these measurements provide no evidence for or against the possiblity of a vanishingly small
    source-motion correction, as predicted by the path amplitude calculation.  
  \par Precise measurement of the temperature dependence of the line widths
    of excited atoms in a gaseous source then constitutes another stringent experimental
    test of path amplitude predictions. In practice it was very difficult in the past
    to disentangle
    conjectured `Doppler effect' and actual `pressure broadening' effects in the data. Both are expected to
    result in larger linewidths at higher temperatures for a constant volume gaseous
    source. One possible approach is to measure $\tau_S$ in a Michelson interferometer,
     using a constant temperature gaseous source, as a function of pressure at low pressure.
    Extrapolating to zero pressure then gives a line width with contributions only from
   the natural width and possible source motion effects, as $\tau_P$ is infinite at
    zero pressure. The prediction of the Feynman path amplitude calculation is that the same
    extrapolated value of $\tau_S$ should be obtained, by this procedure, independently of the temperature
    of the source.
  
 \SECTION{\bf{Spatially Dependent Interference Effects in Quantum Systems with Two Probability Amplitudes.}}
     In this section the Feynman path amplitude description will be applied to some analogous physical
    systems: the Young double-slit experiment (YDSE) using either photons or electrons, quark flavour
    oscillations in the neutral kaon system and neutrino oscillations. In all cases the total probability
  amplitude for the system, $A_{FI}$, decomposes into the sum of two probability amplitudes, $A_{FI}^A$
   and $A_{FI}^B$:
    \begin{equation} 
    A_{FI} = A_{FI}^A + A_{FI}^B = |A_{FI}^A|e^{i\phi_A}+|A_{FI}^B|e^{i\phi_B}
    \end{equation}
    For the YDSE experiments, $A_{FI}^A$ and $A_{FI}^B$ describe experiments where either one of the 
   slits is closed. For the neutral kaon or neutrino oscillation experiments, these amplitudes correspond to 
   flavour mixing scenarios where only a single mass eigenstate is produced.
    The experimental observable, $P_{FI}$, is the probability to observe one of the set of states F:
    $|f_m\rangle$, $m=1,2,...$ given a prepared state in the set I:  $|i_l\rangle$, \ $l=1,2,...$ (see
    Section 2 above). According to Feynman's principles I and III ((2.1) and (2.3)):
    \begin{eqnarray}
    P_{FI} & = & |A_{FI}^A+A_{FI}^B|^2  \nonumber \\
           & = & |A_{FI}^A|^2 +|A_{FI}^B|^2+2Re[A_{FI}^{A~\ast} A_{FI}^B] \nonumber \\
           & = & |A_{FI}^A|^2 +|A_{FI}^B|^2+2 |A_{FI}^A||A_{FI}^B|\cos(\phi_B-\phi_A)
     \end{eqnarray}
    The interesting physical phenomenon, in every case, is described by the interference term in (9.2) that
   depends on the phase difference $\phi_B-\phi_A$. The origin of the latter is different in all the four
  examples under discussion. In a YDSE, the phase difference is a consequence of different path lengths
   corresponding to the two probability amplitudes just as for the Michelson interferometer discussed in
   the previous section. However the physical mechanism that actually generates the phase difference is
    quite different in the two cases. For the photon experiment it is the space-time propagator of
   the excited source atom. For the electron experiment it is the  space-time propagator of
    the electron itself. For neutral kaon oscillations the path difference is constant and
 the phase difference is a result of the different space-time propagators of the ${\rm K}_S$ and 
   ${\rm K}_L$ mesons. The propagators are different because of the different masses of these states.
   As previously pointed out~\cite{JHF1,JHF2,JHF3}, for the case of neutrino oscillations, 
   both the propagator of the unstable source particle whose decay produces the neutrino, and 
   the propagators of the neutrino mass eigenstates $\nu_1$ and $\nu_2$ give important
  contributions to the phase difference. 
   In each case a realistic experiment is described in which all relevant physical parameters
  are discussed. In most previous discussions only simplifed models have been used. These models give 
   essentially the same results as the path amplitude calculations for both the YDSE 
   as well as for neutral
   kaon oscillations, but a markedly different one for neutrino oscillations.

   \par {\bf \Large Young Double Slit Experiment with Photons}
\begin{figure}[htbp]
\begin{center}
\hspace*{-0.5cm}\mbox{
\epsfysize5.0cm\epsffile{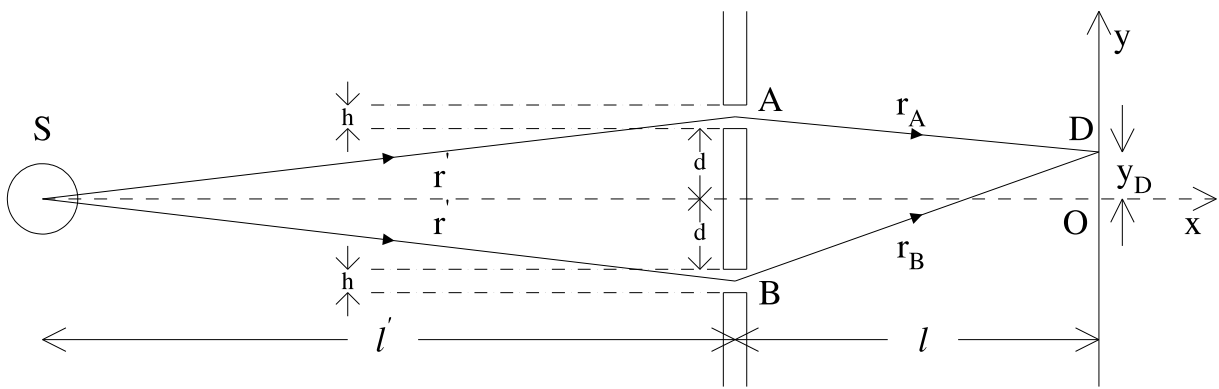}}
\caption{{\sl Geometry of a Young double slit experiment. Particles are 
  produced at the source S and detected at D. For the photon experiment, S is a 
  single excited atom. For electrons S is the exit of an electrostatic 
  accelerator producing a narrow electron beam with the Gaussian momentum
  profile of (9.13). }} 
\label{fig-fig10}
\end{center}
\end{figure}

   \par A schematic layout of a typical YDSE is shown in Fig.10, where the important geometrical
    parameters are defined. The photon source is a single excited atom at S. The photon passes through
   the slits A or B in correspondence with the probability amplitudes $A_{DS}^A(t_D)^{\gamma}$
    or  $A_{DS}^B(t_D)^{\gamma}$, and is detected at D at time $t_D$.  Fig.10 shows the projections of the alternative
   paths of the photon in the $xy$ plane, which is perpendicular to the long edges of the slits.
    In order for the photon to arrive at D it must be Fresnel diffracted, in the $xy$ plane, at
  one of the slits. It is assumed that the height, $h$, of the slits is sufficiently small
  for this to occur, and also that the y-dependence of the detection rate, resulting from diffraction
   (i.e. due to integrating the contributions corresponding to paths passing at different positions
   within the slit) may be neglected for small values of $y_D$ (see Fig. 10). The widths of the slits, $w$, 
    are assumed to be sufficiently long that the photon follows an essentially rectilinear trajectory 
   in the $xz$ and $yz$ planes. The probability amplitudes $A_{DS}^A(t_D)^{\gamma}$
    and  $A_{DS}^B(t_D)^{\gamma}$ are given directly by (4.8) above (compare Fig.1 and Fig.10):
     \begin{eqnarray}
 A_{DS}^A(t_D)^{\gamma} &  = &  \frac{\tilde{{\cal A}}}{r' r_A} {\cal A}_{diff} h w
   \exp\left[ -\frac{i}{\hbar}(\kappa c-i\frac{1}{2 \tau_S})(t_D-t_0- \frac{r_A+r'}{c})\right] \\
 A_{DS}^B(t_D)^{\gamma} &  = &  \frac{\tilde{{\cal A}}}{r' r_B} {\cal A}_{diff} h w
   \exp\left[ -\frac{i}{\hbar}(\kappa c-i\frac{1}{2 \tau_S})(t_D-t_0- \frac{r_B+r'}{c})\right] 
 \end{eqnarray} 
  Neglecting the angular dependence of $ {\cal A}_{diff}$ and setting $r_A = r_B = {\it l}$ in the 
   denominator of the RHS of (9.3) and (9.4) (a good approximation for small values of
   $y_D$), substituting the path amplitudes in (9.2) and integrating, with suitable limits, over $t_D$,
    as done above the the similar case of the Michelson interferometer, gives the result,
   similar to (8.7)   above:
   \begin{equation}
    P_{FI} \equiv P_{FI}(t_D < \infty) = 2 \tau_S |A_0^{\gamma}|^2\{1+\exp\left[-\frac{(r_B-r_A)}{2 c \tau_S}\right]
   \cos\kappa(r_B-r_A)\}
   \end{equation}
   where
   \begin{equation}
  A_0^{\gamma} \equiv - \frac{i\tilde{{\cal A}} \kappa h w}{2 \pi {\it l} r'}   
  \end{equation}
   Assuming that $y_D, d \ll  {\it l}$, it follows from the geometry of Fig.11 that:
   \begin{equation}
  \Delta r \equiv  r_B-r_A \simeq \frac{2(d +\frac{h}{2})y_D}{{\it l}}
   \end{equation}
   giving in (9.5):
    \begin{eqnarray}
   P_{FI}(y_D) & = & 2 \tau_S |A_0^{\gamma}|^2\left[1+\exp\left[-\frac{(d +\frac{h}{2})y_D}{c \tau_S {\it l}}\right]
   \cos \left(\frac{2\kappa(d +\frac{h}{2})y_D}{{\it l}}\right)\right]     \nonumber \\
     & \simeq &  4 \tau_S |A_0^{\gamma}|^2 \cos^2 \left( \frac{2 \pi (d +\frac{h}{2})y_D}{\lambda_{\gamma}^0 {\it l}}
    \right)
 \end{eqnarray}
  where in the last line of (9.8) the small exponential damping correction to the interference term has
  been neglected. This is identical with the well-known result derived using the classical wave theory.
  The spacing, $\Delta y_D$, between adjacent interference fringes is give by (9.8) as:
   \begin{equation}
  \Delta y_D = \frac{\lambda_{\gamma}^0 {\it l}}{2(d +\frac{h}{2})}
   \end{equation}  
    For typical experimental values: {\it l } = 10cm, $d+h/2 = 1$mm, $\lambda_{\gamma}^0 = 5893$\AA, $\Delta y_D$ is
    29 $\mu m$. For the Sodium-lines $\tau^{nat}_S = 5.4$ns, so that the damping correction in (9.5)
    is only $\exp(-1.8 \times 10^{-11} n)$ for the $n$th interference fringe.

     \par {\bf \Large Young Double Slit Experiment with Electrons}
      \par The actual realisation of Feynman's gedankenexperiment, discussed at length in chapter 1 of Vol III
    of~\cite{Feyn2} was not done until the early 1980s~\cite{TIKH}. A closely analagous experiment was
    performed by Tonomura {\it et al} ~\cite{TEMK}.
    In this electron biprism experiment, the double slits of a YDSE are replaced by a region of electric 
     field around a thin wire that provides an equivalent deflection of the electron trajectories to that
     provided by Fresnel diffraction at the slits in a YDSE. 
    \par In electron quantum interference experiments the electrons are typically 
    produced by a `direct current' source using thermionic of field emission.
    There is thus no source with a well-defined quantum state yielding a known time-dependent
   phase as in the optical experiments considered above, or the neutrino oscillation experiments to 
   be discussed below. 
   \par Referring to Fig.10 and using (3.11) for the space-time propagator of the electron,
   the probability amplitudes analagous to (9.3) and (9.4) for the electron case are, on making similar
    geometrical approximations to those made in (9.5) above:
  \begin{eqnarray}
    A_{DS}^{A~e} & = &  A_0^e
   \int \exp\left[-\frac{i}{\hbar}[E_e \Delta t_A-p_e(r'+r_A)]\right] f(p_e)d p_e \\
   A_{DS}^{B ~e} & = &  A_0^e
   \int \exp\left[-\frac{i}{\hbar}[E_e \Delta t_B-p_e(r'+r_B)]\right] f(p_e)d p_e
  \end{eqnarray}
   where 
    \begin{equation}
      A_0^e \equiv  \frac{A_D \beta_e^2  {\cal A}_{diff} h w A_S}{r' {\it l}}
    \end{equation}  
   In these equations, $A_S$ is the electron production amplitude, ${\cal A}_{diff}$, the amplitude for
    diffraction at a slit and $A_D$ the amplitude for the electron detection process. The function
    $f(p_e)$ is the normalised distribution of electron momenta at the detector. In the following
   a Gaussian form will be used:
    \begin{equation}
     f(p) = \frac{1}{\sqrt{2 \pi}\sigma_p}\exp\left[-\frac{(p-\langle p \rangle)^2}{2 \sigma_p^2}\right]
   \end{equation} 
    \par It can be seen that the two important differences between a photon and  an electron YDSE.
     are that, (i) in the electron case the phases of the probability
   amplitudes are determined by the space-time propagator of the electron, whereas in the photon
    case the phase is determined entirely by the coherent source, the contribution from the
   photon propagator vanishing, and (ii) unlike on-shell photons, the electrons do not
    have a constant velocity, so that their actual  velocity or momentum distribution must be taken into
   account in order to construct the probability amplitudes (9.10) and (9.11).
     Because of the photon's constant velocity it must be produced at different 
    times in amplitudes corresponding to paths of different lengths. However, the electron
    can be produced at the same time in both probability amplitudes (i.e. $\Delta t_A = \Delta t_B$ in (9.10) and (9.11))
    and produce the same detection event, at a well defined time, due to different
   velocities along each path. In this case, unlike for the photon YDSE above and neutrino oscillations,
    to be discussed below, there is no contribution to the interference phase from the production
    amplitude $A_S$.
    \par The Lorentz-invariant character of the phase of the space-time propagator of a particle (see (3.11) above)
      enables the phases of the complex exponentials in (9.10) and (9.11) to be written in the following four
     equivalent ways:
  \begin{equation}
   \phi = -\frac{mc^2 \tau}{\hbar} = -\frac{mc^2 t}{ \gamma \hbar} = 
 -\frac{(mc^2)^2  r}{ E v \hbar} = -\frac{(mc)^2 r}{ p \hbar }
 \end{equation}
    Where $\gamma$ is the usual relativistic parameter $1/\sqrt{1-\beta^2}$, $\beta \equiv v/c = pc^2/E = r/(ct)$. 
   Substituting the appropriate values of $r$ and $p$ then gives for the phase difference in (9.2):
   \begin{equation}
   \phi_B -\phi_A = \frac{(mc^2)^2}{c^2 \hbar}\left(\frac{r'+r_A}{p_A}-\frac{(r'+r_B)}{p_B}\right)
   \end{equation}
    For equal production times of the electron in the paths A and B the condition:
   \begin{equation}
      t =   t_A = t_B =\frac{r'+r_A}{v_A} = \frac{r'+r_B}{v_B}
  \end{equation}
    is respected. It is shown in Appendix F that, in this case, the RHS of (9.15)
    may be written, to first order in the small quantity $\Delta r/r' = (r_B-r_A)/r'$, as:
      \begin{equation}
   \phi_B -\phi_A = \frac{\overline{p} \Delta r}{\hbar} =  \frac{2 \pi \Delta r}{ \lambda_e^{DB}}
   \end{equation}
   where 
    \begin{equation}
    \overline{p} \equiv \frac{p_A+p_B}{2}
   \end{equation}  
     and
    \begin{equation} 
     \lambda_e^{DB} \equiv  \frac{h}{\overline{p}}
   \end{equation}
   Which is de Broglie's formula for the quantum wavelength of a particle. As further discussed 
  in Appendix F, the equal time condition (9.16) is essential to obtain this relation
  in the Feynman path amplitude approach. The relation (9.16) also implies the following
  velocity difference for the electron for the paths A and B:
   \begin{equation} 
 v_B -v_A = \frac{c^2(mc^2)^2}{\overline{E}^3}(p_B-p_A)  = \frac{\overline{v} \Delta r}{r'+\overline{r}}
  +O[(\Delta r)^2]
   \end{equation}
 where $\overline{E}$,$\overline{v}$ and $\overline{r}$ are defined similarly to $\overline{p}$ in (9.18).
 \par Taking the modulus squared of (9.10) or (9.11), and performing the momentum integration, gives:
  \begin{eqnarray}
 |A_{DS}^{A,B~e}|^2 & = & |A_0^e|^2 \int \int e^{i(\phi_{A,B}(p)-\phi_{A,B}(p'))}f(p)f(p')\delta(p-p') dp dp'
  \nonumber \\
   & = &  \frac{1}{2 \sqrt{\pi} \sigma_p}|A_0^e|^2
  \end{eqnarray}
   The equal time condition gives, for the interference term in (9.2):
  \begin{eqnarray}
  I_{AB} & \equiv & 2 |A_{FI}^A||A_{FI}^B|\cos(\phi_B-\phi_A)\nonumber \\
  & = & 2 |A_0^e|^2 Re \left\{ \int \int e^{-\frac{i(p_A+p_B)\Delta r}{2 \hbar}}f(p_A)f(p_B)\delta(p_B-p_A-\Delta p)
   dp_A dp_B \right\}\nonumber \\
    & = &  \frac{|A_0^e|^2}{\pi \sigma_p^2} Re  \left\{ \int   e^{-\frac{i(p_A+\frac{\Delta p}{2})\Delta r}{\hbar}}  
    e^{-\frac{(p_A-\langle p \rangle)^2}{2\sigma_p^2}}  e^{-\frac{(p_A+ \Delta p-\langle p \rangle)^2}{2\sigma_p^2}}
     dp_A  \right\}
  \end{eqnarray}
 where, from (9.20),
    \begin{equation}
   \Delta p \equiv p_B-p_A = \overline{p} \left(\frac{\overline{E}}{mc^2}\right)^2\frac{\Delta r}{r'+\overline{r}}
    +O[(\Delta r)^2]
   \end{equation}
 The $p_A$ integral is readily evaluated by the method of `completing the square' to yield the result:
   \begin{equation} 
 I_{AB} =\frac{|A_0^e|^2}{\sqrt{\pi} \sigma_p}\exp[-(\frac{\Delta p}{2 \sigma_p})^2] \exp[-(\frac{\sigma_p \Delta r}{2 \hbar})^2]
  \cos\left(\frac{\langle p \rangle  \Delta r}{ \hbar}\right)
   \end{equation}
   The first exponential damping term takes into account the fact that a non-vanishing spread in electron
   momentum is necessary for interference to occur -- a consequence of the imposed condition of equal production
   times for the electron in the two paths. The second exponential describes damping due to the width of the
    momentum distribution. If it is very wide the probability to satisfy the condition (9.20) becomes very small.
    Inserting (9.21) and (9.24) in (9.2), as well as the expression (9.7) for $\Delta r$, gives finally:
   \begin{eqnarray}
      P_{FI}(y_D)^e & = & \frac{|A_0^e|^2}{\sqrt{\pi} \sigma_p}\left[1+ \exp\{-[(\frac{\Delta p}{2 \sigma_p} )^2+
(\frac{\sigma_p \Delta r}{2  \hbar })^2]\} \cos \left(\frac{2 \langle p \rangle (d+h/2)y_D}{{\it l} \hbar}\right)\right]
  \nonumber \\
    & \simeq & 2 \frac{|A_0^e|^2}{\sqrt{\pi} \sigma_p} \cos^2\left(\frac{2 \pi(d+h/2) y_D}{\lambda_e^{DB} {\it l}}\right)
  \end{eqnarray}
     where in the last line the typically small interference damping corrections
     \footnote{For example, the damping corrections for the $n$th interference fringe of the experiment
    of Reference~\cite{TEMK} where $ p = 229$MeV/c, $\sigma_p/p = 6.0 \times 10^{-7}$ (length of wavepacket 1$\mu$m) and
    $r'+\overline{r} \simeq 2$m are: $\Delta p/(2 \sigma_p) = n \overline{\gamma}^2 h/[\sigma_p(r'+\overline{r})] =
     n 1.7 \times 10^{-9}$ and $\sigma_p \Delta r/(2 \hbar) = n \pi \sigma_p/p = n 1.9 \times 10^{-6}$.} have been neglected.
   It can be seem by comparing (9.8) and (9.25) that, in this approximation, photons and electrons with
    $\lambda_e^{DB} = \lambda_{\gamma}^0$ are predicted
    to  produce identical interference patterns in geometrically indentical YDSE.

   \par {\bf \Large Strange Quark Flavour Oscillations}
    \par In this case the Feynman path amplitude description of a specific experiment~\cite{DSDQ} will be considered.
     This is done to ensure that the effects of all possibly relevant physical parameters are properly taken 
    into account in the discussion. In the experiment, ${\rm K_S}$ or ${\rm K_L}$ mesons were produced via the 
    processes: $\pi^- p \rightarrow \Lambda ({\rm K_S},{\rm K_L})$ by a 1.01 GeV/c pion  beam.
    The initial state proton was either free within a chemically bound hydrogen atom, or bound in a carbon nucleus,
    of the polythene/plastic scintillator target. Roughly 50$\%$ of the interactions occured on bound protons.
      Semi-leptonic decays ${\rm K_S},{\rm K_L} \rightarrow \pi^{\pm}e^{\mp} \nu$ were observed, and their 
    proper time intervals calculated. In this experiment the paths A and B in (9.2) correspond to
    the space-time propagation of the  ${\rm K_S}$ or ${\rm K_L}$ mesons respectively, so in the following
    the labels A,B are replaced by S,L. It is assumed that either meson is detected at a fixed distance,
    $L$, from its production point.
    \par   The detailed physics underlying the production amplitudes of the mesons is not well understood,
     but is not important for the description of the flavour oscillation phenomenon. An $s\overline{s}$ quark
   pair is produced by the strong interaction. The $s$ quark is bound in the $\Lambda$, while the $\overline{s}$
    undergoes a flavour-changing weak charged-current interaction which produces one of the mass eigenstates
    $|{\rm K_S}\rangle$, $|{\rm K_L}\rangle$ of the neutral kaon system. These states contain within their
    quark substructure both  $\overline{s}d$ ( $|{\rm K_0}\rangle$) and  $s\overline{d}$ ( $|{\rm \overline{K}_0}\rangle$)
    components. Neglecting small CP-violating contributions,  $|{\rm K_S}\rangle$ and $|{\rm K_L}\rangle$
    are eigenstates of CP:
    \begin{eqnarray}
     |{\rm K_S}\rangle &  = & \frac{1}{\sqrt{2}}(|{\rm K_0}\rangle-|{\rm \overline{K}_0}\rangle)~~~{\rm CP}=+1 \\
     |{\rm K_L}\rangle &  = & \frac{1}{\sqrt{2}}(|{\rm K_0}\rangle+|{\rm \overline{K}_0}\rangle))~~~{\rm CP}=-1
    \end{eqnarray}
  where the states  $|{\rm K_0}\rangle$ and  $|{\rm \overline{K}_0}\rangle$ are related by the CP operator as:
   \[{\rm CP} |{\rm K_0}\rangle = - |{\rm \overline{K}_0}\rangle,
   ~~~{\rm CP} |{\rm \overline{K}_0}\rangle = -|{\rm K_0}\rangle\]
    The  $\overline{s}$ quark produced in association with the $\Lambda$ couples only to the  $|{\rm K_0}\rangle$
    component of  $|{\rm K_S}\rangle$ and $|{\rm K_L}\rangle$. It then follows from (9.26) and (9.27) that, neglecting
     CP-violating effects, the  $|{\rm K_S}\rangle$ and $|{\rm K_L}\rangle$ production amplitudes are equal:
     \begin{equation}
      \langle{\rm K_S}\Lambda|T|\pi^- p \rangle =  \langle{\rm K_L}\Lambda|T|\pi^- p \rangle \equiv A_P
     \end{equation}
     The decay processes  ${\rm K_S},{\rm K_L} \rightarrow \pi^{\pm}e^{\mp} \nu$ have been found, experimentally, to 
    respect the `$\Delta S = \Delta Q$ Rule', which is predicted in the Standard Electroweak Model, by consideration
   of the W-boson exchange diagrams that mediate these transitions. This rule predicts that the  $|{\rm K_0}\rangle$
    components of  $|{\rm K_S}\rangle$ and $|{\rm K_L}\rangle$, decay only into the channel $\pi^- e^+ \nu$ while 
    the $|{\rm \overline{K}_0}\rangle$ components decay only into $\pi^+ e^- \nu$. The equations (9.26) and
     (9.27) then lead to the following relations between the semileptonic decay amplitudes:
      \begin{equation}
  \langle \pi^- e^+ \nu |T|{\rm K_S} \rangle = \langle \pi^- e^+ \nu |T|{\rm K_L} \rangle 
    = - \langle \pi^+ e^- \nu |T|{\rm K_S} \rangle =  \langle \pi^+ e^- \nu |T|{\rm K_L} \rangle \equiv A_{eK}
     \end{equation}
     On the assumption that the  ${\rm K_S}$, ${\rm K_L}$ mesons are produced at space-time points
    $x_{PS}$,  $x_{PL}$ respectively, and decay at  $x_D$, the probability amplitudes for the production
    of $e^+$, $e^-$ via propagation of ${\rm K_S}$, ${\rm K_L}$ are given by (2.2) and (2.3) as:
     \begin{eqnarray}
     A_{DS}^{e^+}({\rm K_S}) & = & A_{eK}\int \langle D |{\rm K_S}| PS  \rangle f(p_S) d p_S   \\
      A_{DS}^{e^+}({\rm K_L}) & = & A_{eK}\int \langle D |{\rm K_L}| PS  \rangle f(p_L) d p_L  \\
    A_{DS}^{e^-}({\rm K_S}) & = & - A_{eK}\int \langle D |{\rm K_S}| PS  \rangle f(p_S) d p_S   \\
       A_{DS}^{e^-}({\rm K_L}) & = & A_{eK}\int \langle D |{\rm K_L}| PS  \rangle f(p_L) d p_L  
  \end{eqnarray}

    The space-time propagators of the mesons are given by (3.12) as:
      \begin{equation}
   \langle D |{\rm K_i}| Pi  \rangle = \frac{\beta_i}{L}\exp\left[-i \frac{[m_i-i\Gamma_i/(2 c^2)]
    c^2 \Delta \tau_i}{\hbar} \right]~~~i = S,L 
  \end{equation}
    The amplitudes $f(p_S)$,  $f(p_L)$ describe the momentum distribution of the mesons.
    As discussed below the dominant source of momentum smearing is radiative corrections
    rather than variation of the physical masses $W_L$ and $W_S$, of order $\Gamma_L/c^2$, $\Gamma_S/c^2$,
    of ${\rm K_L}$ and ${\rm K_S}$,
   due to their unstable nature. It is shown below that the precise form of  $f(p)$  does not affect the
   final result for the detection probability. For convenience,  $f(p)$ is normalised so that
   \begin{equation}
        \int_{0}^{\infty}f(p)^2 dp = 1
  \end{equation}

    Since the decay widths are relatively large, indeed $\Gamma_S = 2.1 (m_L-m_S)$, the kinematical effects
    of the variation of the physical masses of the mesons are of a similar size to those, generated by the
    difference of pole masses, that underly the whole `flavour oscillation' phenomenon and so cannot be, {\it prima facie},
    neglected. This variation of the physical mass, $W$, modifies the exact expression, (9.14), for the propagator phase
   in the following way:
  \begin{equation}
    \phi = -\frac{mc^2 \tau}{\hbar} = -\frac{mc^2  t}{ \gamma \hbar} = 
 -\frac{(mc^2)^2 L}{ E(W) v(W) \hbar} = -\frac{mc^2 W L}{ p \hbar }
 \end{equation}
   where 
   \[ E(W) \equiv \sqrt{(Wc^2)^2+(pc)^2},~~~ v(W) = \frac{pc^2}{E(W)} \]
    The phase then depends, in general, on both the physical mass and the velocity (or momentum) of the
   particle. Unlike that of the photon in physical optics the particle velocity is variable, and 
   unlike the electron in the YDSE discussed above, the  velocity is not fixed by the value of 
   the momentum. 
    \par In order to proceed further it is necessary to discuss the space-time structure
    of the production and detection events corresponding to the probability amplitudes. Only the spatial distance $L$
     between the production and detection events is necessarily constant in the probability amplitudes
     (9.30)-(9.33). To give some feeling for the variation of the kinematical quantities and time 
     intervals involved one may ask the following questions: (1) What is the difference of momentum
    necessary to compenstate the change in velocity due to the  ${\rm K_S}$--${\rm K_L}$ mass difference? (2)
    In case of equal momenta for the   ${\rm K_S}$ and ${\rm K_L}$ what is the difference in their production
     times in order to arrive simultaneously at the detection event? If the velocities of the ${\rm K_S}$ and ${\rm K_L}$ 
     are equal, their momenta $p_L$ and $p_S$ must satisfy the condition:
    \begin{equation}
     \frac{v_S}{c} = \frac{1}{\sqrt{1+(\frac{W_S c}{p_S})^2}} =
 \frac{v_L}{c} = \frac{1}{\sqrt{1+(\frac{W_L c}{p_L})^2}}
    \end{equation}
   that is
    \begin{equation}
     \frac{W_S}{W_L} = \frac{p_S}{p_L}
   \end{equation}
   so that
    \begin{equation}
     \frac{\langle \delta p \rangle}{\langle p \rangle} = \frac{(\langle W_L \rangle -  \langle W_S \rangle)c}
     {\langle p \rangle} = \frac{(m_L -m_S)c}{\langle p \rangle} = 1.8 \times 10^{-14}
   \end{equation} 
    where the value $\langle p \rangle = 194$ MeV/c, corresponding to the centre-of-mass momentum 
   of the  ${\rm K_S}$ and ${\rm K_L}$ in the experiment~\cite{DSDQ},
    and $m_L-m_S = 3.49 \times 10^{-12}$ Mev/$c^2$ have been used.
       This is many orders of magnitude smaller than than the momentum smearing of the
     ${\rm K_S}$ and ${\rm K_L}$  due to initial state photon radiation. This is estimated, using the 
    soft photon radiatve correction formalism of~\cite{RK} to be:
   \begin{equation}
     \frac{\langle \delta p_{rad} \rangle}{\langle p \rangle} = 4.2 \times 10^{-2}
   \end{equation}
    for the experiment~\cite{DSDQ}.
    In these circumstances the path amplitudes in (9.30)-(9.33) for different momenta and velocities,
   within the appropriate range given in (9.39), must contribute equally to the 
    sum over intermediate states in (2.3), so that no damping of
   the interference term is to be expected from momentum or velocity variation. This is verified in the following
   calculation.
    \par To answer the question (2), the difference
    of production times: $\Delta t_{SL} \equiv t_{PS}- t_{PL}$  for ${\rm K_S}$ and ${\rm K_L}$ of fixed momentum,
     in order to arrive simultaneously at the typical distance $c \beta \gamma \tau_S$ from their production point is calculated:
     \begin{equation}
      \Delta t_{SL} = \frac{(m_L-m_S) \tau_S}{\overline{E}}
    \end{equation} 
     Some values of $\Delta t_{SL}$ for different assumed values of $\overline{p} $ and $\overline{E}$ are 
    presented in Table 2. Associating these with the lifetime of a hypothetical unstable source particle
    gives decay widths of such a particle in the range from 1 to 200 MeV. whereas the `characteristic time'
    \footnote{This is an interesting concept that has not been addressed, to date, by any physical theory.
     Presumably it takes some non-vanishing time for the quark rearrangement and creation processes
     that convert, in the present example, the $\pi^-p$ system into $\Lambda{\rm K_S}$ or  $\Lambda{\rm K_L}$,
     to occur. The possibility to observe such a time-interval in decay processes by measurable deviations from
    the exponential decay law or the Breit-Wigner line shape of an unstable particle has been suggested~\cite{DLR}.} 
   of the strong/weak interaction process that produces the $\Lambda {\rm K_S}$ or $\Lambda {\rm K_L}$
    systems must be much shorter than this as no such resonant state is actually produced. In the following 
   it is assumed that the `characteristic time' is so short in comparison with the values of $\Delta t_{SL}$
   shown in Table 2, that the production time of the ${\rm K_S}$ and ${\rm K_L}$ in the probability amplitudes
   (9.30)--(9.33) is the same in every case, i.e. $t_{PS} = t_{PL} = t_P$. Thus, since the detection
   event occurs some definite time, the ${\rm K_S}$ and ${\rm K_L}$ are assumed to have equal velocities
   in the alternative probability amplitudes. Note here the
   difference with the electron YDSE. There also the equal production time assumption is made, leading
   to the conventional de Broglie wavelength for the associated `matter wave', but in this case due to the 
   unique physical mass of the electron, and the different path lengths in the YDSE, different velocities are necessary.
  
    \begin{table}
   \begin{center}
   \begin{tabular}{|c|c c c c c |} \hline  
     $\overline{p}$ (GeV/c)   & 0.01  & 0.1  &  1.0 & 10.0 & 100.0 \\
    $\overline{E}$ (GeV)   & 0.498  & 0.508 &  1.117 & 10.1 & 100.0 \\
    $\Delta t_{SL}$ (sec) & $6.27 \times 10^{-22}$  & $6.14 \times 10^{-22}$  & $2.8 \times 10^{-22}$ &  $3.1 \times 10^{-23}$
   & $3.1 \times 10^{-24}$   \\             
  \hline
  \end{tabular}
   \caption[] { Momentum and energy dependence of $\Delta t_{SL}$,
  the difference in production times of ${\rm K_S}$ and ${\rm K_L}$ of
   equal momentum in order
 to arrive simultaneously at distance  $c \beta \gamma \tau_S$  from their
   production point.} 
  \end{center}
  \end{table}  
   
   \par Substituting (9.34) into (9.30)-(9.33) and using the expression for $\Delta \tau_i$ given 
 by the second and last members of (9.36), it follows that: 
     \begin{eqnarray}
    |A_{DS}^{e^+}({\rm K_{S,L}})|^2 &  = & |A_{DS}^{e^-}({\rm K_{S,L}})|^2 \nonumber \\
   &  = &  \left|\frac{A_{eK} \beta_{S,L} A_P }{L}\right|^2
     \int f(p'_{S,L})  f(p_{S,L}) \delta(p_{S,L}- (p'_{S,L}) dp'_{S,L} dp_{S,L} \nonumber \\
     & \simeq &  \left|\frac{A_{eK}\overline{\beta} A_P }{L}\right|^2
    \exp \left[-\frac{\Gamma_{S,L} W_{S,L}m_{S,L} L}{\langle p \rangle \hbar}\right] \nonumber \\
     & \equiv &  \left| A_{ep}\right|^2 
 \exp \left[-\frac{\Gamma_{S,L} W_{S,L}m_{S,L} L}{\langle p \rangle \hbar}\right]
    \end{eqnarray}
    where the approximations $\beta_L \simeq \beta_S = \overline{\beta}$ and 
    $W_L \simeq W_S = \overline{m} \equiv (m_L+m_S)/2$ have been made and $\langle p \rangle$ denotes 
   the average value of $p_L$ or $p_S$. 
 
    Replacing the path labels A and B in Eqn(9.2) by $S$ and $L$ respectively, the interference terms
    $I_{SL}^{e^{\pm}}$ for the detection of $e^{\pm}$ are given by (9.30)-(9.36) as :
      \begin{eqnarray}
  I_{SL}^{e^{\pm}} & = & \pm 2 Re \left| A_{ep} \right|^2 
   \int \int \exp\left\{-\frac{1}{2 \hbar}\left[\frac{\Gamma_S W_S}{p_S}+\frac{\Gamma_L W_L}{p_L}\right]\right\}
     \exp\left\{\frac{ic^2}{\hbar}\left[\frac{m_S W_S}{p_S}-\frac{m_L W_L}{p_L}\right] L \right\}   \nonumber \\     
     &  & \times  f(p_S)  f(p_L) \delta(W_S p_L-  W_L p_S) dp_L dp_S
     \end{eqnarray}
     where the $\delta$-function imposes the equal velocity condition (9.38) and the approximation 
     $\beta_S \beta_L \simeq  \overline{\beta}^2$ has been made. Performing the $p_S$ integral in
     (9.43) gives:
      \begin{eqnarray}
  I_{SL}^{e^{\pm}} & = & \pm 2 Re \left| A_{ep} \right|^2 
   \int \exp\left\{-\frac{(\Gamma_S+\Gamma_L)}{2 \hbar}\frac{ W_L L}{p_L}\right\}
     \exp\left\{\frac{ic^2 \Delta m_{LS}}{\hbar}\frac{W_L}{p_L}L \right\}   \nonumber \\     
     &  & \times  f(\frac{W_S p_L}{W_L})  f(p_L) dp_L
     \end{eqnarray}
     where $ \Delta m_{LS} \equiv m_L-m_S$. Since the ratio $W_L/W_S$ differs from unity only by
     quantities of order $\Gamma_S/c^2 \overline{m} \simeq (3.49 \times 10^{-12} {\rm MeV})/( 498{\rm MeV})
      = 7.0 \times 10^{-15}$, then, to a very good approximation, the replacements: $W_L/W_S =1$, $W_L = \overline{m}$
     may be made in (9.44) giving:
    \begin{equation}
    I_{SL}^{e^{\pm}} = \pm 2 Re \left| A_{ep} \right|^2 
      \exp\left\{-\frac{(\Gamma_S+\Gamma_L)}{2 \hbar}\frac{ \overline{m} L}{\langle p \rangle}\right\}
    \cos\left(\frac{ \overline{m} c^2 \Delta m_{LS} L}{ \hbar \langle p \rangle}\right)
     \end{equation}
    where the normalisation condition (9.35) has been used. 

   Subsitituting (9.42) and (9.45) into (9.2) gives finally for the probability,
  $ P(e^{\pm},L,\langle p \rangle)$ to detect $e^{\pm}$ at distance
  $L$ from the production point of ${\rm K_S}$, ${\rm K_L}$ with momentum $ \langle p \rangle$:
      \begin{eqnarray}
  P(e^{\pm},L,\langle p \rangle) &  =  & |A_{ep}|^2\left\{ \exp\left[-\frac{\Gamma_S \overline{m}c^2 L}
   { \hbar \langle p \rangle}\right]
  +\exp\left[-\frac{\Gamma_L \overline{m}c^2 L}{ \hbar \langle p \rangle}\right] \right. \nonumber \\
   &  &\pm 2 \left. \exp\left[-\frac{(\Gamma_S+\Gamma_L)}{2 \hbar }\frac{\overline{m}c^2 L}{ \langle p \rangle }
  \right] \cos \left(\frac{\overline{m}c^2 \Delta m_{LS} L}{\hbar \langle p \rangle} \right)\right\}    
      \end{eqnarray} 
  On the assumption that the  ${\rm K_S}$ and ${\rm K_L}$ mesons have the same velocity this probability
 may be expressed in terms of the proper time interval $\tau$ between their production and decay via the 
  relation (see (9.36):
  \begin{equation}
    \tau = \frac{W L}{ \langle p \rangle } \simeq \frac{\overline{m} L}{ \langle p \rangle}
  \end{equation}
  so that 
      \begin{eqnarray}
  P(e^{\pm},\tau) &  =  & |A_{eP}|^2 \left\{ \exp \left[-\frac{\Gamma_S \tau}{\hbar}\right]
  +\exp \left[-\frac{\Gamma_L \tau}{\hbar}\right] \right. \nonumber \\
   &  &  \pm 2 \left. \exp \left[-\frac{(\Gamma_S+\Gamma_L) \tau}{2 \hbar}
  \right] \cos \left(\frac{\Delta m_{LS} c^2 \tau}{\hbar} \right)\right\}    
      \end{eqnarray}                    
 This is just the formula that has been used till now to interpret such quark flavour oscillation
  experiments~\cite{DSDQ,JACK,CPLEAR}. Thus the Feynman path amplitude calculation,
 where careful account is taken of all relevant physical parameters, particularly with 
 respect to the space-time structure of the production and detection events, gives the
 same result as a simple calculation of the phase difference
 $\phi_S-\phi_L$, neglecting all such considerations, using the time-dependent Schr\"{o}dinger equation in the
 respective rest frames of the ${\rm K_S}$ and ${\rm K_L}$ mesons..

\par {\bf \Large Neutrino Oscillations}

 \par Previous discussions of neutrino oscillations using the Feynman path amplitude formalism
  can be found in~\cite{JHF1,JHF2,JHF3}. The experiment to be discussed here is detection of the
  process: $ \nu n \rightarrow e^- p$, where the incoming neutrino is produced by pion decay at rest:
   $\pi^+ \rightarrow \mu^+ \nu$. The result found generalises in a straightforward manner to 
  any neutrino oscillation experiment where the neutrino source is at rest.  The experiment
  is shown in more detail in Fig.11. A slow $\pi^+$ comes to rest at time $t_0$ in a stopping
  target T (Fig.11 a)). For simplicity, the case of only two neutrino flavours is considered. The neutrino
  mass eigenstates $|\nu_1\rangle$ and $|\nu_2\rangle$ have pole masses $m_1$ and $m_2$ where
   $m_1 > m_2$\footnote{ Note that the subscripts `1' and `2' are arbitary, and do not necessarily
   correspond to fermion generation number. In the presently favoured interpretation of
    atmospheric neutrino oscillations~\cite{Kays04}, where there is an important
   contribution from pion decay neutrinos, the mass eigenstates are those 
  associated with the second and third generations}.
   Fig.11 b) and c) show two alternative histories for the stopped pion. 
   In Fig11 b), the decay  $\pi^+ \rightarrow \mu^+ \nu_1$ occurs at time $t_1$, in Fig.12 c) 
    the decay $\pi^+ \rightarrow \mu^+ \nu_2$ occurs at a later time $t_2$. With a suitable 
   time difference $t_2-t_1$, the neutrinos corresponding to the two alternative
  paths arrive at the detection event at the same time, $t_D$ and are thus indistinguishable
  (Fig.11 d)). The corresponding probability amplitudes $A_{e\mu}^{(1)}$,  $A_{e\mu}^{(2)}$
   then add, as in (9.1), to give the total probability amplitude for the experiment.
   The labels `A' and `B' in (9.1) and (9.2) are here replaced by `(1)' and `(2)' corresponding
    to the mass eigenstates $\nu_1$ and  $\nu_2$.

\begin{figure}[htbp]
\begin{center}
\hspace*{-0.5cm}\mbox{
\epsfysize15.0cm\epsffile{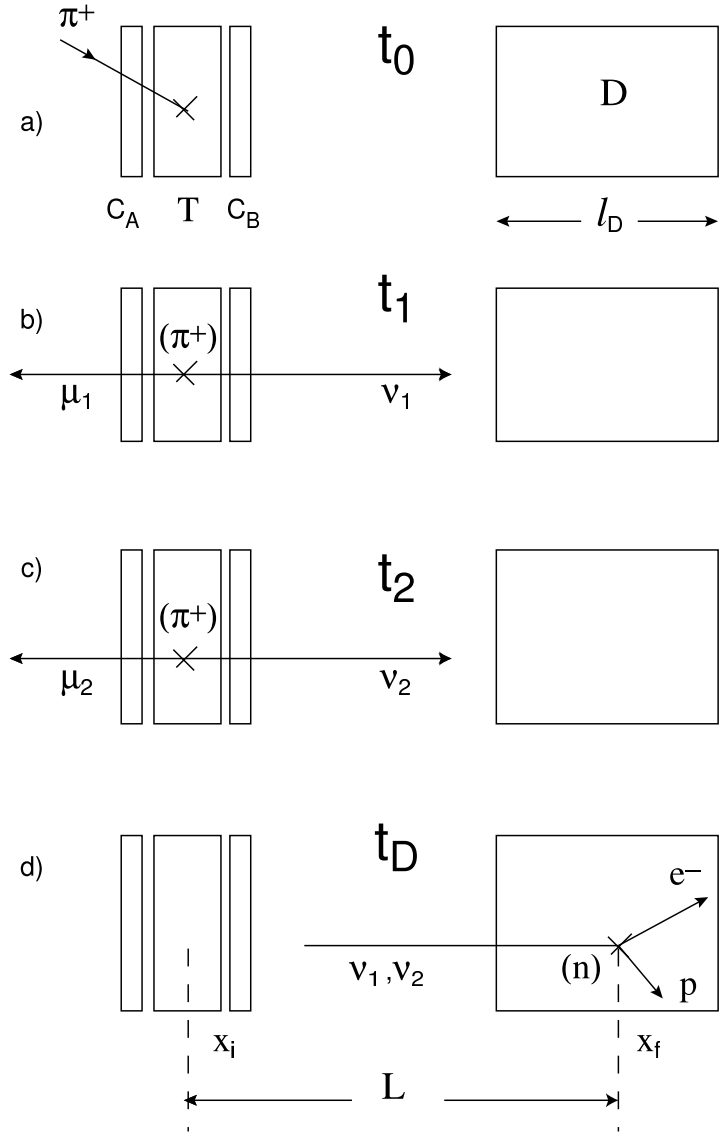}}
\caption{{\sl An experiment to measure neutrino oscillations
  following pion decay. a) a $\pi^+$ comes to rest in a stopping target $T$ at time
  $t_0$.  b) and c) show alternative histories of the stopped pion. In b) the decay $\pi^+ \rightarrow
  \mu \nu_1$ occurs at time $t_1$. In c) the decay $\pi^+ \rightarrow
  \mu \nu_2$ occurs at time $t_2$. If $m_1 > m_2$, and with a suitable value of $t_2-t_1$, the 
    neutrinos may arrive at the same time $t_D$ in the alternative histories
  at the the detection event, as shown in d). Since the detection event ($\nu_1,\nu_2)n \rightarrow p e^-$
  does not distinguish the neutrino flavour, the amplitudes corresponding to the different histories
  in b) and c) must be added, as in (9.1), to give the probability amplitude for the experiment. 
  Note the similarity of the sequence of space-time events to that in the Michelson interferometer of
  Fig.8. }} 
\label{fig-fig11}
\end{center}
\end{figure}

    \par  Note that the physical situation is
    strictly analagous to that of the Michelson Interferometer or the photon YDSE.
   In these experiments the photon must be emitted at different times by the source, in
  the interfering amplitudes, in order to arrive in time coincidence at the
   detection event. In these two experiments the time difference is necessary to 
   compensate the non-equal path lengths. In a neutrino oscillation experiment,
   both paths have the same length, and the time difference of neutrino emission
   is necessary to compensate for the different neutrino velocities. 
   \par Introducing the process amplitudes for the $\pi$-decay and neutrino detection
   processes, and the space-time propagators of the pion source and the neutrinos
   according to (3.12), the probability amplitudes for propagation of the states
   $|\nu_1\rangle$ and $|\nu_2\rangle$ are: 
   \begin{equation}
    A_{e \mu}^{(i)} = \langle e^- p |T|\nu n \rangle U_{e i} \langle D |\nu_i| i \rangle U_{\mu i}
 \langle \nu \mu^+ |T|\pi^+ \rangle \langle i |\pi^+| 0 \rangle~~~~~(i = 1,2)
  \end{equation}
  the space-time propagators of the source pion (assumed to be at rest so that $\Delta \tau = \Delta t$)
   and the neutrinos are:
  \begin{eqnarray}
   \langle i |\pi^+| 0 \rangle & = & \exp \left[-i\frac{[m_{\pi}-i\Gamma_{\pi}/(2 c^2)] c^2(t_i-t_0)}
     {\hbar} \right]~~~~~(i = 1,2) \\
   \langle D |\nu_i| i \rangle & = & \exp \left[-i\frac{m_i c^2 (\tau_D-\tau_i)}  {\hbar} \right]~~~~~(i = 1,2) 
   \end{eqnarray}
 The process amplitudes for pion decay and neutrino detection are written as products of `reduced' process
  amplitudes and elements of the Maki-Nakagawa-Sakata~\cite{MNS} matrix $U_{\ell i}$ that describes the strength
 of the charged-current coupling between lepton $\ell$ and the neutrino mass eigenstate $|\nu_i\rangle$:
   \begin{eqnarray}
  \langle e^- p |T|\nu_i n \rangle & = &  U_{e i} \langle e^-  p |T|\nu n \rangle~~~~~(i = 1,2) \\
   \langle \nu_i \mu^+ |T|\pi^+ \rangle  & = & U_{\mu i} \langle \nu \mu^+ |T|\pi^+ \rangle~~~~~(i = 1,2)
   \end{eqnarray}
  In the case of only two charged lepton and neutrino flavours,
    all the elements of the MNS matrix are real and are specified
  by a single mixing angle $\theta_{12}$:
    \begin{equation}
   U_{e1} = U_{e2} = \cos \theta_{12},~~~ U_{e2} = -U_{\mu 1} = \sin  \theta_{12}
    \end{equation}
  \par Denoting the phase of  $A_{e\mu}^{(i)}$ as $\phi^{(i)}$, (9.49)-(9.51) give:
    \begin{equation}
   -\frac{\hbar \phi^{(i)}}{c^2} =  m_i (\tau_D-\tau_i)+ m_{\pi}(t_i-t_0)~~~(i=1, 2) 
   \end{equation}                         
   Using the last member of (9.14) to re-write the first term on the RHS of (9.55) and noting that:
    \begin{equation}  
    t_i - t_0 = t_D-t_0-\frac{L}{v_i}~~~(i=1, 2)
    \end{equation}
    gives
     \begin{eqnarray}
   -\frac{\hbar \phi^{(i)}}{c^2} & = & \frac{m_i^2 L}{p_i}+ m_{\pi}(t_D-t_0) -\frac{ m_{\pi} L}{v_i}
      \nonumber \\
       & = & \frac{m_i^2 L}{p_0}\left[1-\frac{c m_{\pi}}{2 p_0}\right]+ m_{\pi}(t_D-t_0-\frac{L}{c})
     + O[m_i^4]~~~(i=1, 2) 
   \end{eqnarray}
     where the neutrinos are assumed to be ultra-relativistic so that:
       \begin{equation} 
    v_i = c\left(1-\frac{c^2 m_i^2}{2 p_0^2} \right) + O[m_i^4]~~~(i=1, 2)
    \end{equation}
   In (9.57) and (9.58) $p_0$ is the centre-of-mass momentum of a massless neutrino produced
   in the decay $ \pi \rightarrow \mu \nu$
    \begin{equation}
    \frac{p_0}{c} = \frac{m_{\pi}^2-m_{\mu}^2}{2 m_{\pi}} = 29.8~{\rm MeV/c}
    \end{equation} 
    The replacements $A_{FI}^{A,B} \rightarrow A_{e \mu}^{(1),(2)}$ in (9.2) give then,
    for the probability
   of electron detection at distance $L$ from the neutrino source:
      \begin{eqnarray}
       P_{e \mu} & = & \int_{t_{min}}^{\infty}d t_D \left|\frac{\langle e^-  p |T|\nu n \rangle
        \langle \nu \mu^+ |T|\pi^+ \rangle}{L}\right|^2\exp\left[-\frac{\Gamma_{\pi}(t_D-t_0)}{\hbar}\right]
      \nonumber \\
   &   &  \times \left\{ |U_{e 1} U_{\mu 1}|^2 \exp\left[\frac{\Gamma_{\pi} t_1^{fl}}{\hbar} \right]  
     +  |U_{e 2} U_{\mu 2}|^2 \exp\left[\frac{\Gamma_{\pi} t_2^{fl}}{\hbar}\right] \right. \nonumber \\
   &   & \left. +2 U_{e 1}U_{e 2}U_{\mu 1} U_{\mu 2}\exp\left[\frac{\Gamma_{\pi}(t_1^{fl}+t_2^{fl})}{2 \hbar}\right]
        \cos\left(\frac{\Delta(m_{12})^2 c^2}{p_0}\left[\frac{c m_{\pi}}{2 p_0}-1\right]\frac{L}{\hbar}\right) \right\}
   \end{eqnarray}
  where
 \[\Delta(m_{12})^2  \equiv m_1^2-m_2^2  \]
  In (9.60) the neutrino times-of-flight, $t_i^{fl}$, have been introduced:
     \begin{equation}
     t_i^{fl} \equiv t_D-t_i = \frac{L}{v_i}= \frac{L}{c}\left(1+\frac{c^2 m_i^2}{2 p_0^2} \right) + O[m_i^4]~~~(i=1, 2)
     \end{equation}
 and the unobserved detection time $t_D$ has been integrated out.
  Consideration of Fig.12 shows that the values of $t_{min}$ are: $t_0+t_1^{fl}$ for the propagation
  of neutrinos of mass $m_1$ only, $t_0+t_2^{fl}$ for the propagation
  of neutrinos of mass $m_2$ only, and $t_0+t_1^{fl}$ for the interference term in (9.60). This is because
  both neutrino paths must be possible if interference is to occur, and $t_1^{fl}>t_2^{fl}$. Performing the 
   $t_D$ integrations in (9.60) and substituting the values of the MNS matrix elements from (9.54) then
   gives the final result:
      \begin{eqnarray}
       P_{e \mu} & = & \frac{2 \hbar}{\Gamma_{\pi}}\left|\frac{\langle e^-  p |T|\nu n \rangle
        \langle \nu \mu^+ |T|\pi^+ \rangle}{L}\right|^2 \sin^2 \theta_{12} \cos^2 \theta_{12} \nonumber \\
   &  &  \times \left\{ 1- \exp\left[-\frac{\Gamma_{\pi}c \Delta(m_{12})^2 L}{4 \hbar p_0^2}\right]
         \cos\left(\frac{\Delta(m_{12})^2 c^2}{p_0}\left[\frac{c m_{\pi}}{2 p_0}-1\right]\frac{L}{\hbar}\right) \right\}
    \end{eqnarray}
     The exponential damping correction of the interference term due to the non-vanishing pion lifetime
    is completely neligible\footnote{Setting $\Delta(m_{12})^2 c^2 L/(p_0 \hbar)$ to unity, so that the argument
    of the cosine in (9.62) is of the same order, as necessary for an observable neutrino oscillation
    effect, gives for the damping term: $\exp[-\Gamma_{\pi}/(4 p_0)] = \exp[-4.0 \times 10^{-16}]$.}
    \par  The prediction (9.61) is readily generalised to the case of an arbitary two-flavour neutrino oscillation
     experiment with a stationary source,$S$:
    \[    S \rightarrow X_S+ \overline{\ell}_j + \nu_i,~~~ \nu_i+ T_D \rightarrow X_D \ell_k \]
     to yield the prediction:
       \begin{eqnarray}
       P_{k j} & = & \frac{2 \hbar}{\Gamma_{S}}\left|\frac{\langle \ell_k X_D|T|\nu T_D \rangle
        \langle \nu  \overline{\ell}_j X_S|T|S \rangle}{L}\right|^2 \times \nonumber \\   
  &  &  \left\{ |U_{k 1} U_{j 1}|^2+|U_{k 2} U_{j 2}|^2 \right. \nonumber \\ 
  &  &   \left.  +2 U_{k 1}U_{k 2}U_{j 1} U_{j 2}\cos\left[\frac{\Delta(m_{12})^2}{m_S}\left(\frac{R_m}{1-R_m^2}\right)^2
     \frac{L}{\hbar} \right] \right\}   
     \end{eqnarray}
      Here $\ell_i$ ($\overline{\ell}_i$) denotes a charged lepton (antilepton) of fermion generation
     $i$ and 
  \[ R_m \equiv \frac{m_R}{m_S},~~~ (cm_R)^2 \equiv (p(X_S) + p(\overline{\ell}_j))^2 \] 
   Here $p(X_S)$ and $p(\overline{\ell}_j)$ are 4-vectors, $m_S$ is the mass of the source
   particle and $m_R$ is the effective mass of the particle, or particle system, $ X_S+ \overline{\ell}_j$, recoiling
   from the neutrino mass eigenstate. For the case of $\beta$-decay, where:
  \[ E_{\beta} = E(\nu_i) + E(\overline{\ell}_j) = {\rm constant} \]
   the argument $\phi^{(1)}-\phi^{(2)}$ of the cosine in (9.62) is replaced by~\cite{JHF1,JHF2}:
  \begin{equation} 
 \phi^{(1)}-\phi^{(2)} = \frac{\Delta(m_{12})^2}{p_{\nu}}\left[\frac{E_{\beta}}{2p_{\nu}}-1\right]
   \end{equation}
   \par It is instructive to now compare in more detail the neutral kaon oscillation and neutrino 
   oscillation cases. In Table 2 is shown the difference in production times of ${\rm K_S}$ and ${\rm K_L}$ mesons
   of various fixed momenta in order that they arrive simultaneously, in the alternative
    histories corresponding to either ${\rm K_S}$ or ${\rm K_L}$ propagation,  at a detection event at a typical
   decay distance $c \beta \gamma \tau_S$. A similar comparison is now made for neutrino oscillations
   by considering the difference of production time $\Delta t_{12} = t_2-t_1$ in order that the neutrinos
   $\nu_1$ and $\nu_2$ arrive simultaneously at the detection event, as shown in Fig.11 d). 
   The distance, $L(\pi)$, between the production and detection events is chosen so that the 
  phase of the cosine term in (9.62) is $\pi$ rad. Thus:
    \begin{equation}
      L(\pi)  = \frac{h p_0}{\Delta(m_{12})^2 c^2 (\frac{c m_{\pi}}{p_0}-2)} =
      \frac{13.8 \rm{m}}{\Delta(m_{12})^2 c^4 ({\rm eV})^2)}
    \end{equation}
 The corresponding production time difference is:
    \begin{equation}
     \Delta t_{21} =  L(\pi)\left(\frac{1}{v_1}-\frac{1}{v_2}\right) 
       = \frac{L(\pi)}{c} \frac{\Delta(m_{12})^2 c^2}{2 p_0^2}+ O[m_i^4]
     \end{equation}
     where (9.57) has been used. Combining (9.65) and (9.66) gives:
     \begin{equation}
      \Delta t_{21} = \frac{h}{4 c(\frac{m_{\pi} c}{2}-p_0)} = 8.22 \times 10^{-24} sec
      \end{equation}
 Thus $ \Delta t_{21}$ depends only on the kinematics of the neutrino production process 
   and is independent of the neutrino mass difference. For neutrinos produced in pion decay,
  $ \Delta t_{21}$ is a factor of $3 \times 10^{-16}$ smaller than the pion mean lifetime
   of $2.6 \times 10^{-8}$ sec. This implies strictly equal amplitudes for the decay processes
   shown in Fig.11 b) and c) with completely negligible lifetime damping effects as discussed 
  above. Further estimates of the latter may be found in ~\cite{JHF2}. The situation is thus
  dramatically different from the neutral kaon flavour oscillation case where, instead, the
  time differences for simultaneous arrival of equal momentum  ${\rm K_S}$ and ${\rm K_L}$ mesons
  at the detection event shown in Table 2 are much larger than the characteristic time of
  the production process. Thus the ${\rm K_S}$ and ${\rm K_L}$ are produced at essentially
  the same time. Because of the tiny ${\rm K_S}$--${\rm K_L}$ mass difference:
  \[  \Delta m_{12}/\overline{m} = 3.49 \times 10^{-12} {\rm~ MeV/c^2}/
       497.7 {\rm~ MeV/c^2} = 7.0 \times 10^{-15} \]
    momentum and velocity smearing effects due to variation of the physical masses of the
     ${\rm K_S}$ and ${\rm K_L}$ and radiative corrections ensure that the  probability amplitudes
     for  ${\rm K_S}$ and ${\rm K_L}$ of the same velocity are closely equal, leading to the 
     conventionally used formula (9.47). In the neutrino case however inspection of
      equal velocity condition analgous to (9.38):
      \begin{equation}
     \frac{m_1}{m_2} = \frac{p_1}{p_2}
      \end{equation}
   shows that it may be physically impossible for $\nu_1$ and $\nu_2$ to have the same velocity
   in the interfering probability amplitudes. 
   \par The current analysis of atmospheric neutrino oscillations where pion decays give
    an important contribution estimates the mass difference to be~\cite{Kays04}:
       \begin{equation}
      \Delta(m_{12})^2 c^4 \simeq 2 \times 10^{-3} (\rm{eV})^2
        \end{equation}
    In order to respect the equal velocity condition (9.68) the smearing, $\delta p$, of the
    neutrino momentum must be such that:
       \begin{equation}
       \frac{\delta p}{\overline{p}} = \frac{\Delta m_{12}}{\overline{m}} = \frac{m_1-m_2}{\overline{m}} 
       \end{equation}
      Now,
        \begin{equation}
   \Delta(m_{12})^2 = (m_1-m_2)(m_1+m_2) = 2 \Delta m_{12} \overline{m}
      \end{equation}
     So, combining (9.70) and (9.71), equal velocities require that:
         \begin{equation}
       \frac{\delta p}{\overline{p}} = \frac{\Delta(m_{12})^2}{ 2(\overline{m})^2}  
       \end{equation}
    Assuming $\overline{m} = 0.5$ eV/$c^2$, which is consistent with all direct upper limits
 on neutrino masses~\cite{PDG04}, and inserting the experimental value of $\Delta(m_{12})^2$
  from (9.69) gives:
      \begin{equation}
       \frac{\delta p}{\overline{p}} = 4 \times 10^{-3} 
       \end{equation}  
    The dominant source of momentum smearing in pion decay is radiation of real photons.
 Using the soft photon formalism of~\cite{RK} to calculate the mean momentum smearing,
   $\langle \delta p_{rad} \rangle$ due to soft photon emission gives:
     \begin{equation}
       \frac{\langle \delta p_{rad} \rangle}{\langle p \rangle} = 3.5 \times 10^{-4} 
       \end{equation} 
    which is incompatible with the equal velocity condition (9.73). Indeed the latter condition
   can be consistent with (9.74) only if $\overline{m}$ is close to the current direct upper limit
   on $m_{\nu_1}$ of 3 eV/$c^2$~\cite{PDG04}. In this case $\delta p/{\overline{p}}\simeq 1.0 \times 10^{-4}$,
   which is just compatible with (9.74). There is still, however, no physical reason why 
   probability amplitudes with equal velocities and production times should be 
   favoured over those with different velocities and different production times.
   \par The difficulty to justify the equal velocity hypothesis for neutrinos in
    view of the condition (9.68) has been previously pointed out in the literature~\cite{DeLeo,Okun}
 but the authors of these papers still concluded that the `standard' formula for the
  neutrino oscillation phase~\cite{Kays04}, which will now be discussed, was correct.
   \par As pointed out in ~\cite{JHF1}, the derivation of the standard formula requires that
   the neutrino velocities defined kinematically according to the formula $v_{kin} \equiv p c^2/E$ are
    different for different mass eigenstates, but that the space-time velocities: $v_{s-t} \equiv L/t$ are the same.
    These manifestly contradictory hypotheses (if neutrinos are indeed real (on-shell) particules propagating
   in space-time) give, instead of the prediction of (9.63) for the oscillation phase:
   \begin{equation}
    \phi_{12} = \phi^{(1)}-  \phi^{(2)} = \frac{\Delta(m_{12})^2 c}{m_S \hbar} \left(\frac{R_m}{1-R_m^2} \right)^2 L 
     +O[m_i^4]
     \end{equation} 
  the standard formula:
   \begin{equation}
    \phi_{12}^{stand} = \phi^{(1)}-  \phi^{(2)} = \frac{\Delta(m_{12})^2 c^2}{2 p_0 \hbar} L 
     +O[m_i^4]
   \end{equation} 
  This is actually the same result as found here for the neutral kaon oscillation phase
 (9.45) since:
   \[ \overline{m} \Delta m_{12} = \frac{(m_1+m_2)}{2}(m_1-m_2) = \frac{\Delta(m_{12)}^2}{2} \]
    \par A simple derivation of (9.76) can be found in ~\cite{Kays04}. Each neutrino mass
   eigenstate is assumed to evolve temporally in its rest frame according to the time-dependent
   Schr\"{o}dinger Equation. The corresponding phase increment is correctly given by
   the first member of (9.14) above. The Lorentz invariance of the phase is then used to write it
  in terms of laboratory-frame quantities:
  \begin{equation}
    \phi^{(i)} = -i \frac{m_ic^2 \tau_i}{\hbar} \simeq -i\frac{(E_i t - p_i c L)}{\hbar}~~~~(i=1,2)
   \end{equation} 
   It can be seen that it is just at this point in the derivation, that the assumption
  of equal propagation times, $t$, for each mass eigenstate implies equal space-time velocities
  $L/t$ whereas the kinematical velocities $p_i c^2/E_i$ are assumed to be different.
  As discussed in detail in ~\cite{JHF1} it is this unphysical assumption in the
   approximation made in the last member of (9.77) that leads
 to the factor of 2 difference between (9.76) 
   and the neutrino propagation phase given by the exact relation (9.14). The standard formula (9.76)
   is derived from $ \phi^{(1)}-  \phi^{(2)}$, 
  as given by (9.77), on making the further assumptions that the neutrinos are ultra-relativistic,
   that $L/t = c$, and retaining only the leading $\Delta(m_{12})^2$ neutrino mass terms.
   Evidently, since (9.77) assumes equal production times for different mass eigenstates,
   there is no contribution to the 
   oscillation phase from the source particle in the standard formula. 
   \par The standard formula (9.76) is still universally employed by the neutrino physics
   community. Claims that the Feynman path amplitude formula (9.75) is incorrect 
   have been made~\cite{Giunti} and rebutted~\cite{JHF1,JHF4}. A critical discussion of the
  {\it ad hoc} `wavepacket' approach, that is frequently used in discussions in the
  literature of the quantum mechanics of neutrino oscillations, that results,
   -- from the point-of-view of Feynman's formulation of quantum mechanics embodied
   in the laws I-V presented in Section 2 -- an
   artificial and unphysical
  blurring of the space-time structure of production and detection events, can be found in~\cite{JHF1,JHF2}.
   \par The formula for the production time difference, (9.66) becomes, in the general case
   described by (9.67):
    \begin{equation}
    \Delta t_{21} = \frac{h}{2m_S c^2 R_m^2}
   \end{equation} 
   Thus, when the mass of the system recoiling against the neutrino becomes very small in comparison
   with the mass of the source particle, $ \Delta t_{21}$ becomes very large, as does also
   the `oscillation length, $L_{osc}$, defined by writing the argument of the cosine
   interference term in (9.63) as $2 \pi L/L_{osc}$ so that:
     \begin{equation}
      L_{osc} \equiv \frac{m_S h}{\Delta(m_{12})^2 c} \left(\frac{1-R_m^2}{R_m}\right)^2 
   \end{equation} 
      This dependence of the oscillation length on the kinematics of the neutrino production
   process gives a straightforward  experimental method to discriminate between the path amplitude
   prediction for the oscillation length, (9.79) and that predicted by the standard formula (9.76)
   which is: 
       \begin{equation}
      L_{osc}^{stand} \equiv \frac{ 2 p_0 h}{\Delta(m_{12})^2 c^2}
     \end{equation}   
   As proposed in ~\cite{JHF3}, if a terrestrial long linebase `$\nu_{\mu}$ disappearence' experiment
 such as K2K~\cite{K2K} is performed using neutrinos produced in the process: ${\rm K} \rightarrow \mu \nu$, rather 
  than $\pi \rightarrow \mu \nu$ as in the existing experiment, the path amplitude formula predicts
   that the oscillation length is a factor of $\simeq 28$ times longer for a kaon source than
   for a pion one yielding the same neutrino momentum. This implies a strong suppression
  of the `$\nu_{\mu}$ disappearence' phenomenon when using a kaon rather than a pion source. The
   standard formula predicts the same oscillation length for pion and kaon sources giving
  neutrinos of the same momentum.
   \par The four different `two probability amplitude' experiments that have just been discussed
 are summarised in Table 3. Each experiment is first classified in terms of its space-time properties.
 These are specified by the differences between the path lengths ($\delta r$), the
 times-of-flight ($\delta t$) and the particle velocities ($\delta v$) for the two paths. This information is
 given as a Yes/No answer to the questions:  $\delta r = 0$?,  $\delta t = 0$? and  $\delta v = 0$?.
 Similarly the contributions to the interference phase from the source and the propagating
 particles are specified by the differences $\delta \phi^{source}$ and $\delta \phi^{part}$ for the two paths
 and given as answers to the questions $\delta \phi^{source} = 0$? and  $\delta \phi^{part} = 0$?.
 In the last two rows the values of the interference phase and the ratio $\lambda^{eff}/\lambda^{DB}$
 are reported.  The effective wavelength, $\lambda^{eff}$ is defined for neutral kaon 
  abd neutrino oscillations in terms of the
  phase difference $\Delta \phi$ for paths of length $L$ by the relation:
 \begin{equation}
  \lambda^{eff} \equiv \frac{2 \pi L}{|\Delta \phi|} 
 \end{equation}
 while $\lambda^{DB}$ is the conventional de Broglie wavelength of the propagating particle as given by
 (9.19).
   
     \begin{table}
   \begin{center}
   \begin{tabular}{|c||c| c| c| c|} \hline  
   Experiment   & Photon YDSE  & Electron  YDSE  &  ${\rm K}_0$ Osc. &  $\nu $ Osc. \\
  \hline        
  \hline
   $\delta r = 0$? & No & No & Yes & Yes \\
  $\delta t = 0$? & No & Yes & Yes & No \\
 $\delta v = 0$? & Yes & No & Yes & No \\
 $\delta \phi^{source} = 0$? & No & Yes & Yes & No \\
$\delta \phi^{part} = 0$? & Yes & No & No & No \\
$\hbar(\phi_B-\phi_A)$ & $-\overline{p} \Delta r$ & $-\overline{p} \Delta r$ &
  $-\frac{\Delta(m_{LS})^2 c^2}{2 \overline{p}}L$ & $-\frac{\Delta(m_{12})^2c}{m_S}\left(\frac{R_m}{1-R_m^2}\right)^2L$  \\
$\lambda^{eff}/\lambda^{DB}$ & 1 & 1 &  $2\left(\frac{\overline{p}}{m_S c}\right)^2$ &
  $\frac{\overline{p}}{m_S c}\left(\frac{1-R_m^2}{R_m}\right)^2$    \\
 \hline
  \end{tabular}
   \caption[] {{\sl Comparison of the four `two probabilty amplitude' experiments.
 The characteristics of each experiment are specified by the quantities: $\delta r$,  $\delta t$,
   $\delta v$, $\delta \phi^{source}$ and $\delta \phi^{part}$ which are the diffrences of;
   path length, time-of-flight, velocity, source phase and particle phase for the two interfering
 amplitudes. Also shown is the interference phase $\phi_B-\phi_A$ in (9.2) and the ratio of
  the effective wavelength defined in (9.80) to the de Broglie wavelength. }} 
  \end{center}
  \end{table}    

  \par The puzzling and difficult-to-understand concept of `wave-particle duality' does not survive the detailed
  analysis of the superficially analogous, but in fact very different, experiments presented in Table 3.
   The `wave' associated
  with the particle is different in every case. Although both the YDSEs may be simply (and correctly) interpreted
  in terms of the de Broglie wavelengths of the photon or the electron, the physical origin of the `wave'
  is quite different in the two cases. For the photon it is the phase of the propagator of the excited
  source atom; for the electron it is the phase of the propagator of the latter. The effective
  wavelength for neutral kaons, although originating entirely from their propagators is quite 
  different from that of the electron in the YDSE experiment. This is a result of the different
 space-time properties of the experiments as detailed in Table 3. For neutrino oscillations
   $\lambda^{eff}$ is again different and depends on the propagators of both the source particle
  and the neutrinos. 
  \par The conclusion to be drawn from Table 3 is clear. In a strict, ontological, sense only the
   source and propagating particles exist as physical entities, and it is their particular
   properties together with the space-time structure of the production and detection events that determine
   $\phi_B-\phi_A$. The associated `wave' is only a (sometimes useful, sometimes not) mathematical
   abstraction, as clearly stated long ago by E.J.Williams~\cite{Williams}
   \par {\tt The electron is, of course, a particle. The wave is in the mathematics.}
    \par Although its inventor doubted it until the end of his days\footnote{`What are light quanta?
    Today every scoundrel believes he knows the answer, but he is wrong.'~\cite{EinBess}} the photon is 
    also a particle. The realisation of this is enough remove Einstein's mysterious and counter-intuitive postulate
    concerning the constancy of the velocity of light from the foundations of special relativity~\cite{JHF5},
    and, by comparing photon properties with those of classical electromagnetic waves, to understand,
    in a new way, the basic concepts of quantum mechanics~\cite{JHF6}. Also, all the mechanical aspects of
    classical electromagnetism may be quantitatively explained, at the quantum level, as an effect of
    the exchange of space-like virtual photons~\cite{JHF7}.

 \SECTION{\bf{Summary and Outlook}}
  In this paper Feynman's path amplitude formulation of quantum mechanics has been applied to several
  specific problems. In particular many of the optical experiments discussed in a qualitative way in
  Feynman's book `QED'~\cite{Feyn1} have been calculated in detail. The problems of physical optics
  addressed include: the calculation, from first principles, of the refractive index of a 
  transparent medium, the laws of refraction and reflection of light at the interface of two transparent 
  media, the rectilinear propagation of light in a homogeneous transparent medium, Fermat's
    Principle, the reflection
   coefficient of light at normal incidence from the interface between two transparent media,
   and temporal dependence of the fringe visibility in a Michelson interferometer. The path amplitude
  method is also used to analyse quantum interference effects in Young double slit experiments (YDSE)
  using photons or electrons, in neutral-kaon quark flavour oscillations and in neutrino
  flavour oscillations. In all cases, except neutrino oscillations, agreement is found
  with the predictions of the classical wave theory of light or `wave particle duality' 
 (i.e. introducing the de Broglie wavelength of a massive particle in analogy with 
  the wavelength of the photon), for the YDSEs, or, 
   in the case of neutral kaon or neutrino flavour oscillations, solving the 
   time-dependent Schr\"{o}dinger Equation in the particle rest frame.
   \par All of the calculations presented are straightforward applications
   of Feynman's five laws of quantum dynamics, that provide a description
   of causally-related elementary processes that succeed each other in space-time. These laws, 
  presented and discussed in Section 2, are:
  \begin{itemize}
   \item[(I)] The probabilistic interpretation of the Probability Amplitude (2.1).
  \item[(II)] The law of Sequential Factorisation (2.2).
  \item[(III)] The law of Quantum Mechanical Superposition (2.3). 
  \item[(IV)] The law of Composite Factorisation (2.4)
  \item[(V)] The Feynman Path Integral (2.9)
  \end{itemize}
  The dynamics of the theory is entirely contained in V. The laws II-IV 
  describe how the probabilty amplitude is constructed, while I gives its
  physical interpretation. The probability amplitude is built, according to
  II, as the product of a series of amplitudes describing elementary physical 
   processes ordered in time. For the calculation of space-time dependent
  interference effects, the most important of these process amplitudes are the
  Green functions, of space-time propagators, of particles in motion or at rest.
  The derivation of these propagators from, respectively, the Feynman Path Integral
 and the time-dependent Schr\"{o}dinger Equation, is described in Section 3.
  \par Before addressing specific problems of quantum interference, the physical
  interpretation of Feynman's formulation of quantum mechanics is discussed in
  some detail in Section 2. This interpretation is indeed quite different from
  that of previous formulations since the primary physical concept,
  the path amplitude, is neither a Hilbert-space state vector nor 
  a wavefunction, and the concept of a particle, localised in space-time,
  as in classical physics, is also essential. Indeed consideration of classical 
  propagation of particles localised in space-time is necessary for the correct
  calculation of the phases of the path amplitudes that are the physical 
  basis of all space-time quantum interference phenomena. In this theory, particles,
  once created, propagate in space-time in a classical manner within each path
  amplitude. In the limit 
  $\hbar \rightarrow 0$ the sum over paths is replaced by a single path, the one
  predicted by Hamilton's Principle of classical mechanics,
   and Feynman's laws I and II become a statement
  of Bayes theorem for the combination of conditional propabilities. 
  \par Shortcomings
  of the conventional text book `wave packet' representation of particles and the associated
  misinterpretation of the space-time Heisenberg Uncertainty Relation are pointed out.
  The relation of Feynman's formulation to other `interpretations' of quantum mechanics:
  Consistent Histories, The `Many Worlds' interpretation and de Broglie's Pilot
  Wave theory are discussed. The much debated `measurement problem' of the
  Heisenberg and Schr\"{o}dinger formulations of quantum mecanics, seems, at least superficially, to be
  absent from Feynman's formulation. Perhaps new `problems' remain to be discovered; but in any case,
   to the writer's best knowledge, no
  discussion of the `philosophy' of Feyman's formulation  yet exists.
  \par In Section 4 it is demonstrated that Feynman's path amplitude formalism, in which
   the dynamics is entirely contained within the Green functions of particles, predicts, in
   a completely general manner, the classical wave theory of light, under a certain well-defined
   condition. The latter is that the lifetime of the photon source must be much
    longer than the photon path differences, in the experiment under consideration,
    divided by the velocity of light. This is the case for essentially all 
   quantum optics experiments using  a laser as light-source. All such experiments
   can then be correctly analysed using the classical wave theory of light, i.e. as done in well-known
    text books on optics~\cite{BW1,MW1}, in terms only
   of spatial intervals and photon wavelengths, without taking into account the times of
    the events corresponding to different observed physical processes. 
    Although Feynman's formulation predicts the existence of an effective `wave field' $U$
    that satisifies Hemholtz' Equation (4.13), from which all predictions for interference and diffraction
   phenomena of the classical wave theory may be derived, the physical source
  of the phase of $U$ is solely the temporal propagator of the excited source atom, not the
  propagator of the photon itself. Thus any `wave particle duality' in physical optics is
  between the photon (`particle') and its source (`wave') not between the `particle' and
  `wave' aspects of the photon itself. Indeed the latter does not exist in 
   the path amplitude formulation. Thus, starting only from the quantum 
   properties of particles (their Green functions) their `wave' properties may be
    derived in full. It can seen that the latter are then, albeit useful, mathematical abstractions
  that do not exist in the same strictly ontological sense as the particles.
   Given the {\it a priori} existence of particles and Feynman's rules for their
  quantum behaviour, all their so-called `wave' properties may be derived. The inverse procedure
  is evidently not possible. The fundamental physical entities are then the particles,
  not the `waves'. In consequence there is in fact no `wave particle duality' in the physical 
  description, though by chance, in some cases such as in Davisson and Germers's original `matter wave'
  experiment~\cite{DG} or the electron Young double slit experiment, a naive application
  of this idea gives the same prediction as the full path amplitude calculation.
   \par Finally in Section 4 it is demonstrated that the amplitude for forward diffractive scattering
    derived from the Helmholtz Equation by means of Green's theorem, can also be derived by direct
    summation of path amplitudes. The integration procedure used is similar to that employed
   in Section 5 to calculate the refractive index of a transparent medium in terms of the
   scattering amplitude of light by the atoms of the medium.
   \par Section 5 contains three different calculations of the refractive index.
  In the first two it is assumed that the photon which interacts with the refractive medium
 is produced sufficently long after after the production time of the excited source atom
  that all geometrically allowed paths of the photon are possible. The first calculation 
  assumes a sheet of refractive material so thin that, to a good approximation, the refractive index
  may be calculated by considering a single scattering of the photon from the atoms 
  of the material. For this condition to be satisfied the sheet must be very thin --a small fraction
  of the photon wavelength. The second calculation removes this restriction and considers a block
  of uniform refractive material of arbitary thickness. In this case any number of scatterings of the photon from the
  atoms of the material are taken into account. The formula obtained for the refractive index is
  the same (5.9) as that obtained in the single scattering case. The third calculation
   considers photons produced and detected promptly after the production of the excited atom. In this case,
   causality  forbids photon trajectories that deviate significantly from the classical straight 
  line path between the source and the detector, reducing the number of possible interactions
   with the atoms of the refractive medium and therefore strongly reducing the refractive 
  index --the so-called `refraction annulment' effect. A simple experiment to detect this phenomenon
   (if indeed it exists) is shown in Fig.4.
   \par In Section 6 the laws of refraction, reflection and rectilinear propagation of light
  are derived by imposing the condition that the phases of the probabilty amplitudes of
   suitably defined experiments are stationary for variations of the spatial location 
   of the photon path from its source to the detector. The generalisation of the stationary
   phase condition to the motion of an arbitary physical object gives the quantum mechanical
   explanation of Hamilton's Principle of classical mechanics. Since the phase of the probability
   amplitude is proportional to the effective propagation time of a photon from its 
   source to the detector Fermat's principle of least time is also derived.
 By considering the second 
   derivatives of the phase of the probability amplitude, with respect to the positions
  and angles of the corresponding paths, the the size of the spread of photon trajectories
   around the classical path corresponding to a stationary phase, is estimated.
   \par In Section 7 the experiment used in ~\cite{Feyn1} to exemplify the application
  of quantum mechanics to optics, the reflection of light at normal incidence from the boundary
  of two transparent media with different refractive indices is analysed in detail for the 
  case of a realistic experimental set-up. It is essentially the same as the first 
   quantitative quantum mechanical experiment ever performed --Newton's
   study of his `Rings'~\cite{Newton}. The result for the reflection coefficient at a vacuum/medium
   interface, (7.8) is significantly different from the formula (7.10), first derived by Fresnel, that
   is to be found in all text books on optics. For a refractive index $n= 1.5$, the reflection
   coefficient predicted by the path amplitude calculation is 0.028, some 40$\%$ smaller than the
    prediction, 0.040, of the Fresnel formula. Experimental discrimination between these two possiblities
    should be simple and straightforward. The present writer has been unable to find either in
    text books or in the published research literature any account of experimental verification of the
    Fresnel formula (7.10) for photons in the optical region of wavelength.
     \par Fresnel originally derived his formula for the reflection coefficient from
    an elastic-solid model of light waves. The same formula is obtained in 
    classical electromagnetism by applying surface
    boundary conditions to the electric and magnetic fields of the different electromagnetic waves
     in refractive
    media of different refractive indices. The problem with this approach may be that
   although a classical electromagnetic wave is, certainly, a high density beam of
    monochromatic photons, a single photon is, equally certainly, {\it not} a classical
   electromagnetic wave, so it makes no sense to associate with it classical
  electric and magnetic fields. If the Feynman space-time picture of reflection is correct,
   the essential physics of the process relates not to the interface
   between media but resides in the amplitude ${\cal A}_{scat}$ for elastic scattering of
   photons by the atoms in the interior of the medium, as is manifest in the derivation
   of the  formula (5.7) for the
   refractive index. This formula, and the formula (7.8) for the reflection coefficient
  are derived in exactly the same way by by integrating the path amplitudes corresponding
   to the scattering of the photon over {\it all of the atoms} of the refractive
   material probed by the photon in the experiment. The interface occurs only geometrically
  as a limit on this spatial integration.
  \par  This being said, it is certainly possible to envisage an experiment where the Fresnel
  formula would be appropriate and correct. For example, the reflection of a beam
   of microwave photons of sufficiently high spatial density from the surface 
  of a dielectric. In this case the description of the beam as an electromagnetic wave will be 
  appropriate, and the molecules of different dielectrics will be differently polarised, over macroscopic
  distances,
  by the electric fields associated with the microwave beam. The waves in different media will
  have different characteristics related by the boundary conditions on the electric fields
  of the waves, so that the Fresnel formula for the reflection coefficient should be 
  correct, and the scattering amplitude of individual microwave photons from individual
  molecules less directly relevant. On the other hand, the macroscopic dielectric polarisation effects
  for a single optical photon entering a block of glass are clearly negligible and a purely
  quantum mechanical description is mandatory. For a general discussion of the different circumstances
  in which either the classical electromagnetic wave (high photon density) or the quantum
  mechanical (low photon density) description is appropriate, see ~\cite{JHF6}

  \par The experiment described in Section 8, measurement of the temporal dependence of the
   fringe visiblity in a Michelson interferometer is one in which path length differences
  are deliberately chosen to be of the order of, or greater than, $c \tau_S$ where 
   $\tau_S$ is the mean lifetime of the excited atomic state that produces the observed
  photon. The classical wave theory of light, in which only spatial intervals are considered
  is therefore unable, in principle, to describe such an experiment. Thus new predictions
  specific to the path amplitude method are obtained. These predictions for the fringe visiblity
  as a function ot the time elapsed after the production of the excited source atom are shown,
   for different path length differences, in Fig.9. The path amplitude calculation of the
  effects of random source motion on the time-integrated fringe visibility found in Appendix E predicts
  that there should be no effect on the fringe visiblity from this source. This is in contradiction
  with the prediction, (8.14), of Rayleigh, based on the classical wave theory, that such
  `Doppler broadening' effects should be large\footnote{Indeed, as shown by the entries in the 
   first and last rows of Table 1, the damping predicted by the Rayleigh formula
   is too large to explain Michelson's fringe visiblity measurements for the H$_r$ 3p-2s
    and  Na D 3p-3s transitions.}. In order to test the path amplitude prediction of a
   vanishing effect due
  to source motion, experiments are required that clearly discriminate this effect from
  that of pressure broadening. So far this has not been done.
  \par In Section 9 analagous `two probability amplitude' experiments in photon or electron
  optics, neutral kaon oscillations and neutrino oscillations are analysed and compared. 
  The understanding of the different results obtained, summarised in Table 3, requires 
  careful consideration of the nature of the particle source and the particles 
   themselves as well as the space-time 
  structure of the production and detection events of the different path amplitudes.
  Only for the photon and electron YDSE does the simple
  `wave particle duality' concept, that is the introduction of the de Broglie
   wavelength of the photon or the electron, give the same result as the path 
   amplitude analysis. For the neutral kaon and neutrino oscillation cases, quite
   different effective wavelengths are found for the associated `matter waves'.
   For  optical interference experiments the source of the `wave' is solely the
  propagator of the excited source atom. As neutrino oscillation experiments
   (unlike the electron YDSE and neutral kaon oscillation experiments)  
  have a similar excited source particle, it is clear, if the path
   amplitude description is indeed correct, that the source particle must give 
  an important contribution to the oscillation phase also in this case. In addition there is the
  contribution of the propagators of the neutrinos themselves, which is analogous
  to the sole source of the interference phase in the electron YDSE and neutral
   kaon oscillation experiments. Before the work reported in ~\cite{JHF1,JHF2,JHF3}
   only this last contribution to the neutrino oscillation phase was taken into account.
   \par Although all optical and massive particle diffraction and interference experiments that
   have been performed
    to date
   and found to be agreement with the predictions of the classical wave theory of light,
   or `wave particle duality'
   can be interpreted, following the arguments given in Section 4 and 9, as evidence for
    the correctness
  of Feynman's formulation of quantum mechanics, the present paper contains several new
  predictions. If any of these predictions is not confirmed by experiment,
   Feynman's space-time formulation of quantum mechanics is not of general validity
  and answers must be sought to the question why, in some cases, cited above,
   the predictions agree with experiment, and in others not. 
   These new predictions are:
 \begin{itemize}
  \item[(i)] The refraction annulment effect (Section 5).
 \item[(ii)] The reflection coefficient of light
            at normal incidence (Section 7).
  \item[(iii)] Time dependence of fringe visiblity in the
               Michelson interferometer (Section 8)
 \item[(iv)] Absence of source motion damping of fringe visiblity in the
               Michelson interferometer (Section 8).
  \item[(v)] Different oscillation lengths for `$\nu_{\mu}$ disappearence' using $\pi \rightarrow \mu \nu$
       or ${\rm K} \rightarrow \mu \nu$ as neutrino sources.
 \end{itemize}
 \par It is clearly of interest to apply the methods developed in the present paper to 
  other related physical problems. Some examples are:
  \begin{itemize}
   \item[(A)] The coefficients of reflection and refraction of light at non-normal incidence.
    For this it is necessary to take into account photon polarisation. This requires, for the 
    path amplitude method, a complete quantum description of polarisation effects
    in both the photon production process (electric dipole radiation from a polarised source 
    atom) as well as photon scattering processes (Rayleigh scattering of polarised photons).
    Another interesting related problem is the path amplitude description of bi-refringence.

    \item[(B)] Heavy quark flavour oscillations for entangled systems. Examples are:
   \[ \phi \rightarrow {\rm K_L} {\rm K_S}~~~~~~~\Upsilon(4S) \rightarrow {\rm B_H} {\rm B_L}  \]
      The latter is of considerable current interest due to the on-going 
    experimental programs of the BABAR and BELLE b-factories. Unlike the `unentangled' process
     $\pi^- p \rightarrow \Lambda ({\rm K_L}, {\rm K_S})$ described above, both the $\phi$
   and the $\Upsilon(4S)$ are coherent sources, similar to an excited atom in physical optics.
    The corresponding values of $\Delta t_{SL}$, $\Delta t_{LH}$ calculated using (9.41) 
    are $6.12 \times 10^{-25}$s, $9.55 \times 10^{-26}$s  repectively, where the value
    $m_H-m_L = 3.28 \times 10^{-10}$ Mev/c$^2$~\cite{PDG04} has been used for the neutral
   b-meson mass difference. These time differences may be compared with the mean lifetimes 
   of the $\phi$ and the $\Upsilon(4S)$ source particles of
      $1.5 \times 10^{-22}$s and $4.7 \times 10^{-23}$s~\cite{PDG04}, which are factors 
    of 24.5 and 492 times greater than  $\Delta t_{SL}$ and  $\Delta t_{LH}$. Thus, unlike for the 
    case of  $\pi^- p \rightarrow \Lambda ({\rm K}_L, {\rm K}_S)$ neutral K- or b-mesons
    can, in principle, have the same detection time in alternative histories with different
    mass eigenstates, provided that, as in physical optics or neutrino oscillations, the particles
    are produced at different times in the alternative histories. In view of more stringent
   space-time restrictions imposed by the requirement of observation of both final state particles
   it is possible that the off-shell nature of the propagating particles is more important, in this case, in
   enabling interference, rather than production time differences or velocity smearing due to 
   radiative effects. Only an actual calculation can show if this is the case or not.
  \item[(C)] Effective refractive indices for neutrinos. 
  It is important, in view of the great current interest in neutrino oscillation 
    experiments to repeat the refractive index calculations of Section 5 above for the case of
  neutrinos interacting with matter. The model used till now to derive the effective
   refractive indices of neutrinos in bulk matter ~\cite{Wolfenstein,MS} was constructed
  by analogy with that used to used to describe coherent regeneration~\cite{PP} of neutral kaons
   for which (see Table 3 above) simple `wave particle duality' is a good approximation
   to path amplitude calculations. However, in the neutrino case, the path amplitude calculation
   suggests that this analogy breaks down, and an important contribution to the interference
 phase originates in the propagator of the source particle. This effect has not, to date, been included
  in the calculation of neutrino refractive indices.
  \end{itemize}
  \par One of the precepts of the standard `Copenhagen Interpretation' of quantum mechanics is that our direct
  exerience of the world can only be understood, in a rational manner, by an appeal to classical concepts
  \par{\tt The quantum theory is characterised by the acknowledgement of a \newline fundamental limitation in the
   classical physical ideas when applied to \newline atomic phenomena. The situation then created is of a peculiar
   nature, \newline since our interpretation of the experimental material rests essentially \newline upon the
   classical concepts
    \cite{BohrCI}.}
     \par Is it indeed true that `...our interpretation of the experimental material
      rests essentially upon the classical concepts'? In the case of one `experiment' that is performed
   by every human being\footnote{Colour-blind people excepted} every day of her (or his) life
     without she (or he) being consciously aware of it,
   this statement seems to be quite untrue. 
     The `experiment' in question is our perception of colour.
    The photon detector D, forming an essential part of all the optical experiments discussed in Sections 2-8
     above, was never precisely specified, but could well be one of the three types of cone receptor cells
    in the retina of the human eye. The detection efficiency of these receptors for photons is similar
    to that of a photo-diode with a flat spectral response in the optical region preceded 
    by wide-band wavelength filters with roughly Gaussian acceptance profiles centered at $\lambda_{\gamma}
    \simeq 440,~540$ and 570 nm with full widths at half-maximum $ \simeq 200,~200$ and 300nm~\cite{PD92}.
     What we perceive as the colour `green' corresponds not always to some discrete or narrow range of
    values of  $\lambda_{\gamma}$ but in general to a weighted average (by the human nervous system)
     of the intensities of photons in different spectral regions that are incident on the three
    types of receptors. The `interpretation of the experimental material' mentioned above by Bohr
    when this material is, say, the perception of the colour green or the colour red, varies 
     enormously according to the circumstances, but in all cases  this interpretation is possible only in
    terms of quantum, not classical, concepts. The sources of the photons for all the colour perception
    experiments that will now be discussed are excited atoms in the Sun. Myriads of different spectral
  lines produced by spontaneous decay of these atoms are responsible for the `white light' arriving at the
   surface of the Earth in daytime. The experiments differ in exactly how certain of these photons come to impinge
    on the cone receptors of the retina.
    \par I step out on to my balcony and look at a geranium. I see that its leaves are green and its flowers
    red. Why? The explanation is the different atomic structures of the matter of the leaves and flowers.
    Photons of some wavelengths are absorbed, others re-emitted as fluorescent radiation by these atoms.
    The re-emitted radiation, incident on the retinal receptors and suitably re-weighted gives the perception
   of green for the leaves and red for the flowers. In this case the physics explanation of colour vision
    is simple, the different parts of the geranium filter, in a different way, the sunlight photons that are incident
    on it.
    \par Nearby there is a shower of rain, and the Sun is lowering towards the western horizon. I look at the
    eastern sky and, at certain angles, see the green and red components of a rainbow. In this case
    the physics explanation of my perception of these colours is quite different. Now the photons responsible
  for the sensations of `green' and `red' have in each case closely similar wavelengths. The reason
   why they are observed in different spatial directions is explained by the phenomena analysed in Sections 5 and 6
   above --refraction and reflection by spherical rain-drops. The angles at which the green and red bows are 
    observed are controlled by the values of the scattering amplitude ${\cal A}_{scat}$ and the 
    wavelength $\lambda_{\gamma}^0$ in (5.9) causing different wavelength photons to follow diffent paths
    according to Snell's law of refraction, (6.11).
    \par Now I take a compact disc (CD) and hold it in the direct sunlight, turning it in my hand. 
     At certain angles of the face of the CD relative to the direction of the Sun I observe vivid flashes of
    green or red light. As for the rainbow, the photons striking the cone receptors in my retina,
    when I perceive the green or red colour, occur in narrow bands of wavelength. However, the physical
    source of this sensation of colour is yet again different. The face of the CD constitutes a diffraction
    grating. The bright flashes of colour correspond to constructive interference of the path amplitudes
    of photons for particular intervals of wavelength. The quantum mechanical description is essentially
    the same as for the photon YDSE described in Section 9, except that, not just two, but many, probability
    amplitudes interfere constructively~\cite{JWDG} to give large values of the combined
    probability amplitude $A_{fi}$ in (2.1) and therefore a large signal in the output
    neurones of the cone cell receptors. Where is now the physical origin of the visual perception
   of the different colours? It is not simply the grooves on the surface of the CD. These serve simply to
    define the lengths of the photon paths from their begining, at an atom in the Sun, to
    their end, at a cone cell in my retina. Constructive interference corresponds to a constant
    phase difference between the amplitudes of photon paths reflected between successive grooves on
    the CD. In Feynman's interpretation of quantum mechanics, these phase diffences are those between
    alternative paths of the {\it same photon} reflected from different places on the surface of the 
     CD in its alternative histories. Constructive interference occurs when, for each adjacent pair 
     A,B of paths, $\phi_B-\phi_A = 2 \pi n$ where $n$ is an integer. The actual physical source
    of this phase difference is, uniquely, the temporal propagator (2.12),(3.20) of the source atom.
    Thus, as I move the CD to observe either green or red interference fringes I select photons
    from different sets of source atoms in the Sun, and am, in fact, observing the temporal phase
    advances of the wavefunctions of these different atoms (Feynman's `stopwatch hand'~\cite{Feyn1}) as probed
     by alternative paths followed by each single photon that is absorbed in the cone cells
     of my retina. These phase advances actually occur some 8.3 minutes before
    I perceive the fringes!
     \par The last experiment on the perception of colour to be described is the one featured on the
     dust-jacket of the first edition of Feynman's book~\cite{Feyn1}, that provided the primary motivation
     for the writing of the present paper. In order to observe this effect --thin film interference,
     under the best conditions, an overcast, rainy, day is desirable. This is for two reasons,
     firstly because wet tarmac road surfaces favour the production of thin films when small oil-drops
     fall upon them from passing vehicles, and secondly because diffuse sunlight, multiply scattered
    and refracted from the water vapour droplets of clouds, ensures that photons with paths that satisfy conditions
    for constructive or destructive interference can arrive at the eye from all parts of the
    thin oil-patch. In this case, the perceived colour is affected both by refraction, as in the case
   of the rainbow, and by path differences as in the case of the photon YDSE, or the CD experiment.
    The angular positions at which the differently coloured `Newton's Rings' appear depend on
     both the thickness of the oil film and, as Newton noted~\cite{Newton}, on the refractive
    index of the oil. However, the basic physical mechanism, in Feynman's interpretation, is the same as for
    the CD experiment  --temporal phase advances of the wavefunctions of atoms in the Sun,
    several minutes before the photons from the decay of those atoms strike the receptor
    cells of the retina of the observer. 
    \par Indeed, quantum mechanics is essential not only for `the interpretation of the experimental
      material' but to the very physical mechanism of colour vision itself in the eye and the brain. 
     The cone cell receptors are quite analagous to the leaves and flowers of the geranium i.e. they function
   essentially as wavelength filters. They do this because of the different structures of atomic
   energy levels to be found in the three different types of cone receptor cells. The world of our
   every-day experience not only requires quantum mechanics for its rational interpretation
   but also for enregistering the raw data from which that experience is derived.

 \newpage
  {\bf Appendix A}
 \par The case of a rectangular boundary, of sides $L_Y$, $L_Z$ for the refractive medium in Fig.2 or Fig.3
   is shown in Fig 12a. The Ox axis, along which are situated the excited source atom and the 
   photon detector, intersects the $YZ$ plane at O$_P$ and has cartesian coordinates ($Z$,$Y$)
 relative to the center, O$_R$, of the rectangular boundary. The axes OY and OZ are parallel to the sides
  of the rectangle $B_1 B_2 B_3 B_4$  whose plane is perpendicular to Ox. The
  typical boundary point, $B$, has polar coordinates ($R_1$, $\phi_1$) relative to O$_P$
   The corresponding value of $r_1^{max}$ in (5.2) (see also Fig.2) is given by:
 
  \[ r_1^{max} = \sqrt{x_1^2 + R_1^2} = x_1 + \frac{R_1^2}{2 x_1} + O(\frac{R_1^4}{x_1^3})~~~~~~(A1)\]

 The angles $\Phi_1$, $\Phi_2$, $\Phi_3$ and $\Phi_4$, of the line segments $O_P B_1$, $O_P B_2$,
  $O_P B_3$ and  $O_P B_4$, joining the corners of the rectangle to O$_P$, are 
  defined, in an anti-clockwise sense, relative to the axis OZ,.
 \par The function $r_1^{max}(d_i, \phi_1)$ in (5.2) then has the following form,
  where corrections of $ O(R_1^4/x_1^3)$ are neglected:
  \par
  \[\Phi_4 < \phi_1 < 2\pi +\Phi_1\]
  \[ r_1^{max}(x_1,L_Z,Z, \phi_1) = x_1 + \frac{(L_Z/2-Z)^2}{2 x_1 \cos^2 \phi_1}~~~~~~(A2)\]       
    \par
  \[\Phi_1 < \phi_1 < \Phi_2\]
  \[ r_1^{max}(x_1,L_Y,Y, \phi_1) = x_1 + \frac{(L_Y/2-Y)^2}{2 x_1 \sin^2 \phi_1}~~~~~~(A3)\]
      \par
  \[\Phi_2 < \phi_1 < \Phi_3\]
  \[ r_1^{max}(x_1,L_Z,Z, \phi_1) = x_1 + \frac{(L_Z/2+Z)^2}{2 x_1 \cos^2 \phi_1}~~~~~~(A4)\]
        \par
  \[\Phi_3 < \phi_1 < \Phi_4\]
  \[ r_1^{max}(x_1,L_Y,Y, \phi_1) = x_1 + \frac{(L_Y/2+Y)^2}{2 x_1 \sin^2 \phi_1}~~~~~~(A5)\] 
  
  \par The case of a circular boundary is shown in Fig 12b. By symmetry, it is sufficient to consider
   the displacement, $Y$, of of the projection, O$_P$, of the Ox axis from the center of the circle, O$_C$,
   whose plane is perpendicular to Ox. Choosing the OY axis parallel to  $O_C O_P$, and a 
  boundary point, $B$, with polar coordinates ($R_1$, $\phi_1$) relative to O$_P$, the function 
  $r_1^{max}(d_i, \phi_1)$ is
   found, from the geomery of Fig.10b, to be:
  \[  r_1^{max}(x_1,R_B,Y,\phi_1) = x_1 + \frac{(\sqrt{R_B^2-Y^2 \cos^2 \phi_1}-Y \sin\phi_1)^2}{2 x_1}~~~(A6)\]
    where $R_B$ is the radius of the circular boundary.\

 \begin{figure}[htbp]
\begin{center}
\hspace*{-0.5cm}\mbox{
\epsfysize10.0cm\epsffile{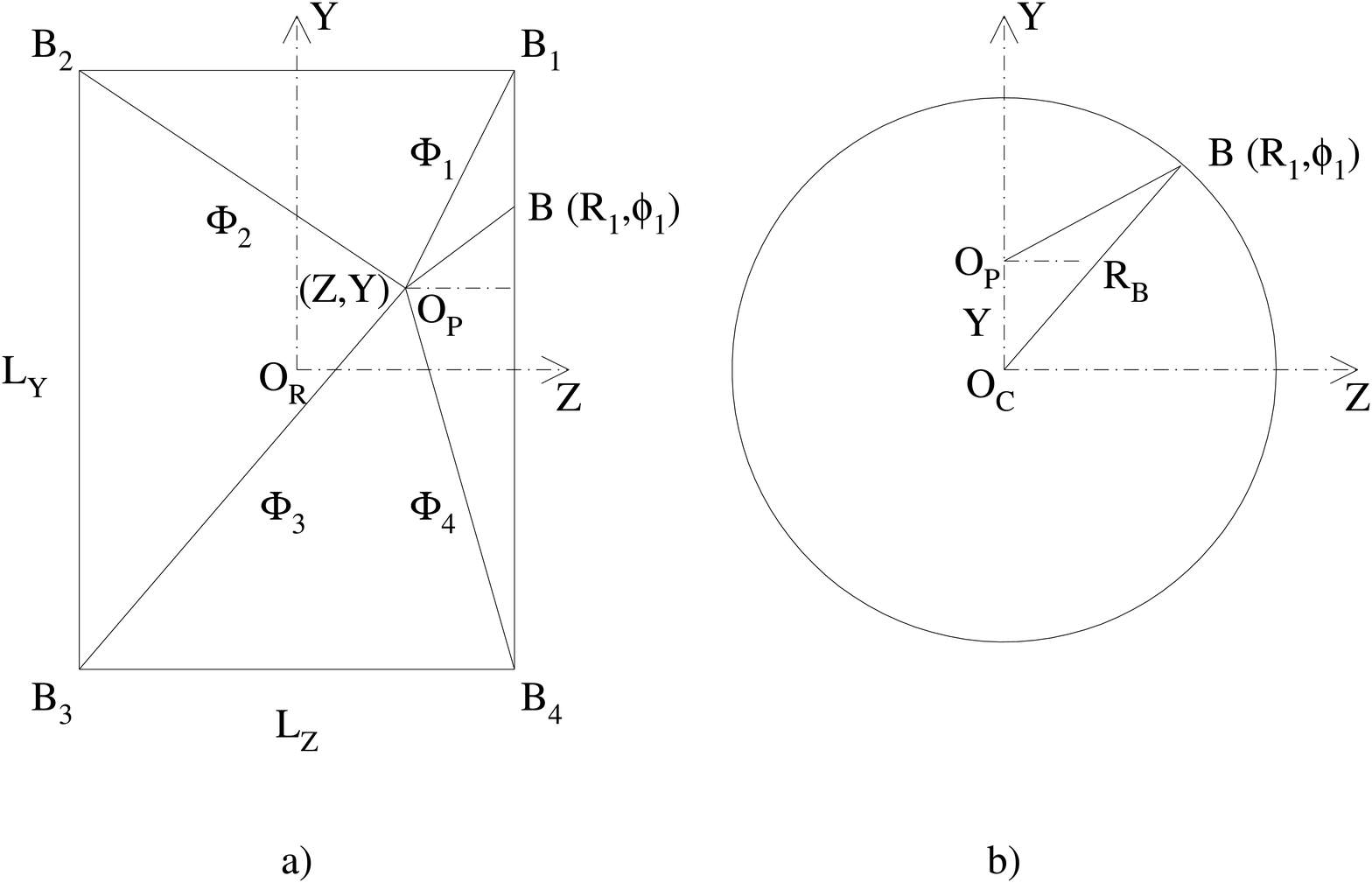}}
\caption{{\sl  Dependence of the upper limits of the $r_1$ integral in (5.2)
   on the azimuthal angle $\phi_1$ for different
   boundary geometries. $O_P$ is the projection of the classical rectilinear path joining the source 
   and detector on the YZ (transverse) plane. Various geometrical parameters used to specify
    $r_1^{max}$ are defined for the case of rectangular, a), and circular, b), boundaries.  }} 
\label{fig-fig12}
\end{center}
\end{figure}
  
 \par    The following contributions of the upper limit of the $r_1$ integral in (5.2) are then 
    found :
  \[ I_U = -\int_0^{2 \pi} \exp[i\kappa r_1^{max}(d_i, \phi_1) d \phi_1 \]
  \[ ~~~ =  -\int_0^{2 \pi}\left[ \cos\frac{2 \pi  r_1^{max}(d_i, \phi_1)}{\lambda_{\gamma}^0} +
  i\sin \frac{2 \pi  r_1^{max}(d_i, \phi_1)}{\lambda_{\gamma}^0}\right] d \phi_1 ~~~(A7) \]
  where
\[~~~d_i = x_1, L_Y, L_Z, Y, Z~~~~~ {\rm rectangular~boundary} \]
\[ d_i = x_1, R_B, Y~~~~~{\rm circular~boundary} \]
  In the typical case that $r_1^{max} \gg \lambda_{\gamma}^0$, a very rapid phase variation 
  of the integrand in (A7) occurs as a function of $\phi_1$, so that the integral $I_U$ 
   over this variable vanishes.
   \par An possible exception to this is the case $Y = 0$ in (A6), in which case:

  \[  r_1^{max}(x_1,R_B,0,\phi_1) = x_1 + \frac{R_B^2}{2 x_1}~~~~~~(A8)\]
     which is independent of  $\phi_1$, so that no cancellations due to rapid phase variation
    as a function of this variable occur.
   However, even in this case, both the position of O$_P$ must be specified, and the boundary radius
     $R_B$ must be constant, to within a fraction of $\lambda_{\gamma}^0$ in order to avoid the 
    cancellation of $I_U$ due to rapid phase variation of the sine and cosine 
    functions in (A7).

 \newpage
  {\bf Appendix B}

  \par The nested $n$ fold integral ${\cal I}_n(L)$ is defined as:
 \[ {\cal I}_n(L) =  \int_{x_2}^{\bar{x} +\frac{L}{2}} d x_1  \int_{x_3}^{\bar{x} +\frac{L}{2}} d x_2
  ... \int_{\bar{x} -\frac{L}{2}}^{\bar{x} +\frac{L}{2}} dx_n~~~(B1) \]
  Consider the incremental change: $L \rightarrow L + \delta L$. Substitution in (B1) gives:
  \begin{eqnarray}
  {\cal I}_n(L+\delta L) & = & \left( \int_{x_2}^{\bar{x} +\frac{L}{2}}d x_1
    +  \int_{\bar{x} +\frac{L}{2}}^{\bar{x} +\frac{L+\delta L}{2}}d x_1 \right)
   \left( \int_{x_3}^{\bar{x} +\frac{L}{2}}d x_2
    +  \int_{\bar{x} +\frac{L}{2}}^{\bar{x} +\frac{L+\delta L}{2}}d x_2 \right)
 \nonumber \\
      &   & ...\left( \int_{\bar{x} -\frac{L+\delta L}{2}}^{\bar{x} -\frac{L}{2}}d x_n
      +  \int_{\bar{x} -\frac{L}{2}}^{\bar{x} +\frac{L}{2}}d x_n
    +  \int_{\bar{x} +\frac{L }{2}}^{\bar{x} +\frac{L+\delta L}{2}}d x_n \right) 
   \nonumber \\
    & = &   {\cal I}_n(L) +i_1 + i_2 +...+i_j+ ...+i_n + i_- +O(\delta L^2) \nonumber ~~~(B2)
 \end{eqnarray}
   where
  \begin{eqnarray}
  i_1 & = & \frac{\delta L}{2} \int_{x_3}^{\bar{x} +\frac{L}{2}} d x_2 
   \int_{x_4}^{\bar{x} +\frac{L}{2}} d x_3...
    \int_{\bar{x} -\frac{L}{2}}^{\bar{x} +\frac{L}{2}} dx_n
     = \frac{\delta L}{2}{\cal I}_{n-1}(L) \nonumber~~~(B3) \\ 
    i_j & = & \int_{x_2}^{\bar{x} +\frac{L}{2}} d x_1 
   \int_{x_3}^{\bar{x} +\frac{L}{2}} d x_2...
 \int_{\bar{x} +\frac{L}{2}}^{\bar{x} +\frac{L+\delta L}{2}} d x_j...
    \int_{\bar{x} -\frac{L}{2}}^{\bar{x} +\frac{L}{2}} dx_n \nonumber~~~(B4) \\
    i_- & = & \int_{x_2}^{\bar{x} +\frac{L}{2}} d x_1 
   \int_{x_3}^{\bar{x} +\frac{L}{2}} d x_2...
 \int_{x_n}^{\bar{x} +\frac{L}{2}} d x_{n-1}
    \int_{\bar{x} -\frac{L+\delta L}{2}}^{\bar{x} -\frac{L}{2}} dx_n \nonumber~~~(B5)
  \end{eqnarray}
  Because of the linear dependence of each integral on its upper and lower limits, the 
  following replacement may be made in (B4):
   \[... \int_{x_j}^{\bar{x} +\frac{L}{2}} d x_{j-1} 
   \int_{\bar{x} +\frac{L}{2}}^{\bar{x} +\frac{L+\delta L}{2}} d x_j...
    \rightarrow ... \int_{\bar{x}_j}^{\bar{x} +\frac{L}{2}} \frac{\delta L}{2}... \]
      where
     \[\bar{x}_j = \bar{x} +\frac{L}{2}+ \frac{\delta L}{4} \]
     similarly
 \[ ... \int_{\bar{x}_{j-1}}^{\bar{x} +\frac{L}{2}} d x_{j-2}
\int_{\bar{x}_j}^{\bar{x} +\frac{L}{2}} \frac{\delta L}{2}
 ...   \rightarrow... \int_{\bar{x}_{j-1}}^{\bar{x} +\frac{L}{2}}
   \frac{\delta L}{4} \frac{\delta L}{2}...\]
      where   
    \[\bar{x}_{j-1} = \bar{x} +\frac{L}{2}+ \frac{\delta L}{8} \]
   It follows that $i_2$, $i_3$,...$i_n$ are at least of order $\delta L^2$.
   Similar considerations show that:
 \[    i_-  =  \int_{x_2}^{\bar{x} +\frac{L}{2}} d x_1 
   \int_{x_3}^{\bar{x} +\frac{L}{2}} d x_2...
 \int_{\bar{x}_n}^{\bar{x} +\frac{L}{2}} d x_{n-1} \frac{\delta L}{2}~~~(B6) \]
  where
  \[\bar{x}_n = \bar{x} -\frac{L}{2}- \frac{\delta L}{4} \]
  or 
   \begin{eqnarray}
    i_- &  =  & \int_{x_2}^{\bar{x} +\frac{L}{2}} d x_1 
   \int_{x_3}^{\bar{x} +\frac{L}{2}} d x_2...\left(
 \int_{\bar{x} -\frac{L}{2}}^{\bar{x} +\frac{L}{2}} d x_{n-1}+ 
    \int_{\bar{x} -\frac{L}{2}- \frac{\delta L}{4}}^{\bar{x} -\frac{L}{2}} d x_{n-1}
    \right) \frac{\delta L}{2} \nonumber \\
   & = &  \frac{\delta L}{2}{\cal I}_{n-1}(L)+ O(\delta L^2) \nonumber  ~~~(B7)
  \end{eqnarray}
   Since  $i_2$, $i_3$,...$i_n$ give no $O(\delta L)$ contribution, combining
 (B2),(B3) and (B7) then gives:
  \[{\cal I}_{n}(L+\delta L) = {\cal I}_{n}(L) + {\cal I}_{n-1}\delta L + O(\delta L^2)~~~(B8)\]
   or, from the Taylor expansion of ${\cal I}_{n}(L+\delta L)$,
   \[ \frac{ d{\cal I}_{n}(L)}{dL} =  {\cal I}_{n-1}(L)~~~(B9)\]
   hence
   \[ {\cal I}_{n}(L) = \int_{0}^{L} {\cal I}_{n-1}(L) dL~~~(B10)\]
    Since
   \[  {\cal I}_{1}(L) = \int_{\bar{x} -\frac{L}{2}}^{\bar{x} +\frac{L}{2}}d x_1 = L~~~(B11) \]
    then 
     \begin{eqnarray}
     {\cal I}_{2}(L) &  = &  \int_{0}^{L} LdL = \frac{L^2}{2} \nonumber \\
     {\cal I}_{3}(L) &  = &  \int_{0}^{L} \frac{L^2}{2} dL = \frac{L^3}{3!} \nonumber \\
      {\cal I}_{4}(L) &  = &  \int_{0}^{L} \frac{L^3}{3!} dL = \frac{L^4}{4!} \nonumber \\
           &   &....................................... \nonumber \\
           &   &....................................... \nonumber \\
    {\cal I}_{n}(L) &  = &  \int_{0}^{L} \frac{L^{n-1}}{(n-1)!} dL = \frac{L^n}{n!}
    \nonumber ~~~(B12)
    \end{eqnarray}
     which is (5.16) of the text.

 \newpage
  {\bf Appendix C}
  
  The nested $r_1$,$r_2$,$r_3$ integral in (4.30) is:
  \[ {\cal I}_{3}(r)= \int_{x_1-x_2}^{\Delta s-r_2-r_3+x_1} e^{i\kappa r_1}  d r_1
   \int_{x_2-x_3}^{\Delta s-r_3+x_2} e^{i\kappa r_2}  d r_2
   \int_{x_3}^{\Delta s+x_3} e^{i\kappa r_3}  d r_3 \]
    Performing the $r_1$ integral:
   \[ {\cal I}_{3}(r)= \frac{e^{i\kappa x_1}}{i\kappa} e^{i\kappa r_1}  d r_1
   \int_{x_2-x_3}^{\Delta s-r_3+x_2}\left[ e^{i\kappa(\Delta s-r_3)}
     -   e^{i\kappa(r_2- x_2)}\right] d r_2
   \int_{x_3}^{\Delta s+x_3} e^{i\kappa r_3}  d r_3 \]
    Performing the $r_2$ integral:
      \[ {\cal I}_{3}(r)= e^{i\kappa x_1}\int_{x_3}^{\Delta s+x_3}\left[
       \frac{(\Delta s-r_3+x_3)}{i\kappa} e^{i\kappa \Delta s}
      -\frac{(e^{i\kappa \Delta s}-e^{i\kappa(r_3-x_3)})}{(i\kappa)^2} \right] d r_3 \]
    Finally, performing the $r_3$ integral:
      \[ {\cal I}_{3}(r) =  e^{i\kappa x_1} \left(\frac{i}{\kappa}\right)^3
      \left[\left(-\frac{(i\kappa \Delta s)^2}{2}+i\kappa \Delta s -1\right)e^{i\kappa \Delta s}
        +1 \right]~~~~~~(C1) \]
      The factor $\exp[i\kappa x_1]$ is cancelled by the factor  $\exp[-i\kappa x_1]$ in (5.30)
      so that the integrand of the $x_1$, $x_2$ and $x_3$ integrals in this equation is constant.
      The integral is then given by (5.16) of the text as the $r$ and $x$ integrals
      factorise. The replacement $\kappa \Delta s = \Delta \Phi$ in (C1) yields (5.31) of the
      text. 
      \par The general expression for the nested $n$-fold $r$ and $x$ integrals
       obtained in a similar manner to the three-fold case just discussed is, in the 
      notation of (5.33):
      \[ {\cal I}_{n}(x,r) =  \frac{(i\beta L)^n}{n!}
       \left\{\left((-1)^n \frac{(i \Delta \Phi)^{n-1}}{(n-1)!}
       - (-1)^n \frac{(i \Delta \Phi)^{n-2}}{(n-2)!}+...
       -\frac{(i \Delta \Phi)^2}{2}+i \Delta \Phi -1\right) e^{i \Delta \Phi}+1 \right\}~~(C2)\]

 \newpage
  {\bf Appendix D}
  \par Multiplying $\exp[i\Delta \Phi]$ by the polynomials in $i\Delta \Phi$
   at each order in $\beta L$, the expresion in large curly brackets of (5.33) may be
    written as:
   \begin{eqnarray}
 F_{ref} & = & 1-i\beta L \left[i\Delta \Phi +\frac{(i\Delta \Phi)^2}{2!}+\frac{(i\Delta \Phi)^3}{3!}
   +...\right] \nonumber \\
     & + & \frac{(i\beta L)^2}{2!}\left[(i\Delta \Phi)^2 \left( \frac{1}{1!}-\frac{1}{2!}\right)+
     (i\Delta \Phi)^3 \left(\frac{1}{2!}-\frac{1}{3!}\right)+...\right]  \nonumber \\
     & + & \frac{(i\beta L)^3}{3!}\left[(i\Delta \Phi)^3 \left(-\frac{1}{2!}+\frac{1}{2!}-\frac{1}{3!}\right)+
     (i\Delta \Phi)^4 \left(-\frac{1}{2!2!}+\frac{1}{3!} -\frac{1}{4!} \right)+...\right]  \nonumber \\     
   & + & \frac{(i\beta L)^4}{4!}\left[(i\Delta \Phi)^4 \left(\frac{1}{3!}-\frac{1}{2!2!}
    +\frac{1}{3!}-\frac{1}{4!}\right)+
     (i\Delta \Phi)^5\left(\frac{1}{3!2!}-\frac{1}{2!3!} +\frac{1}{4!}- \frac{1}{5!}  \right)+...\right] +... \nonumber \\
   & + & \frac{(i\beta L)^n}{n!}\left[(i\Delta \Phi)^n\left(...-\frac{1}{(n-2)!2!}+\frac{1}{(n-1)!} 
    -  \frac{1}{n!}\right) \right. \nonumber \\
   & + & \left. (i\Delta \Phi)^{n+1}\left(...-\frac{1}{(n-1)!2!}+\frac{1}{n!} -\frac{1}{(n+1)!} \right)+...\right]
  \nonumber \\
   & +  & ... ~~~~~~~~~(D1) \nonumber
 \end{eqnarray}

   Separating out the real and imaginary parts at each order in  $\beta L$ the gives the expression:
   \begin{eqnarray}
 F_{ref} & = & 1+\beta L \left[\Delta \Phi -\frac{(\Delta \Phi)^3}{3!} +...
         +i\left( \frac{(\Delta \Phi)^2}{2!}-\frac{(\Delta \Phi)^4}{4!} +...\right) \right]
      \nonumber \\
      & + & \frac{(\beta L)^2}{2!} \left[\Delta \Phi( \Delta \Phi-\frac{(\Delta \Phi)^3}{3!}+...)
        -\frac{(\Delta \Phi)^2}{2!}+\frac{(\Delta \Phi)^4}{4!}-...\right. \nonumber \\
      & + & \left. i\left(\frac{(\Delta \Phi)^3}{3!}-\frac{(\Delta \Phi)^5}{5!}+...
           + \Delta \Phi(\frac{(\Delta \Phi)^2}{2!}-\frac{(\Delta \Phi)^4}{4!}+...\right)\right]
       \nonumber \\
      & + &  \frac{(\beta L)^3}{3!} \left[\frac{(\Delta \Phi)^3}{3!}-\frac{(\Delta \Phi)^5}{5!}+...
        +  \Delta \Phi(-\frac{(\Delta \Phi)^2}{2!}+\frac{(\Delta \Phi)^4}{4!}-...)
        + \frac{(\Delta \Phi)^2}{2!}(\Delta \Phi-\frac{(\Delta \Phi)^3}{3!}+...) \right.
       \nonumber \\
     & + & \left.  i\left(\frac{(\Delta \Phi)^4}{4!}-\frac{(\Delta \Phi)^6}{6!}+...
        + \Delta \Phi(-\frac{(\Delta \Phi)^3}{3!}+\frac{(\Delta \Phi)^5}{5!}-...)
        + \frac{(\Delta \Phi)^2}{2!}(\frac{(\Delta \Phi)^2}{2!}-\frac{(\Delta \Phi)^4}{4!}+...)\right)
       \right]   \nonumber \\
      & + &  \frac{(\beta L)^4}{4!}\left[-\frac{(\Delta \Phi)^4}{4!}+\frac{(\Delta \Phi)^6}{6!}-...
            +\Delta \Phi(\frac{(\Delta \Phi)^3}{3!}-\frac{(\Delta \Phi)^5}{5!}+...) \right.
         \nonumber \\
       & + & \frac{(\Delta \Phi)^2}{2!}(-\frac{(\Delta \Phi)^2}{2!}+\frac{(\Delta \Phi)^4}{4!}-...)
           +\frac{( \Delta \Phi)^3}{3!}(\Delta \Phi-\frac{(\Delta \Phi)^3}{3!}+...)    \nonumber \\
       & + & + i\left(-\frac{(\Delta \Phi)^5}{5!}+\frac{(\Delta \Phi)^7}{7!}-...
        + \Delta \Phi(\frac{\Delta \Phi)^4}{4!}-\frac{(\Delta \Phi)^6}{6!}+...) \right.
          \nonumber \\
       & + & \left. \left. \frac{(\Delta \Phi)^2}{2!}(-\frac{(\Delta \Phi)^3}{3!}+\frac{(\Delta \Phi)^5}{5!}-...)
           +\frac{( \Delta \Phi)^3}{3!}(\frac{(\Delta \Phi)^2}{2!}-\frac{(\Delta \Phi)^4}{4!}+...)\right) \right]
         +... \nonumber ~~~~~~(D2)
       \end{eqnarray}     
  Noting that all the infinite series in $\Delta \Phi$ in (D2) are sine or cosine series, possibly with sequences
  of missing low order terms,
  enables the right side to be written as in the large curly brackets of (5.35).

  \newpage

   Factoring out
  the terms proportional to $S \equiv \sin \Delta \Phi$ and $C \equiv \cos \Delta \Phi$ at each order in $\beta L$,
  D2 may then be re-written as:
  \begin{eqnarray}
 F_{ref} & = & 1+\beta L [S +i(1-C)] \nonumber \\
      & + & \frac{(\beta L)^2}{2!} \left[C-1+S\Delta \Phi+i(S-C \Delta \Phi)\right] \nonumber \\
      & + &  \frac{(\beta L)^3}{3!} \left[-S(1-\frac{(\Delta \Phi)^2}{2!})+C\Delta \Phi 
             +i\left(C(1-\frac{(\Delta \Phi)^2}{2!})+S\Delta \Phi-1\right)\right]   \nonumber \\
      & + &  \frac{(\beta L)^4}{4!}\left[1-C(1-\frac{(\Delta \Phi)^2}{2!})-S(\Delta \Phi-\frac{(\Delta \Phi)^3}{3!})
               \right.  \nonumber \\
      & + & \left. i \left(C(\Delta \Phi-\frac{(\Delta \Phi)^3}{3!})-S(1-\frac{(\Delta \Phi)^2}{2!})\right) \right] \nonumber \\
     & + &  \frac{(\beta L)^5}{5!}\left[-C(\Delta \Phi-\frac{(\Delta \Phi)^3}{3!})+S(1-\frac{(\Delta \Phi)^2}{2!}
     + \frac{(\Delta \Phi)^4}{4!}) \right.  \nonumber \\
      & + & \left. i \left(1-S(\Delta \Phi-\frac{(\Delta \Phi)^3}{3!})-C(1-\frac{(\Delta \Phi)^2}{2!}
      + \frac{(\Delta \Phi)^4}{4!})\right)\right]  \nonumber \\
     & + &... \nonumber ~~~~~~~~~~~~~~~~~~~~~(D3)
 \end{eqnarray}
  The polynomials in $\Delta \Phi$ that multiply $S$ and
    $C$ at each order in $\beta L$ are the truncated sine and cosine series
    $S_j$ and $C_j$ respectively, defined in (5.36) and (5.37).
    Substitution of these series in (D3) yields then (5.38) and (5.39) of the text.

 \newpage
  {\bf Appendix E}
  \par It is assumed in the discussion of the Michelson interferometer in Section 8 that the source atom
  is at rest. In the case that the source is in motion in the laboratory frame the time
   factors in the complex exponentials of (8.1) and (8.2) must be scaled by the factor $1/\gamma$
 where $\gamma \equiv 1/\sqrt{1-(v/c)^2}$ and $v$ is the source atom velocity, to take into
  account relativistic time dilatation. This is because the correct argument of the space-time
  propagator (see (3.11)) is the proper time $\tau$, and $\Delta \tau = \Delta t/\gamma$
  where $t$ is the laboratory time.
  \par Assuming the Maxwellian distribution:
  \[ \frac{dN}{dp} = Cp^2 e^{-\frac{p^2}{\overline{p}^2}} ~~~~~~~~~~~~~~~~(E1)\]
  for the momentum, $p$, of the source atom, of mass, $M$, at absolute temperature, $T$,
   where $\overline{p}^2 = 2MkT$,
   enables the distribution of the
  relativistic parameter $\gamma$ to be calculated via the relation:
  \[ \gamma = \frac{E}{M} = \frac{\sqrt{M^2+p^2}}{M} = 1+ \frac{1}{2} \frac{p^2}{M^2}+ ...~~~~~~~~~~~(E2)\]
  Modifying the argument of the cosine in
   (8.7) to take into account the motion of the source atom gives
\[ \cos(2 \kappa d) \rightarrow  \cos(\frac{2 \kappa d}{\gamma}) \simeq  \cos(2 \kappa d- \frac{\kappa p^2 d}{M^2})
   ~~~~~~~~~~~~~(E3) \]
 Performing the average over $p$ using the distribution (E1):
 \[ \langle  \cos(\frac{2 \kappa d}{\gamma}) \rangle = Re~ e^{2i \kappa d} \left\{
  \frac{\int_0^{\infty} e^{-\frac{p^2}{\overline{p}^2}}  e^{-\frac{i \kappa  p^2 d}{M^2}} dp}
    {\int_0^{\infty}  e^{-\frac{p^2}{\overline{p}^2}} dp} \right\}~~~~~~~~~(E4) \]  
 Performing the integrals (see~\cite{Jeans}) gives the result:
  \[ \langle  \cos(\frac{2 \kappa d}{\gamma}) \rangle = Re~\frac{k^{-\frac{3}{2}} e^{2 i \kappa d}}
    {\overline{p}^3}~~~~~~~~~~~~~(E5) \]
  where 
  \[ k = \frac{1}{\overline{p}^2}[1+ i \kappa\left(\frac{\overline{p}}{M}\right)^2 d] \]
  Since $\kappa\left(\frac{\overline{p}}{M}\right)^2 d \ll 1$, it follows that:
 \[ k^{-\frac{3}{2}} \simeq \overline{p}^3[1-\frac{3}{2}i \kappa\left(\frac{\overline{p}}{M}\right)^2 d]
  \simeq   \overline{p}^3 \exp\left[-\frac{3}{2}i \kappa\left(\frac{\overline{p}}{M}\right)^2 d\right]~~~~~~~~~(E6) \]
  Combining (E5) and (E6) gives finally:
 \[ \langle  \cos(\frac{2 \kappa d}{\gamma}) \rangle = \cos \left[ 2 \kappa d(1- 
  \frac{3}{4}\left(\frac{\overline{p}}{M}\right)^2)\right]~~~~~~~~~~~(E7) \]
  Thus the only effect of random Maxwellian motion of the source atom is the small relative change $-(3/4)(\overline{p}
   /M)^2$ of the phase of the interference term. There is no damping of the fringe visibility due
    to Doppler smearing of the wavelength of light as predicted by the classical wave theory.
     The numerical size of the correction term $3/4(\overline{p}/M)^2$ for a sodium atom at NTP is
     $1.6 \times 10^{-12}$.

  \newpage
  {\bf Appendix F}
  \par   For convenience, units with $\hbar = c =1$ are employed in this Appendix.
   The following variables are used to characterise the velocities $v_A$ and $v_B$ of the electron in
   path A or path B:
  \[ \alpha \equiv \frac{ t_A}{ t_B},~~~\beta \equiv  t_A- t_B ~~~~~~~~~~~~~(F1) \]
    The velocities are given, in terms of these variables and the path lengths $r'+r_A$ and $r'+r_B$ by
 the relations:
   \[ v_A =\frac{(1-\alpha)(r'+r_A)}{\beta}~~,~~ v_B =\frac{(1-\alpha)(r'+r_B)}{ \alpha \beta}~~~~~~~~(F2) \]
   The physical ranges of $\alpha$ and $\beta$ are:
    \begin{eqnarray}
     1 <  & \alpha & < \frac{(r'+r_B)}{(r'+r_A)} \equiv \alpha_{max} \nonumber~~~~~~~~~(F3) \\
     0 < & \beta & < \infty  \nonumber~~~~~~~~~~~~~~(F4)
   \end{eqnarray}
   The limit $\beta \rightarrow \infty$ corresponds to vanishingly small velocity.  The values $\alpha = 1$, $\beta = 0$
   (in which case the ratio $(1-\alpha)/\beta$ remains finite) correspond to equal production times for any allowed velocity,
    whereas $\alpha = \alpha_{max}$, for any $\beta$, corresponds to equal velocities. The equations (F2) give
    for the velocity difference:
    \[ v_B -v_A = \frac{r'(1-\alpha)}{\beta}\left[(\frac{1}{\alpha}-1)+ \frac{1}{r'}(\frac{r_B}{\alpha}-r_A)\right]~~~~~~(F5) \]
      and for the average velocity:
     \[ \overline{v} = \frac{v_B +v_A}{2} = 
     \frac{r'(1-\alpha)}{2 \beta}\left[(\frac{1}{\alpha}+1)+ \frac{1}{r'}(\frac{r_B}{\alpha}+r_A)\right]~~~~(F6) \]
     Combining (F5) and (F6):
    \begin{eqnarray}   
     v_B -v_A & = &   2  \overline{v}\frac{(\frac{1}{\alpha}-1)+ \frac{1}{r'}(\frac{r_B}{\alpha}-r_A)}
      {(\frac{1}{\alpha}+1)+ \frac{1}{r'}(\frac{r_B}{\alpha}+r_A)}  \nonumber \\
      & = & \overline{v} \frac{(r'+r_A)(\frac{\alpha_{max}-1}{\alpha})}{(r'+\overline{r})} + O[(\Delta r)^2]
       \nonumber \\
      & \equiv & 2 \delta_v + O[(\Delta r)^2]~~~~~~~~~~~~~~~~~~(F7) \nonumber
    \end{eqnarray}
    To first order in $\Delta r$ :

    \[ v_A = \overline{v}-\delta_v~~~,~~~  v_B = \overline{v}+\delta_v~~~~~~~~~~~~~~(F8) \]
    From the relativistic relation:
   \[ v = \frac{p}{E} = \frac{p}{\sqrt{(m^2+p^2)}} \]
   it follows that $\delta_v = (m^2/E^3) \delta_p$ so that
     \[ p_A = \overline{p}-\delta_p~~~,~~~  p_B = \overline{p}+\delta_p~~~~~~~~~~~~~~(F9) \]
     where
    \[ \delta_p = \frac{\overline{E}^3}{m^2} \delta_v = \frac{ \overline{p}(1+(\frac{\overline{p}}{m})^2)(r'+r_A)
  (\frac{\alpha_{max}}{\alpha}-1)}{2(r'+\overline{r})} ~~~~~~~~~(F10) \]
   Substituting now (F9) into the formula (9.15) for the phase difference, retaining only 
   first order terms in $\delta_p$, and using (F10), gives:
     \begin{eqnarray} 
     \phi_B - \phi_A & = & m^2 \left[ -\frac{\Delta r}{\overline{p}}
     +2(r'+\overline{r})\frac{\delta_p}{\overline{p}^2}\right] \nonumber \\
 & = & m^2 \left[ -\frac{\Delta r}{\overline{p}}+\left(\frac{\overline{p}}{m^2}+\frac{1}{\overline{p}}\right)
      (r'+r_A)(\frac{\alpha_{max}}{\alpha}-1) \right]  \nonumber~~~~~~~~~(F11) 
   \end{eqnarray}
    Substituting $\alpha = \alpha_{max}$ in (F11), corresponding to $v_A = v_B$, gives
   \[  \phi_B - \phi_A  = -\frac{ m^2 \Delta r}{\overline{p}}~~~({\rm equal~velocities})~~~~~~~~~~~(F12) \]
  while setting $\alpha = 1$ or $ t_A =  t_B$ gives:
  \[  \phi_B - \phi_A  = \overline{p} \Delta r~~~({\rm equal~production~times})~~~~~~(F13) \]
    For non-relativistic particles for which $\overline{p} \ll m$ the equal velocity relation (F12)
    gives an effective de Broglie wavelength much shorter than (9.19) that is applicable in the
    equal production time case, (F13), and which is in good agreement with experiment~\cite{DG}.
.

\end{document}